\documentclass[12pt]{article}
\usepackage[utf8]{inputenc}
\usepackage[T1]{fontenc}
\usepackage{url}
\usepackage[noend]{algpseudocode}
\usepackage{verbatim}
\usepackage{amsmath,amssymb}
\usepackage{amsthm}
\usepackage{courier}
\usepackage{physics}
\usepackage[pdftex]{graphicx}
\usepackage{booktabs}
\usepackage{natbib}
\usepackage{float}
\usepackage{xcolor}
\usepackage{array}
\usepackage{caption}
\usepackage{subcaption}
\usepackage{setspace}
\usepackage{tabto}
\usepackage{lineno}
\usepackage{bbold}
\usepackage{bm}
\usepackage[shortlabels]{enumitem}
\usepackage{quantikz}

\newcommand*\patchAmsMathEnvironmentForLineno[1]{%
      \expandafter\let\csname old#1\expandafter\endcsname\csname #1\endcsname
      \expandafter\let\csname oldend#1\expandafter\endcsname\csname end#1\endcsname
      \renewenvironment{#1}%
         {\linenomath\csname old#1\endcsname}%
         {\csname oldend#1\endcsname\endlinenomath}}%
    \newcommand*\patchBothAmsMathEnvironmentsForLineno[1]{%
      \patchAmsMathEnvironmentForLineno{#1}%
      \patchAmsMathEnvironmentForLineno{#1*}}%
    \AtBeginDocument{%
    \patchBothAmsMathEnvironmentsForLineno{equation}%
    \patchBothAmsMathEnvironmentsForLineno{align}%
    \patchBothAmsMathEnvironmentsForLineno{flalign}%
    \patchBothAmsMathEnvironmentsForLineno{alignat}%
    \patchBothAmsMathEnvironmentsForLineno{gather}%
    \patchBothAmsMathEnvironmentsForLineno{multline}%
    }
    \include{def}
\def\dispmuskip{\thinmuskip= 3mu plus 0mu minus 2mu \medmuskip=  4mu plus 2mu minus 2mu \thickmuskip=5mu plus 5mu minus 2mu}
\def\textmuskip{\thinmuskip= 0mu                    \medmuskip=  1mu plus 1mu minus 1mu \thickmuskip=2mu plus 3mu minus 1mu}
\def\beq{\dispmuskip\begin{equation}}    \def\eeq{\end{equation}\textmuskip}
\def\beqn{\dispmuskip\begin{displaymath}}\def\eeqn{\end{displaymath}\textmuskip}
\def\bea{\dispmuskip\begin{eqnarray}}    \def\eea{\end{eqnarray}\textmuskip}
\def\bean{\dispmuskip\begin{eqnarray*}}  \def\eean{\end{eqnarray*}\textmuskip}

\newcounter{mntcomm}

\newcounter{alcomm}

\def\tr{\text{\rm tr}}

\def\argmin{\text{\rm argmin}}
\def\argmax{\text{\rm argmax}}

\def\i{{\rm i}}

\usepackage[margin=2cm]{geometry}

\begin{document}

\title{An Introduction to Quantum Computing for Statisticians and Data Scientists}

\author{Anna Lopatnikova\thanks{Discipline of Business Analytics, the University of Sydney Business School. The research was partially supported by the Australian Research Council's Discovery Project DP200103015, the ARC Centre for Data Analytics for Resources and Environments (DARE), the ARC Centre of Excellence for Mathematical and Statistical Frontiers (ACEMS) and a USYD Business School research support scheme. Corresponding to \texttt{minh-ngoc.tran@sydney.edu.au}.}
\and Minh-Ngoc Tran\footnotemark[1]
\and Scott A. Sisson\thanks{ UNSW Data Science Hub and School of Mathematics and Statistics, University of New South Wales, Sydney. SAS is supported by the ARC through the Discovery Project scheme FT170100079 and ACEMS.}}

\date{2nd version: March 2022}
\maketitle
\begin{abstract}
    Quantum computers promise to surpass the most powerful classical supercomputers when it comes to solving many critically important practical problems, such as pharmaceutical and fertilizer design, supply chain and traffic optimization, or optimization for machine learning tasks.  Because quantum computers function fundamentally differently from classical computers, the emergence of quantum computing technology will lead to a new evolutionary branch of statistical and data analytics methodologies.   This review provides an introduction to quantum computing designed to be accessible to statisticians and data scientists, aiming to equip them with an overarching framework of quantum computing, the basic language and building blocks of quantum algorithms, and an overview of existing quantum applications in statistics and data analysis.  Our goal is to enable statisticians and data scientists to follow quantum computing literature relevant to their fields, to collaborate with quantum algorithm designers, and, ultimately, to bring forth the next generation of statistical and data analytics tools.     \\
    
\end{abstract}


\tableofcontents


\section{Introduction}

Quantum computing has emerged as the next computing technology paradigm, currently in the state of development reminiscent of classical transistor-based computers in the 1950s.  Quantum computers in existence today are small and noisy, but they are capable of proof-of-concept computation, with cloud-based access available to academics and industrial users.  The capacity of quantum computers is expanding every year.   The library of quantum algorithms is growing and currently includes efficient solutions to linear systems of equations, optimization, Markov chain Monte Carlo, principal component analysis, machine learning models with correlations not possible to model classically, and other methods of interest to statisticians and data scientists.  These exciting developments will enable us in the next decade and beyond to solve problems we are currently too computationally constrained to solve.    

The development of quantum computers will lead to new statistical methodologies, because quantum computers behave very differently from classical computers.   Modern statistical methods co-evolved with classical computing technologies.   Markov chain Monte Carlo, gradient descent, or re-sampling optimize the computational properties of classical computers, such as the ability to copy any data structure or to reset individual bits.  Quantum computers do not possess these capabilities, which, through our daily experience with classical computers, we have come to take for granted.  Instead, they offer quantum superposition, quantum parallelism, quantum interference, and quantum entanglement, unique properties quantum computers exploit as a resource, promising to deliver game-changing computational power for many classes of important problems.

Because quantum computers differ fundamentally from classical computers, statistical methodologies taking full advantage of them will differ from classical statistical methodologies.   Development of quantum statistical methods will require a collaboration between statisticians with a deep understanding of the requirements, end goals, and tradeoffs of existing statistical methods and quantum algorithm designers adept at harnessing the power of quantum computers despite their considerable quirks.   To be successful, these collaborations will require a shared language and a shared core of frameworks.

The objective of this review is to equip statisticians with the language and basic frameworks of quantum computing to enable them to understand, at a high level, the rapidly evolving state of the art of quantum computing and engage in successful collaborations with quantum algorithm designers.  It is our hope that these collaborations will bring forth an exciting next generation of statistical methods -- quantum statistical methods. 

The review is designed to be accessible and self-contained, emphasizing quantum theory's linear algebra foundations, deeply familiar to statisticians and data scientists.   This extended version of the review discusses quantum algorithms and their building blocks in greater detail.  

The first part of the review, Sections \ref{sec:theory} through \ref{sec:programming_qcs}, sets out the concepts necessary to understand the state of the art of quantum computing.  Section~\ref{sec:theory} starts with a birds-eye view of quantum theory that serves an overarching framework for the building blocks of quantum computing. Section~\ref{sec:buildingblocks} lays out the necessary details of quantum theory and the critical properties of quantum computers, with concrete examples throughout to illustrate theoretical concepts.  Section~\ref{sec:overview} provides a framework for quantum algorithm design and includes an high overview of quantum algorithms (detailed in Sections~\ref{sec:grover_section} through \ref{sec:qsvt}) and a review of Quantum Machine Learning, a vibrant area of quantum algorithm research.  Section~\ref{sec:programming_qcs} provides a guide to accessing and programming physical quantum computers and includes an extended discussion of quantum gates and other quantum programming primitives.  
 
The second part of the review, Sections \ref{sec:grover_section} through \ref{sec:qsvt}, lays out quantum algorithms of interest to statisticians and data scientists.  Quantum computing is a rapidly evolving field where every year new, more efficient methods surpass the previous year's cutting edge.  To address this problem of rapid obsolescence, when selecting algorithms for this review, we have sought to provide a balance between algorithms considered seminal and the promising recent algorithms.  

Section \ref{sec:grover_section} starts with Grover's search algorithm -- a seminal algorithm, whose core idea gave rise to many widely-used quantum routines including Quantum Amplitude Amplification and Quantum Amplitude Estimation. We describe how these routines can be used to speed up calculation of statistical descriptive quantities such as sample mean and sample median. We then describe the Quantum Monte Carlo method for estimating the probability expectation of a function. This method, an application of Quantum Amplitude Estimation, offers a provable quadratic speedup over the classical Monte Carlo method.

Section \ref{sec:quantum_walks} presents quantum Markov chains (known in the literature as quantum walks).  Similar to classical Markov chains, they power many statistical applications, including quantum Markov chain Monte Carlo.

Section \ref{sec:Quantum Linear Systems} reviews quantum algorithms for linear algebra computations including solving a system of linear equations and Principle Component Analysis, central to statistics and machine learning. 

Section \ref{sec:Hamiltonian Simulation} presents Hamiltonian simulation. While Hamiltonian simulation may not be of direct interest to statisticians, it is a critical subroutine for quantum linear algebra (Section \ref{sec:Quantum Linear Systems}) and is at the heart of potentially transformational applications of quantum computing in vital fields such as chemistry, agriculture, and energy.

Section \ref{sec:Quantum Optimization} discusses quantum algorithms to speed up optimization, which plays a critical role in quantum machine learning.  The section describes Adiabatic Quantum Computation and the Quantum Approximate Optimization Algorithm (QAOA).   QAOA belongs to the class of hybrid quantum-classical variational algorithms, an important class of algorithms that combine the strengths of quantum and classical computers, which may provide the way to harness the quantum advantage of the current generation of quantum computers.  The section also discusses quantum approaches to gradient descent.

Section \ref{sec:qsvt} describes Quantum Singular Value Transformation (QSVT) -- a cutting-edge algorithm at the time of writing, which enables polynomial transformations of singular values.   QSVT is a promising recent framework that encompasses other popular quantum algorithms -- such as search, Hamiltonian simulation, or systems of linear equations -- as specific instances.  Because of its flexibility and expressivity, we expect QSVT to give rise to other influential algorithms of interest to statisticians and data scientists.

Section \ref{sec:conclusion} concludes.

\section{A Birds-Eye View of Quantum Theory}
\label{sec:theory}

Richard Feynman, one of the best-known Nobel Prize-winning physicists, quipped that nobody understands quantum theory. Because the laws of quantum theory describe what happens at the atomic scale -- i.e.~at distances close to the size of a typical atom -- they do not comport with human intuition developed on the basis of our experience with everyday objects. At the atomic scale, nature is probabilistic, but the objects around us comprise septillions of atoms\footnote{A cup of water contains around $25\times 10^{24}$ atoms – 25 septillion.}. Figuratively speaking, what we experience in our everyday lives are large-sample properties of quantum theory rather than its ``small-sample'' properties.  But even our probability theory intuitions do not furnish solid analogies, because quantum theory has properties, such as quantum interference, that do not fit into classical probabilistic frameworks.  Nevertheless, even though quantum theory is highly counter-intuitive, it describes the behavior of particles at the atomic scale to an astonishing degree of accuracy.   It is humanity’s best-tested scientific theory, withstanding a hundred years of rigorous experimental tests \citep[see, e.g.,][]{Peskin:1995ev}.

The core of quantum theory is linear algebra on complex vector spaces, called \emph{Hilbert spaces}, which can be finite- or infinite-dimensional.   Within this framework, any closed quantum system can be described as a vector in a Hilbert space.   The vector, called a \emph{pure quantum state} (often called a \emph{``quantum state''} as a shorthand), is denoted as $\ket{\psi}$, where the object $\ket{\bullet}$ is called a \emph{ket}:
\begin{align}
 \ket{\psi} \equiv \begin{pmatrix} \psi_1 \\ \psi_2 \\ : \\ \psi_n \\ : \end{pmatrix}.
\end{align}
The coordinates $\psi_m$ are complex numbers, $\psi_m \in \mathbb{C}$, called \emph{amplitudes}.  The quantum state $\ket{\psi}$ can be written as a linear combination of basis vectors, which are also quantum states -- i.e. vectors in the Hilbert space.  In quantum theory, a linear combination of quantum states is called a \emph{quantum superposition}.  Denoting the standard basis vectors as $\ket{\bf{e}_m}$, we have
\begin{align}
    \ket{\psi} = \psi_1\ket{\bf{e}_1}+\psi_2\ket{\bf{e}_2}+\ldots+\psi_n\ket{\bf{e}_n}+\ldots,
\end{align}
i.e.~we expressed the state $\ket{\psi}$ is a \emph{superposition} -- a linear combination -- of the basis states $\ket{\bf{e}_m}$.

The inner product of two vectors $\ket{\phi}$ and $\ket{\psi}$ in a Hilbert space is analogous to the inner product in the Euclidean space and requires the complex conjugation and transposition of the left vector.   The conjugate-transpose of $\ket{\phi}$ is $\bra{\phi} = (\phi_1^*,\phi_2^*,..,\phi_n^*,..)$ (the object $\bra{\bullet}$ is called a \emph{bra}), where the asterisk $*$ indicates complex conjugation.   The inner product between two states  $\ket{\phi}$ and  $\ket{\psi}$ is denoted as $\braket{\phi}{\psi} = \sum_m\phi_m^*\psi_m$.   The quantum states are normalized, so that, for any $\ket{\psi}$, we have $\braket{\psi}=\sum_m|\psi_m|^2 = 1$.     

The normalization equation $\sum_m|\psi_m|^2 = 1$ suggests the interpretation of the (non-negative) squared amplitudes $|\psi_m|^2$ as probabilities and, in fact, the squared amplitudes do play this role in quantum theory.  The squared amplitudes $|\psi_m|^2$ determine the probability of the outcomes of \emph{quantum measurement} -- a critical part of quantum theory, discussed more formally in Section \ref{sec:q-statesandops}.  

Quantum measurement is the way a classical observer characterizes an unknown quantum state.  For an arbitrary quantum state and reference basis, the outcomes of a quantum measurement are probabilistic. For example, if the observer measures $\ket{\psi}$ in the standard basis, the measurement yields an outcome $m$ corresponding to the state $\ket{\bf{e}_m}$ with probability $|\psi_m|^2$.  After the measurement, the quantum state collapses to the basis vector $\ket{\bf{e}_m}$ corresponding to the outcome $m$.   A helpful classical analogy is a closed box with a object of an unknown color and a probability distribution over the set of possible colors.  Once the box is opened and the color of the object is revealed, the probability distribution for the color of the object in the box collapses to certainty for the observed color.  We can characterize the probability distribution over the set of possible colors by performing experiments on identically prepared boxes.   Similarly, we can characterize a quantum state by repeatedly preparing it and taking quantum measurements of it.  

An important aspect of quantum theory, which makes it both counter-intuitive and powerful for developing efficient computation, is that quantum states evolve according to their amplitudes rather than to the corresponding probabilities, i.e.~the squared amplitudes.  Amplitudes can take negative and, more generally, complex values and, as a result, make probability masses cancel out when quantum states undergo transformations.  Quantum transformations are linear.  A quantum state $\ket{\psi^A}$ transforms into a state in the same Hilbert space $\ket{\psi^B}$ as $\ket{\psi^A} = U\ket{\psi^B}$, where $U$ is a linear (more specifically, unitary) operator (Section \ref{sec:q-statesandops}).    The cancellation of amplitudes under linear transformations is called \emph{quantum interference}, a phenomenon that plays a critical role in efficiency of quantum computers, as discussed in Section~\ref{sec:power}. 

A \emph{quantum algorithm} $\mathcal{A}$ is a series of linear transformations that transform the input state $\ket{\psi^{in}}$ into an output state $\ket{\psi^{out}}$:  
\begin{align}
	\ket{\psi^{out}} = \mathcal{A} \ket{\psi^{in}}.
\end{align}
The quantum state $\ket{\psi^{out}}$ encodes the desired output, to be passed on to another quantum or classical algorithm.  The power of quantum computers is that, in many cases, they can perform the transformation $ \mathcal{A} $ very efficiently, promising to dramatically reduce computational time for many classes of problems as we outline in Section~\ref{sec:algos_overview} and detail in Sections~\ref{sec:grover_section} through \ref{sec:qsvt}.

\section{Basic Concepts of Quantum Computation}
\label{sec:buildingblocks}

Having outlined the high-level framework for quantum algorithms in the previous section, we now proceed to lay out the basic concepts of quantum computing necessary to understand how efficient quantum algorithms harness the counter-intuitive properties of quantum theory to deliver computational power.  We start with the most basic unit of a quantum computer -- the qubit.

\subsection{The Simplest Quantum System: a Qubit}
\label{sec:qubit}

The basic unit of information in a classical computer is a bit, taking either 0 or 1 values.
The basic information-storage unit of a quantum computer is a \emph{qubit}, which is a two-dimensional system 
\begin{align}
    \ket{q} = a \ket{0} + b \ket{1},\label{eq:q}
\end{align}
where $\ket{0}$ and $\ket{1}$ form an orthonormal basis of this two-dimensional space.\footnote{The orthonormality is understood in the usual sense of linear algebra.}  The basis state $\ket{0}$ is a shorthand for $\begin{pmatrix} 1 \\ 0 \end{pmatrix}$ and
$\ket{1}$ is a shorthand for $\begin{pmatrix} 0 \\ 1 \end{pmatrix}$. The qubit state $\ket{q}$ in (\ref{eq:q}) represents $\begin{pmatrix} a \\ b \end{pmatrix}$, a normalized vector in the two-dimensional Hilbert space that describes the state of the single-qubit quantum system.  The coefficients $a$ and $b$ are the complex-valued amplitudes, $a,b \in \mathbb{C}$, normalized so that $|a|^2 +|b|^2 =1$.  The state $\ket{q}$ is a linear superposition of the basis states $\ket{0}$ and $\ket{1}$.  

A quantum measurement of $\ket{q}$ in the basis of $\{\ket{0},\ket{1}\}$ yields $0$ with probability $|a|^2$ and $1$ with probability $|b|^2$ (more on quantum measurement in Section~\ref{sec:q-statesandops}).   The state $\ket{q}$ is analogous to a random variable taking the classical bit values $0$ and $1$ with probabilities $|a|^2$ and $|b|^2$ respectively, but there is a crucial difference:  the coefficients $a$ and $b$ can be negative, and more generally, complex. As we discuss below, this property is crucial to the power of quantum computers.

The superposition property extends to collections of multiple qubits, sometimes referred to as \emph{quantum registers}.  For example, a register of three qubits can support quantum states of the form
\begin{align}
    \ket{\psi} = &  \psi_{000}\ket{000}+\psi_{001}\ket{001} + \psi_{010}\ket{010} + \psi_{011}\ket{011} \nonumber\\ 
    & + \psi_{100}\ket{100} + \psi_{101}\ket{101}  +\psi_{110}\ket{110} + \psi_{111}\ket{111}.\label{eq:3q_state}
\end{align}
The states $\ket{b_1b_2b_3}$, where $b_k = \{1,0\}$ for $k = 1,2,3$,  form an orthonormal basis, and the 3-qubit register supports $2^3 = 8$ dimensional quantum states -- superpositions of $8$ basis states.   
The notation $\ket{b_1b_2b_3}$ is a shorthand for $\ket{b_1}\otimes\ket{b_2}\otimes\ket{b_3}$, where $\otimes$ represents the tensor product, so that  $\ket{b_1b_2b_3}$ is a 8-dimensional vector. The tensor form shows that the state $\ket{b_1b_2b_3}$ is \emph{separable}, which means each qubit in the state can be manipulated independently of the other qubits. In general, quantum states formed by multiple qubits cannot be expressed as a tensor product.  The state $\ket{\psi}$ in (\ref{eq:3q_state}) is, in general, not separable.  Non-separable states are called $\emph{entangled}$. 

The superposition property and entanglement enable efficient encoding of information that supports the computational efficiency of quantum computers (see Section~\ref{sec:power}). Generalizing the state $\ket{\psi}$ in (\ref{eq:3q_state}) to $n$ qubits, we see that an $n$-qubit register can encode a $2^n$-dimensional vector with $2^n-1$ independent amplitudes, accounting for the normalization of quantum states.  In contrast, a separable state on $n$ qubits can only encode $n$ independent amplitudes.  Without entanglement, quantum computers lose a significant source of their power \citep{jozsa2003role}.\footnote{Quantum entanglement is not the only source of quantum computing power.  Even without entanglement, quantum computers can perform computations beyond the capabilities of a classical computer \citep{biham2004quantum}.}

\subsection{The Power of Quantum: Superposition, Entanglement, Parallelism, and Interference}
\label{sec:power}

A quantum algorithm is a series of linear operations and quantum measurements performed on a set of quantum registers.  Like their classical counterparts, quantum algorithms are implemented via a series of simple \emph{quantum gates} -- small units of computation, analogous to the familiar classical logic gates, AND, XOR, or NOT.  Quantum gates are elementary quantum transformations that act on one, two, or three qubits; we review popular gates Section~\ref{sec:gates}.

As we introduced in Section~\ref{sec:theory}, a quantum algorithm takes a quantum state as an input and transforms it into an output state that encodes the desired result.  The superposition property suggests that quantum algorithms might offer exponential speedup over classical algorithms in some cases.   Consider an algorithm that checks $100$-bit strings, i.e.~$2^{100} \approx 10^{30}$ possibilities.  A classical computer has to check each of the $2^{100}$ strings. A quantum computer could potentially perform the task in massive parrallel, labelling the correct and incorrect strings using 100 operations -- one for each of the 100 qubits holding the $2^{100}$ strings in a \emph{quantum superposition}.   This natural ability of quantum computers to perform parallel computation is called \emph{quantum parallelism}.   The trouble is, the output of the naive quantum computation is a superposition of all the $2^{100}$ labelled results.  If all results are approximately equally probable, then it would take  $\sim 10^{30}$ measurements to extract the correct answer, negating any benefit from the quantum computation.   To overcome this problem, quantum algorithms leverage \emph{quantum interference} -- the property that quantum amplitudes are complex numbers that can cancel out during computation.  A well-designed quantum algorithm uses quantum interference to suppress the amplitudes of the wrong answers and to amplify the amplitudes of the correct answer(s). The output state of an efficient quantum algorithm is a superposition of the desired answers, so that just a few measurements are sufficient to extract them within required precision. 

The quantum properties of superposition, entanglement, parallelism, and interference drive the next-level computational potential of quantum computers.   Superposition and entanglement enable highly efficient encoding, supporting massively parallel computation by qubit manipulation.  Quantum interference cancels out undesirable by-products of parallel computation and amplifies desired results for efficient readout.  Finding quantum algorithms able to perform parallel computation while providing an efficient way to extract results has proven challenging, but the list of such algorithms is expanding. 
We overview some of these algorithms in Section~\ref{sec:algos_overview} and outline their central ideas in Sections~\ref{sec:grover_section} to \ref{sec:qsvt}.   The building blocks of these algorithms can be combined to solve a variety of practical problems.


\subsection{Quantum States and Quantum Operators}
\label{sec:q-statesandops}

\subsubsection*{Multi-Qubit Quantum States}

As we discussed in Section~\ref{sec:qubit}, a quantum computer stores information using collections of qubits in quantum registers.   Equation~(\ref{eq:3q_state}) provides the general form of a quantum state created on a three-qubit register.  
A convenient shorthand for the expansion of $\ket{\psi}$ in (\ref{eq:3q_state}) interprets the ``bit strings'' of individual qubit basis states as integer numbers, so that to represent, e.g.,~$\ket{101}$, we write $\ket{5}$.  In this notation, a $n$-qubit quantum state is written as
\begin{align}
    \ket{\psi} = \sum_{i=0}^{N-1}\psi_i\ket{i},  \label{eq:psi}
\end{align}
where $N=2^n$, and $i$ in $\ket{i}$ represents the bit string representation of the integer $i$. The 3-qubit quantum state $\ket{\psi}$ in \eqref{eq:3q_state} is fully described by a normalized 8-dimensional vector over complex numbers, $(\psi_0,\psi_1,..,\psi_7)^\top$.   The orthonormal basis formed by states $\ket{i} \equiv \ket{b_1b_2..b_n}$, where $b_j = \{0,1\}\,\,\forall j=1,..,n$, is called the \emph{computational basis}.

\subsubsection*{The Vector Space of Quantum States}

A quantum register of $n$ qubits supports a $N=2^n$-dimensional \emph{Hilbert space}, 
a generalization of Euclidean space over complex numbers in finite or infinite dimensions, introduced briefly in Section~\ref{sec:theory}.  As in Euclidean space, the inner product between two quantum states $\ket{\psi}$ and $\ket{\phi}$ in an $N$-dimensional Hilbert space can be expressed in terms of vector coordinates $\braket{\phi}{\psi} = \sum_{i=0}^{N-1} \phi_i^*\psi_i $, where the asterisk $*$ indicates complex conjugation.  In quantum notation, the conjugate-transpose of a ket $\ket{\phi}$ is denoted as a \emph{bra} $\bra{\phi}$ and decomposed as $\bra{\phi} = \sum_{j=1}^{N-1}\phi^*_j\bra{j}$, where $\bra{j}$ is the conjugate transpose of $\ket{j}$.  For example, in a two-dimensional Hilbert space, if $\ket{0}$ denotes the basis vector $\begin{pmatrix} 1 \\ 0 \end{pmatrix}$, then $\bra{0}$ denotes its conjugate transpose $\begin{pmatrix} 1 & 0 \end{pmatrix}$. The inner product is expressed as $\braket{\bullet}$.   For example, the inner product between two basis states $\ket{i}$ and $\ket{j}$ is $\braket{i}{j}$. Because the basis states are orthonormal:
\begin{align}
    \braket{i}{j} = \delta_{ij} \equiv \begin{cases} 
    1, &\text{if  } i = j \\
    0, &\text{if  } i \neq j
    \end{cases},  \label{eq:braket_ij}
\end{align}
where $\delta_{ij}$ is the \emph{Kronecker delta function} used widely in quantum computing literature.
For two general $N$-dimensional states, the inner product takes the form:
\begin{align}
    \braket{\phi}{\psi} & = \Big(\sum_{j=0}^{N-1}\phi^*_j\bra{j}\Big)\Big(\sum_{i=0}^{N-1}\psi_i \ket{i}\Big) = \sum_{j=0}^{N-1}\sum_{i=0}^{N-1}\phi^*_j\psi_i\braket{j}{i} = \sum_{i=0}^{N-1} \phi_i^*\psi_i, 
\end{align}
where (\ref{eq:braket_ij}) helped simplify the expression.

\subsubsection*{Operators}

Operators in quantum theory are linear.\footnote{\cite{abrams1998nonlinear} demonstrate that, if it were possible to construct non-linear quantum operators, then the computational complexity class NP of problems exponentially hard for classical computers would be equal to the complexity class P -- the class of problems classical computers can solve efficiently (i.e.~in time that scales polynomially with the size of the input to the problem).  While no proof exists that NP $\neq $ P, it is considered highly unlikely that NP $=$ P.}   They can be expressed as matrices of complex numbers acting on vectors in the $N$-dimensional Hilbert space.  Consider the operator $A$ represented by an $N \times N$ matrix with elements $\{a_{ij}\}_{i,j=0}^{N-1}$ with respect to the computational basis.  In the \emph{bra-ket} notation it takes the form
\begin{align}
    A = \sum_{i,j=0}^{N-1}a_{ij}\ketbra{i}{j},
\end{align}
so that, when it acts on state $\ket{\psi} = \sum_{k=0}^{N-1}\psi_k\ket{k}$, the operation yields the expected result
\begin{align}
    A \ket{\psi} = \Big(\sum_{i,j=0}^{N-1}a_{ij}\ketbra{i}{j}\Big) \Big(\sum_{k=0}^{N-1}\psi_k\ket{k}\Big) = \sum_{i,j=0}^{N-1}a_{ij}\psi_j\ket{i} = \sum_{i=0}^{N-1}\Big(\sum_{j=0}^{N-1}a_{ij}\psi_j\Big)\ket{i},
\end{align}
where we used orthonormality of basis vectors $\braket{j}{k} = \delta_{jk}$ (as shown in Eq.~\ref{eq:braket_ij}).
In effect, quantum operators are linear combinations of {\it outer products} $\ketbra{i}{j}$ of basis states in quantum notation.

Because of the constraints in quantum theory, not all linear operators are quantum operators. There are two types of quantum operators: \emph{unitary transformations} and \emph{observables}.  Unitary transformations transform one quantum state into another; observables are related to quantum measurement and will be discussed shortly. 

A linear operator $U$ is {\it unitary} if $U^{-1}=U^\dagger$, where the dagger $\dagger$ denotes the conjugate transpose. Unitary operators preserve the unit norm of quantum states.\footnote{Let $\ket{\psi}$ be the initial state, and $\ket{\phi}=U\ket{\psi}$. Then, $\braket{\phi}{\phi}=\bra{\psi}U^\dagger U\ket{\psi}=\braket{\psi}{\psi}$=1.}

The identity operator $I$ is a unitary operator.    It is commonly used in quantum algorithms to re-express a quantum state in an alternative basis.   Let $\{\ket{a_i}\}_{i=0}^{N-1}$ be an orthonormal set of basis states of an $N$-dimensional Hilbert space and let $\{\ket{b_j}\}_{j=0}^{N-1}$ be an alternative set of orthonormal basis states of the space.  A state $\ket{\psi}$ expressed in terms of basis states $\{\ket{a_i}\}$ can be expressed in terms of states $\{\ket{b_j}\}$ by using the identity operator $I$ written in terms of states $\{\ket{b_j}\}$,
\begin{align}
I  = \sum_{j=0}^{N-1} \ketbra{b_j}, 
\end{align}
as follows
\begin{align*}
    \ket{\psi} & = \sum_{i=0}^{N-1} \psi_i\ket{a_i} = \Big(\sum_{j=0}^{N-1} \ketbra{b_j}\Big) \Big(\sum_{i=0}^{N-1} \psi_i\ket{a_i}\Big) \\ 
    & = \sum_{j=0}^{N-1} \Big(\sum_{i=0}^{N-1} \psi_i\braket{b_j}{a_i}\Big)\ket{b_j} = \sum_{j=0}^{N-1} \tilde{\psi}_j\ket{b_j},
\end{align*}
where $\tilde{\psi}_j=\sum_{i=0}^{N-1} \psi_i\braket{b_j}{a_i}$ are the coordinates of $\ket{\psi}$ in the basis $\{\ket{b_j}\}$.

\subsubsection*{Quantum Measurement}
Quantum measurement, introduced informally in Section~\ref{sec:theory}, is the way for a classical observer characterizes a quantum state.   Quantum measurement is famously probabilistic. The goal of a quantum algorithm is to transform an input quantum state in such a way that a measurement of the resulting state yields the desired result with high probability.

Formally, quantum measurement is a collection of operators $\{M_m\}$ that correspond to outcomes $m$ (where $m$ can be an outcome or the index of an outcome) that occur with probability $p_m$.  A measurement applied to a quantum state $\ket{\psi}$ yields the outcome $m$  with probability $p_m = \bra{\psi}M_m^\dagger M_m\ket{\psi}$.  The state of the system after the measurement is \[\frac{M_m\ket{\psi}}{\sqrt{\bra{\psi}M_m^\dagger M_m\ket{\psi}}}.\]
Because the probability of all possible outcomes adds up to 1, i.e.~$\sum_m p_m =1$, for all quantum states, the measurement operators satisfy the completion relation $\sum_m M_m^\dagger M_m = I$.

For example, quantum measurement of the qubit $\ket{q}$ in (\ref{eq:q}) in the computational basis is a collection of two measurement operators, called \emph{projection measurement operators}, $P_0 = \ketbra{0}$ and $P_1 = \ketbra{1}$, corresponding to outcomes $0$ and $1$ respectively.  These operators satisfy the completeness relation $P_0^\dagger P_0 + P_1^\dagger P_1 = \ket{0}\braket{0}\bra{0} + \ket{1}\braket{1}\bra{1}  = \ketbra{0}+\ketbra{1} = I$.   The probability that the measurement yields $0$ is $p_0 = \bra{q}P_0^\dagger P_0\ket{q} = \braket{q}{0}\braket{0}{q} = |\braket{0}{q}|^2 = |a|^2$ and, similarly, $p_1 = |\braket{1}{q}|^2 = |b|^2$.  Because $|a|^2+|b|^2=1$, the relation $p_0+p_1=1$ is satisfied.  

More generally, the measurement of an $N$-dimensional quantum state $\ket{\psi}$ in the orthonormal basis $\{\ket{u_i}\}_{i=1}^N$ is the set of projection operators $P_i = \ketbra{u_i}$.  The probability $p_i$ that the measurement of state $\ket{\psi}$ yields basis state $\ket{u_i}$ is 
\begin{align}
    p_i = |\braket{u_i}{\psi}|^2.  
\end{align}
This property is called \emph{the Born rule}.

One consequence of the Born rule is that a quantum state is determined up to a complex phase pre-factor $e^{{\rm i}\delta}$, where $\delta \in \mathcal{R}$.  In other words, the states $\ket{\psi}$ and $e^{{\rm i}\delta}\ket{\psi}$ are equivalent. The pre-factor $e^{{\rm i}\delta}$, called the \emph{global phase}, has no physically observable consequences.

\subsubsection*{Observables}
\emph{Observables} are linear operators that do not preserve the norm, but have the property that all their eigenvalues are real.  These operators are self-adjoint, $K^\dagger = K$, and are called \emph{Hermitian}.\footnote{Historically, observables corresponded to physical properties, such as energy or momentum, of quantum states that could be observed in physics experiments.} 
Let $\ket{\kappa_j}$ be eigenvectors of the observable $K$ with (real) eigenvalues $\kappa_j$, so that $K\ket{\kappa_j}=\kappa_j\ket{\kappa_j}$.   The observable $K$ then can take the form
\begin{align}
    K & = \sum_j \kappa_j\ketbra{\kappa_j},
\end{align}
where the states $\ket{\kappa_j}$ form an orthonormal (or orthogonalizable\footnote{In case of degenerate subspaces.}) basis.  Denoting by $P_j = \ketbra{\kappa_j}$ the projection operator onto the subspace spanned by $\ket{\kappa_j}$, we can connect the observable $K$ to the quantum measurement defined by the complete set of projection operators $\{P_j\}$ with outcomes $\{\kappa_j\}$.  This quantum measurement is often called the measurement of the observable $K$.

For a quantum state $\ket{\psi}$, the expectation value of the result of the measurement of $K$ is 
\begin{align}
    \mathbb{E}_\psi[K] & \equiv \langle K \rangle = \sum_j p_j\kappa_j = \sum_j |\braket{\kappa_j}{\psi}|^2\kappa_j,
\end{align}
where $\langle \bullet \rangle$ denotes the expectation value of an observable and $p_j=|\braket{\kappa_j}{\psi}|^2$ is the probability that the measurement of $K$ in the state $\ket{\psi}$ yields the value $\kappa_j$.  Since $ |\braket{\kappa_j}{\psi}|^2 = \braket{\psi}{\kappa_j}\braket{\kappa_j}{\psi}$, the expectation value is commonly written as
\begin{align}
    \langle K \rangle & = \sum_j \braket{\psi}{\kappa_j}\braket{\kappa_j}{\psi}\kappa_j = \bra{\psi}\Big(\sum_j \kappa_j\ketbra{\kappa_j}\Big)\ket{\psi} = \bra{\psi}K\ket{\psi},\label{eq:Kexpectation}
\end{align}
as a consequence of linearity of inner and outer products.  

Similar logic applies to the expectation value of higher moments of $K$, e.g., the variance $\text{Var}(K)$.  Given a Hermitian $K$ and a positive integer $a$, the operator $K^a$ is also Hermitian and can serve as a quantum observable.  The eigenvectors $\ket{\kappa_j}$ of the operator $K$ are also eigenvectors of $K^a$, with eigenvalues $\kappa_j^a$, so that higher moments of $K$ are expressed as:
\begin{align}
    \langle K^a \rangle & = \sum_j\kappa^a_jp_j = \sum_j\kappa^a_j|\braket{\kappa_j}{\psi}|^2 = \bra{\psi}K^a\ket{\psi}, 
\end{align}
with the variance of $K$ equal to $\text{Var} (K) =  \langle K^2 \rangle -  \langle K \rangle^2.$

Observables play an important role in algorithms of interest to data scientists and statisticians, e.g., those for quantum optimization and quantum machine learning, particularly the \emph{hybrid quantum-classical variational algorithms}, which harness the strengths of quantum and classical computers  (see Sections~\ref{sec:quantumML} and \ref{sec:variational}).   In these algorithms, a quantum mechanical observable, which we denote $L$, often represents the cost function.  A quantum variational state $\ket{\psi(\theta)}$ encodes a parameterized model, where $\theta$ is the parameter vector.  The structure of the quantum variational state\footnote{Physics literature uses the word \emph{ansatz} for a fixed-form variational state.} $\ket{\psi(\theta)}$ can vary widely depending on the the nature of the optimization.  For example, the state $\ket{\psi(\theta)}$ could represent a machine learning model with quantum correlations \citep{low2014quantum,rebentrost2018quantum,schuld2019quantum,schuld2020circuit,abbas2020power,bausch2020recurrent,park2020geometry} or a quantum superposition of all possible binary strings of length $n$ for combinatorial optimization \citep{farhi2014quantum}.  The optimization or training problem takes the form
\begin{align}
    \theta^* = \argmin_{\theta} L(\theta) = \argmin_{\theta} \bra{\psi(\theta)}L\ket{\psi(\theta)}.
\end{align}
A popular way to encode the parameter vector $\theta$ into the quantum state $\ket{\psi(\theta)}$ is to parameterize the quantum gates used to prepare the state $\ket{\psi(\theta)}$ \citep[see, e.g.,][and references therein]{farhi2014quantum,schuld2020circuit,cerezo2021variational}.  A classical computer controls the quantum gates; for example, it sets the angle of rotation in controlled rotation gates.  The parameterization of the gates enables a handoff of information from the classical to the quantum computer. A measurement of the observable $L$ in the state $\ket{\psi(\theta)}$ yields an unbiased estimate of $L(\theta)$.    The classical computer collects the results of the repeated measurements of $L$ (using repeated preparations of $\ket{\psi(\theta)}$) and uses this information to update the parameter vector $\theta$ to use in the next iteration, with the ultimate goal of finding the optimal parameter vector $\theta^*$.  This hybrid quantum-classical approach harnesses the strengths of quantum computers, such as superposition and entanglement, while leveraging the strengths of classical computers, such as straightforward resetting of bits or copying of data.  By injecting quantumness into the mature classical computing environment, hybrid quantum-classical algorithms may deliver efficient optimization on the current generation of quantum computers.   For more on optimization and hybrid quantum classical algorithms, see Section~\ref{sec:Quantum Optimization}.

\subsubsection*{Unitary and Hermitian operators}

For any unitary operator $U$ there exists a Hermitian operator $K_U$, such that $U = e^{{\rm i}K_U}$, where $\rm i$ is the imaginary unit.  The operator $K_U$ has a set of eigenstates $\{\ket{u_j}\}$ with eigenvalues $u_j\in \mathbb{R}$, i.e.~for any $\ket{u_j}$: $K_U\ket{u_j} = u_j\ket{u_j}$.  The eigenstates $\{\ket{u_j}\}$ are also eigenstates of the operator $U$, with eigenvalues $e^{{\rm i}u_j}$.  Conversely, for any Hermitian $K$, the operator $U_K = e^{{\rm i}K}$ is unitary.  This connection between unitary and Hermitian operators is widely used in quantum algorithms (see, e.g., Sections~\ref{sec:Quantum Linear Systems} and \ref{sec:Hamiltonian Simulation}).

\subsubsection*{Time-evolution of quantum states and the Hamiltonian of a system}

Because all reversible transformations of quantum states are unitary, the evolution of a quantum state between time $t_1$ and time $t_2$ is a unitary operator $U(t_2,t_1)$: 
\begin{align}
    \ket{\psi(t_2)} & = U(t_2,t_1)\ket{\psi(t_1)}.  
\end{align}
For the time-evolution operator 
$U(t_2,t_1)$, there is a Hermitian operator $K_U(t_2,t_1)$, such that $U(t_2,t_1) = e^{-{\rm i}K_U(t_2,t_1)}$.  If the system is \emph{stationary}, that is its fundamentals do not change over time, then we can write $K(t_2,t_1) = \mathcal{H}\times (t_2-t_1)$, where $\mathcal{H}$ is a Hermitian operator called the \emph{Hamiltonian} of the system.\footnote{Sometimes the inverse of Planck's constant $1/\hbar$ pre-multiplies $\mathcal{H}$ in $K_U(t_2,t_1) = \frac{1}{\hbar} \mathcal{H}\times (t_2-t_1)$, so that eigenvalues of $H$ have the units of energy.  In quantum computing and much of quantum physics literature an assumption is made that $\hbar = 1$, corrected when it becomes necessary to consider relative energy scales.} For a stationary system we have
\begin{align}
    \ket{\psi(t)} = e^{-{\rm i}\mathcal{H}t}\ket{\psi(0)}.
\end{align}

Time evolution of a quantum state plays a central role in quantum algorithms.  The first proposed use for quantum computers was the simulation of quantum systems \citep{benioff1980computer,benioff1982quantum,feynman1981simulating}.  Classical simulations of quantum systems are exponentially hard, quickly running into limitations of current technology.   But quantum computers may be able to simulate the evolution of quantum systems in polynomial time.  The ability to simulate complex physical systems efficiently would revolutionize engineering, allowing us to design new materials, fertilizers, superconductors, or pharmaceuticals at the molecular level.  Furthermore, the ability to simulate Hamiltonians can help us solve problems beyond direct simulations of quantum systems, such as combinatorial optimization problems or problems that can be cast as linear systems of equations. For a detailed discussion of Hamiltonian simulation, see Section~\ref{sec:Hamiltonian Simulation}.

\subsubsection*{Density matrix formulation of quantum states}

So far in this review, we have described quantum states using kets $\ket{\psi }$ -- vectors in a Hilbert space.  The states that can be represented as vectors in a Hilbert space are called \emph{pure} quantum states.  In this section, we briefly introduce an alternative way to describe quantum states using \emph{density operators} aka \emph{density matrices}.   
Density matrices are more general than kets because they enable us to describe not only the pure quantum states but also \emph{mixed} quantum states -- classical probabilistic ensembles of pure quantum states. For example, Alice prepares for Bob the pure state $\ket{0}$ with probability 1/3 and pure state $\ket{1}$ with probability 2/3, then the resulting state is a mixed state $\{(1/3,\ket{0}),(2/3,\ket{1})\}$. Note that this state is different from $1/\sqrt{3}\ket{0}+\sqrt{2}/\sqrt{3}\ket{1}$, which is a pure state, or from $1/3\ket{0}+2/3\ket{1}$ which is not a valid quantum state as the amplitudes are not normalized.  Density matrices, in effect, exist on the continuum between quantum states and classical probability distributions. Because of this, density matrices most commonly appear in the literature concerning noise and de-coherence -- the loss of ``quantumness'' over time -- in physical quantum computers.    However, some important quantum algorithms, such as the quantum Principal Component Analysis algorithm (Section \ref{sec:qpca}), also rely on the density matrix formalism.

The density operator of a pure state $\ket{\psi }$ is defined as $\rho_\psi = \ketbra{\psi}$. 
Consider a system that is in a pure state $\ket{\psi_i}$ with probability $p_i$. The density operator for the system is
\begin{align}
    \rho = \sum_i p_i \ketbra{\psi_i}.
\end{align}
All the quantum postulates can be equivalently reformulated using density operators. For example, given quantum measurement operators $\{M_m\}$, the probability of getting outcome $m$ is
\[p(m)=\bra{\psi}M_m^\dagger M_m\ket{\psi} = \text{tr}(M_m^\dagger M_m\rho_\psi).\] 
When a unitary operator $U$ is applied to a mixed quantum state $\rho$ comprising pure states $\ket{\psi_i}$, the operator acts on each of the states $\ket{\psi_i}$: $U: \ket{\psi_i} \to U\ket{\psi_i}$. The state $\rho$ becomes
\begin{align}
    U: \rho = \sum_i p_i \ketbra{\psi_i} \to \sum_i p_i U\ketbra{\psi_i}U^{\dagger} = U\rho U^\dagger.
\end{align}

The expectation value of an observable in a mixed quantum state equals to the trace of the product of the observable with the density matrix:
\begin{align}
    \langle K \rangle = \tr (K\rho) = \sum_ip_i \tr (K\ketbra{\psi_i}) = \sum_i p_i \bra{\psi_i}K\ket{\psi_i}, \label{eq:trace_rho} 
\end{align}
where we used the linearity and the cyclic property of the trace.  For a pure state $\rho = \ketbra{\psi}$, the expectation value $\langle K \rangle$ equals $\tr (K\rho) =  \bra{\psi}K\ket{\psi}$, just as we have seen in (\ref{eq:Kexpectation}).

An important concept often used 
in the density matrix formulation of quantum computing is 
\emph{partial trace}. 
If the quantum system, described by a density operator $\rho$, is defined over a Hilbert space that is a tensor product of two Hilbert spaces $H^A\otimes H^B$, we can define a partial trace $\tr_B(\rho)$ to obtain a density operator $\rho^A$ of the subsystem $H^A$.
This concept is similar to marginalizing a joint probability distribution to obtain a marginal distribution. 
Let the set of states $\{\ket{u_B}\}$ comprise an orthonormal basis of subspace $H^B$.  Taking a partial trace $\tr_B$ of the density matrix $\rho$ over the subspace $H^B$ results in a reduced density matrix $\rho^A$ on $H^A$:
\begin{align}
    \rho^A = \tr_B \rho = \sum_{u_B}\bra{u_B}\rho\ket{u_B}.
\end{align}

The density matrix is a Hermitian operator, and we can transform density matrices using the quantum algorithmic building blocks that apply to general Hermitian operators.  For example, because $\rho$ is Hermitian, $e^{-i\rho t}$ is unitary -- a property used in the quantum Principal Component Analysis algorithm (Section~\ref{sec:qpca}).

\subsection{Properties of Quantum Computers}
\label{sec:properties}
Efficient quantum algorithms exploit the laws of quantum physics as a computational resource, in ways that can seem strange and unfamiliar.  Even simple operations like copying, erasing, or addition proceed very differently on a quantum vs.~on a classical computer \citep{draper2000addition,haner2018optimizing}.  This section reviews properties of quantum computers that highlight these differences.

\subsubsection*{No Cloning Theorem}

An important property of quantum computers that sets them apart from classical computers is the \emph{No Cloning Theorem} \citep{dieks1982communication,wootters1982single,barnum1996noncommuting}.   The theorem states that for a general unknown quantum state $\ket{\psi}$ there is no unitary operator $O_C$ that makes an exact copy of $\ket{\psi}$:
\begin{align*}
    O_C : \ket{\psi}\otimes \ket{0} \to e^{\alpha(\psi)}\ket{\psi}\otimes \ket{\psi} \text{    does not exist.}
\end{align*}

Exact copying is possible if $\ket{\psi}$ is known, by repeating the process of creating the state in a different register.  However, the No Cloning Theorem states that it is not possible to copy the unknown result of a computation, for example to conduct repeated measurements.   When repeated measurements are required -- as is most often the case when a quantum result needs to be interpreted classically -- the result has to be re-computed after each measurement.  

The No Cloning Theorem does not preclude approximate cloning \citep{buvzek1996quantum} or perfect cloning with some probability of success \citep{duan1998probabilistic}; however, these methods of cloning are rarely used in quantum algorithms because they have limited fidelity \citep{bruss1998optimal,gisin1998quantum}.  For example, the approximate cloning method developed by \cite{buvzek1996quantum}, which has been proven optimal by \cite{bruss1998optimal} and \cite{gisin1998quantum}, has fidelity $5/6$ for copying a single-qubit state -- i.e., the initial state $\ket{\psi_{init}}$ and the cloned state $\ket{\psi_{copy}}$ have a maximum overlap $|\braket{\psi_{init}}{\psi_{copy}}|^2\leq 5/6$.   This fidelity limits the practical ability to apply approximate quantum cloning to multi-qubit states used in quantum algorithms.

\subsubsection*{Reversible computation}

Another property of quantum computers is that all computation with unitary gates is reversible; it neither creates nor destroys information.  Common Boolean gates such as AND or OR used in classical computation are irreversible \citep{vedral1996quantum}. For example, given the result of $a$ AND $b$, it is not possible to recover the values of $a$ and $b$.  An example of a reversible classical gate is the NOT gate:  given NOT $a$, we can recover $a$.  Because classical computers perform irreversible gates with ease, few classical algorithms rely exclusively on reversible gates, even though, in principle, any classical algorithms can be represented in terms of these gates \citep{bennett1989time}.  

An example of a reversible gate is the Fredkin gate, which has three input bits and three output bits.  One of the bits is the control bit; if the control bit is 1, then the values of the other two bits are swapped.   This logic gate is not only universal -- i.e. can be cascaded to simulate any classical circuit -- it is also self-inverse and conservative -- i.e. it conserves the number of 0 and 1 bits.   

Any classical algorithm can therefore be, in principle, implemented on a quantum computer.  First, the classical algorithm is rewritten using reversible gates, then these gates are translated into unitary gates.  Such direct translations, however, are usually inefficient because they do not leverage quantum properties and simply replicate classical ideas on the more expensive and noisy hardware of a quantum computer.  Efficient quantum algorithms often have a structure fundamentally different to that of the classical algorithms designed to achieve similar goals.

\subsubsection*{Uncomputing}

Computation often results in temporary ``garbage'' data.  During classical computation such data can be discarded, but during a quantum computation these ``garbage'' data may be entangled with the main result of the computation making it impossible to reset supplementary registers (often called \emph{auxiliary}, see Section~\ref{sec:gates}), e.g.~using quantum measurement.\footnote{The quantum measurement of the auxiliary register would affect the main result.   Consider a state with two registers, where the first register contains the main result and the second register holds the byproducts of computation $\ket{\psi} = \sum_{x}a_x\ket{x}\ket{y(x)}$, where states $\{\ket{x}\}$ span the Hilbert space on the first register, $a_x$ are the quantum amplitudes, and $y(x)$ are the computational byproducts.  In general, the state $\ket{\psi}$ is entangled.   If we measure the auxiliary register and get an outcome $y_0$, then the state $\ket{\psi}$ will collapse to  $\sum_{x\,\text{s.t.}\, y_x=y_0}a_x\ket{x}\ket{y_x}$, which can be a dramatically different state.}  \emph{Uncomputing} -- running parts of a quantum algorithm ``in reverse'' -- is a way to remove the ``garbage'' data and clear the auxiliary register.  Let algorithm $\mathcal{A}$ be such that $\mathcal{A}\ket*{0} = \ket{\psi}$. The inverse $\mathcal{A}^{-1}$ uncomputes the register containing the state $\ket{\psi}$:  $\mathcal{A}^{-1} \ket{\psi}=\mathcal{A}^{-1}\mathcal{A}\ket*{0}=\ket*{0} $.

\cite{aharonov2007adiabatic}  demonstrate that if quantum computers were able to ``forget'' information, they could solve the NP-complete graph isomorphism problem efficiently.  The uncomputing requirement is consequential -- it limits quantum computers' power.

\section{Quantum Algorithm Design}\label{sec:overview}
This section presents the general structure of quantum algorithms, and highlights the considerations required for the development of successful quantum algorithms. 

A general quantum algorithm often proceeds in three steps: 
\begin{enumerate}
    \item Quantum state preparation.
    \item Quantum computation.
    \item Postprocessing and readout of the resulting quantum state.
\end{enumerate}
If a quantum algorithm is embedded in another quantum algorithm, then steps 1 and 3 may be omitted; however, a quantum algorithm that has classical data as its input and delivers a result for use by classical computers or for human interpretation requires all three steps.  We focus the discussion on steps 1 and 3 in this section, while step 2 is the subject of the subsequent sections.    

\subsection{Data Encoding in a Quantum State}
\label{sec:encoding}

The first thing that a statistician may like to know when using quantum computing in statistics is how to import classical data into a quantum computer. 
Data can be imported into a quantum computer using \emph{quantum state preparation} - the process of encoding data into a quantum state supported by one or more qubit registers.  The qubit registers typically start out initially in the \emph{ground state} -- all qubits are in the 0 state.  Quantum state preparation is a quantum routine that transforms this initial state into a state that encodes the necessary data. 

Efficient loading of data onto a quantum computer is an open area of research \citep{ciliberto2018quantum}. The data loading step can require significant computational resources and, if not carefully thought out, can offset the computational efficiency attained via quantum computation.  Similarly, extracting the result from a quantum state can be a resource-consuming task requiring careful planning as part of the algorithm design.  
The creation of a general quantum state on $n$ qubits can be computationally taxing and requires, at a minimum, $O(\frac{2^n}{n})$ quantum operations \citep[see, e.g.,][]{prakash2014quantum,schuld2018supervised}.  The computational complexity of data preparation is reduced considerably when it is possible to exploit the structure of the data, such as if the data has a functional form.  For example, if the amplitudes represent probability densities of a discretized integrable probability distribution function, loading can be achieved more efficiently \citep{grover2002creating}.  Additional proposals include loading pre-compressed data for analysis; see, e.g.,~\cite{harrow2020small}.

We now describe a few methods for encoding data in a quantum state.

\subsubsection*{Amplitude Encoding}

A quantum state provides several natural ways to encode data.  Consider a general $n$-qubit quantum state in (\ref{eq:psi}).  One of the most common ways to encode data in this state is \emph{amplitude encoding} where data are encoded in the amplitudes $\psi_i$ and the basis vectors $\ket{i}$ serve as indices.  For example, a vector $x \in \mathbb{C}^{2^n}$, normalized so that $\|x\| = 1$, can be encoded as
\begin{align}\label{eq: amplitude encoding}
    \ket{x} = \sum_{i=0}^{2^n-1}x_i\ket{i}.
\end{align}
The basis vectors $\ket{i}$ serve as indices and the amplitudes $x_i$ encode the data.  This type of encoding is widely used in quantum linear systems of equations (Section~\ref{sec:algos_overview}) and related algorithms.   The benefit of this encoding is that it is \emph{qubit efficient}, i.e. a vector of length $N$ needs only $O(\log N)$ qubits \citep{prakash2014quantum,adcock2015advances}.  The downside is that it may be difficult to initialize and to \emph{read out} -- i.e.~to transfer the information from the output quantum state to a classical computer for processing.   Initialization may require an intermediate step, such as quantum Random Access Memory (see below).   Readout within error $\epsilon$ generally requires $O(N/\epsilon^2)$ measurements.  Even though, in some cases more efficient readout is possible, for example through compressive sensing methods (see Section~\ref{sec:readout}), sometimes alternative ways to transfer information to the classical computer are used, such as distilling the results of classification algorithms to few-bit summaries that are more efficient to read out \citep{schuld2016prediction}.

\subsubsection*{Computational Basis Encoding}

An alternative way to encode the data in an $n$-qubit quantum state is to encode information in the basis vectors $\ket{i}$ (as opposed to the amplitudes as in Eq.~\ref{eq: amplitude encoding}).  Consider a data set of $M$ vectors $D=\{x^m=(x_1^m,...,x_N^m)^\top,m=1,...,M\}$, each of a dimension of $N$.
Suppose each vector $x^m$ has been already represented by a binary string with $N\tau$ bits
\[x^m=b_{x_1^m}...b_{x_N^m}\]
where $b_{x_j^m}$ is the binary representation of $x_j^m$ (a string of $\tau$ bits with $\tau$ the precision).
There exists a procedure to prepare data $D$ in the superposition 
\[\ket{D}=\frac{1}{\sqrt{M}}\sum_{m=1}^M\ket{x^m}\]
with $\ket{x^m}$ the basis quantum state corresponding to the binary representation of $x^m$; see, \cite{Ventura2000QuantumAM} or \cite{schuld2018supervised}, Ch.~5.
Technically, $\ket{D}$ can be understood as a superposition state with respect to the computational basis $\{\ket{0},...,\ket{2^{N\tau}-1}\}$,   
where the basis states corresponding to the $\ket{x^m}$ have the amplitude of $1/\sqrt{M}$ and other states have zero amplitude.
This data encoding, known as {\it basis} or \emph{computational basis} encoding, requires $O(N\tau)$ qubits and takes $O(MN)$ operations to initialize.   

The benefit of computational basis encoding is that it enables quantum algorithms to directly leverage quantum parallelism.  For example, let $U$ be an operator that implements a function $f(x^m)$:
\[U:\ket{x^m}\ket{0}\mapsto\ket{x^m}\ket{f(x^m)},\]
then
\[U:\ket{D}\ket{0}\mapsto\frac{1}{\sqrt{M}}\sum_{m=1}^M\ket{x^m}\ket{f(x^m)}.\]
That is, a single application of $U$ gives us $M$ values $f(x^1),...,f(x^M)$ encoded in a superposition quantum state.  Other uses of computational basis encoding include applications in optimization where the quantum optimization algorithms aim to amplify the optimal entry $x^m$ (Section~\ref{sec:Quantum Optimization}).

\subsubsection*{Qsample Encoding and Quantum Sample States}

Another encoding, often called \emph{qsample} encoding \citep{aharonov2003adiabatic}, can be used to encode a probability distribution $P$ on a finite set $\{x^m,m=1,...,M\}$ with probabilities $p(x^m)$
\begin{equation}\label{eq:qsample}\ket{P}=\frac{1}{\sqrt{M}}\sum_{m=1}^M\sqrt{p(x^m)}\ket{x^m}.
\end{equation}
The quantum state $\ket{P}$ in \eqref{eq:qsample} is known as a \emph{quantum sample state}. 
This encoding uses the amplitudes $ \sqrt{p(x^m)}$ to encode the probabilities $p(x^m)$ and the basis vectors to encode the data points $x^m$.  The qsample encoding is appropriate for use in statistics, especially in Monte Carlo methods. For example, as a measurement in the computational basis yields $x^m$ with probability $p(x^m)$, the measurement serves as a sampling technique: it generates samples from the distribution $P$. Also, it is computationally efficient, compared to classical methods, to estimate the expectation of a function with respect to the probability distribution $P$ if it is encoded in $\ket{P}$; see Section \ref{sec:QMC}. As we will see later in Section \ref{sec:qmcmc}, the output of quantum Markov chain Monte Carlo is a quantum sample state in the form of \eqref{eq:qsample}.

\subsubsection*{Data Encoding Using Multiple Qubit Registers}

Splitting the collection of $n$ qubits into multiple registers makes further data structures possible. Consider a collection of $r$ registers of $n_k$, $k=1,..,r$ qubits each, such that the total number of qubits is $n$: $\sum_{k=1}^r n_k = n$.  Each basis vector $\ket{i}$ of the $2^n$-dimensional Hilbert space of $n$-qubits can be expressed as a tensor product of basis states of the $2^{n_k}$-dimensional subspaces spanned by each $n_k$-qubit register: $\ket{i} = \ket{i_1}\otimes\ket{i_2}\otimes\ldots\otimes\ket{i_r} \equiv \ket{i_1}\ket{i_2}\ldots\ket{i_r}$.   In this notation, we can re-express the quantum state in Eq.~(\ref{eq:psi}) as a multi-register state:
\begin{align}
    \ket{\psi} = \sum_{i_1=0}^{2^{n_1}-1}\sum_{i_2=0}^{2^{n_2}-1}\ldots\sum_{i_r=0}^{2^{n_r}-1}\psi_{i_1,i_2,..,i_r}\ket{i_1}\ket{i_2}\ldots\ket{i_r}. \label{eq:psi_registers}
\end{align}

This general form provides a rich set of possibilities for encoding data including \emph{quantum Random Access Memory (QRAM)} and \emph{quantum Read-Only Memory (QROM)} schemes that provide quantum algorithms with efficient access to data.

\subsubsection*{QRAM}

Classical RAM is a scheme which, given an index $i$, outputs the data element $x_i$ stored at the address indexed with the unique binary label $i$.  QRAM is a scheme that, given a superposition of states corresponding to index values in an input register and an empty output data register, outputs the data elements into the data register
\begin{align}
    \ket{\psi^{in}} & = \sum_{i=0}^{N-1}a_i\ket{i}^{in}\ket{0}^{out} \mapsto \ket{\psi^{out}}  = \sum_{i=0}^{N-1}a_i\ket{i}^{in}\ket{x_i}^{out},
\end{align}
where both $i$ and $x_i$ are recorded in the computational basis; $N$ represents the size of the memory; and the coefficients $a_i$ provides the (optional) weights of the various addressed data elements.  For an overview of the method, see \cite{hann2021resilience} and also \cite{giovannetti2008quantum,giovannetti2008architectures}.

The QRAM data structure is suitable for use in some algorithms directly; in others, QRAM is a stepping stone to amplitude encoding, where it is possible to use a controlled rotation to turn the QRAM encoding into amplitude encoding efficiently.

Query complexity in QRAM encoding is $O(\log N)$; however, the method needs $O(N)$ auxiliary qubits.\footnote{Alternatively, it is possible to reformulate QRAM so that its query complexity of $O(N)$ using $O(\log(N))$ auxiliary qubits.} 
Critics of QRAM point out that QRAM requires unphysically high qubit fidelity to work \citep{arunachalam2015robustness}, although \cite{hann2021resilience} have recently demonstrated that QRAM is more robust than had been previously thought.  Additionally, the approach requires a parallel gate architecture; if a classical computer leveraged a similar parallel architecture, it would be able to achieve similar speedups over sequential classical architecture as quantum computers do \citep{aaronson2015read,steiger2016racing,csanky1975fast}. 

\subsubsection*{QROM}

Instead of storing data in a quantum state, it is possible to create a classical data structure that provides efficient quantum access to the data for use in some algorithms, such as those based on quantum singular value transformation (Section~\ref{sec:qsvt}).   One such structure, proposed by \cite{kerenidis2016quantum}  and named \emph{quantum read-only memory (QROM)} by  \cite{chakraborty2018power}, can store a matrix $A \in \mathbb{R}^{M\times N}$ in $O(w\log^2 MN)$ \emph{classical} operations, where $w$ is the number of non-zero elements of $A$.  Once the structure is in place, it is possible to perform the following quantum initializations with $\epsilon$-precision in $O(polylog(MN)/\epsilon)$ time (requiring $O(N)$ gates accessed in parallel), where $polylog(\bullet )$ is a common shorthand for ``some polynomial in $\log(\bullet)$:
\begin{align}
    U: \ket{i}\ket{0} \mapsto \ket{i}\frac{1}{\|A_{i}\|}\sum_{j=1}^NA_{i,j}\ket{j} = \ket{i}\ket{A_i} \\
    V: \ket{0}\ket{j} \mapsto \ket{1}{\|A\|_F}\sum_{i=1}^M \|A_{i}\| \ket{i}\ket{j} = \ket*{\tilde{A}}\ket{j},
\end{align}
where $\ket{A_i}$ is the quantum state encoding the normalized $i$th row of $A$; $\ket*{\tilde{A}}$ is the quantum state such that its inner product with the quantum state corresponding to the row index $\ket{i}$ yields the normalization factor $\|A_{i}\|$:  $\braket*{i}{\tilde{A}}=\|A_{i}\|$.  

\subsection{Result Postprocessing and Readout}
\label{sec:readout}

Reading out the results of a quantum algorithm can be a challenging, resource-consuming step. The result of a quantum algorithm is a quantum state that may be handed off for processing either to a classical or a quantum algorithm.   Post-processing by quantum algorithm is usually more efficient because the quantum-classical readout requires a number of operations that may overwhelm the number of operations required to run the algorithm itself \citep{zhang2021quantum}.   Consider a state $\ket{x} = \sum_{i=0}^{N-1}x_i\ket{i}$ encoding an $N$-dimensional vector $x = (x_0,...x_{N-1})$.  Reading out the elements of vector $x$ within precision $\epsilon$ requires a minimum of $O(N/\epsilon^2)$ measurements \citep{odonnell2016efficient}.  When the vector $x$ is a result of a quantum computation that has a polylogarithmic complexity dependence on $N$, the readout complexity of $O(N/\epsilon^2)$ overwhelms the complexity of the quantum computation to obtain $x$.    Full readout of a quantum state is called quantum \emph{tomography}.

Improvements to readout complexity are only possible assuming prior knowledge of the structure of the quantum state to be read out.   Efficient readout methodologies that exploit the quantum state's structure often involve reparametrization in order to reduce the effective dimensionality of the state.  Methods include compressed sensing \citep{gross2010quantum,kyrillidis2018provable}\footnote{Classical compressed sensing is a method of recovering a sparse vector from a small number of measurements.  Quantum measurement techniques leveraging compressed sensing aim to recover a pure or mostly pure quantum state efficiently.}, permutationally invariant tomography \citep{toth2010permutationally,moroder2012permutationally}, schemes based on tensor networks \citep{cramer2010efficient,baumgratz2013scalable,lanyon2017efficient}, or mapping target states onto highly entangled but structured lower dimensional models, such as restricted Boltzmann machines \citep{torlai2018latent,torlai2018neural}.  

For many applications, full tomography of the quantum state may not be required.  \cite{aaronson2019shadow} proposed \emph{shadow tomography}, a method to predict specific properties of the state, called target functions, without fully characterizing it. In order to predict with high probability an exponential number of target functions, it is often sufficient to have only a polynomial number of copies of the quantum state.  \cite{huang2020predicting} improved the efficiency of shadow tomography to reduce its exponential circuit depth requirements.  The method involves repeated application of random unitaries drawn from a purposefully constructed ensemble followed by a measurement of the resulting state in the computational basis.  The expectation value of the repeated unitary transformations and measurements is, in itself, a transformation of the quantum state.  The inverse of the expectation value of the measurement acts as a snapshot of the quantum state, its classical shadow.  Classical shadows are expressive enough to yield many efficient predictions of the quantum state \citep{huang2020predicting}: A shadow based on $M$ measurements is sufficient to predict $L$ linear functions $O_i$, $i=1,..,L$, of the quantum state up to an error $\epsilon$, provided $M$ exceeds $O(\log L \max_i\|O_i\|^2_{shadow}/\epsilon^2)$, where the norm $\|O_i\|^2_{shadow}$ depends on the distribution of random unitaries used in the construction of the shadow \citep[see][for further details]{huang2020predicting}. It has the property $\|O_i\|^2_{shadow} < 4^n\|O_i\|_\infty$, where $n$ is the number of qubits and $\|\cdot\|_\infty$ denotes the operator norm.

\subsubsection*{Swap Test and Sample Mean Estimation}
One of the methods used in post-processing a quantum result state is the \emph{swap test} \citep[see, e.g.][]{buhrman2001quantum}, used to estimate the inter product $a^\top b$ of two normalized vectors $a$ and $b$, encoded in two states $\ket{a}$ and $\ket{b}$
\[\ket{a}=\sum_ia_i\ket{i},\;\;\;\ket{b}=\sum_ib_i\ket{i}.\]
For example, \cite{schuld2016prediction} use a swap test to perform a prediction by linear regression using a quantum algorithm. The swap test also provides an efficient way for computing a sample mean (see below).

The swap test applies a series of three-qubit \emph{controlled swap gates} (also known as \emph{Fredkin gates}), which swap two qubit states conditional on the state of the auxiliary qubit so that
\begin{align}
    c\text{-}SWAP\Big[\frac{1}{\sqrt{2}}(\ket{0}+\ket{1})\otimes\ket{a}\otimes\ket{b}\Big] = \frac{1}{\sqrt{2}}(\ket{0}\ket{a}\ket{b}+\ket{1}\ket{b}\ket{a}).
\end{align}    
Applying a Hadamard gate to the auxiliary qubit and rearranging the terms results in
\bean
&&\frac{1}{2}\big(\ket{0}(\ket{a}\ket{b}+\ket{b}\ket{a})+\ket{1}(\ket{a}\ket{b}-\ket{b}\ket{a})\big)\\
&&=\frac{\alpha_1}{2}\ket{0}\frac{(\ket{a}\ket{b}+\ket{b}\ket{a})}{\alpha_1}+\frac{\alpha_2}{2}\ket{1}\frac{(\ket{a}\ket{b}-\ket{b}\ket{a})}{\alpha_2}
\eean
with $\alpha_1$ and $\alpha_2$ the norm of $(\ket{a}\ket{b}+\ket{b}\ket{a})$ and $(\ket{a}\ket{b}-\ket{b}\ket{a})$, respectively. It is easy to see that 
\[\alpha_1=\sqrt{2(1+|\braket{a}{b}|^2)}.\]
Hence, the probability of measuring state $\ket{0}$ in the first qubit is
\[p=\frac{\alpha_1^2}{4}=\frac{1}{2}\big(1+|\braket{a}{b}|^2\big),\;\;\;\;\text{  hence, }\;\;\;\;|\braket{a}{b}|=\sqrt{2p-1}.\]
Estimating $p$ by repeated measurement gives us an estimate of the absolute value $|\braket{a}{b}|=|a^\top b|$. 

Its sign can also be determined. Consider two states $\tilde{a}$ and $\tilde{b}$ that encode the vectors $\frac{1}{\sqrt{2}}(a_1,...,a_N,1)^\top$ and $\frac{1}{\sqrt{2}}(b_1,...,b_N,1)^\top$ respectively. Applying the swap test to these two states results in  $|\braket*{\tilde{a}}{\tilde{b}}|=\sqrt{2p-1}$. By noting that $|\braket*{\tilde{a}}{\tilde{b}}|=a^\top b/2+1/2$, we have 
\[a^\top b=2\sqrt{2p-1}-1.\]

Now, suppose that a data vector $(x_0,..,x_{N-1})$ has been encoded into a quantum state $\ket{x}=\sum_ix_i\ket{i}$.
Applying the swap test to $\ket{x}$ and the uniform superposition
state $\ket{u} = \frac{1}{\sqrt{N}}\sum_{i=0}^{N-1}\ket{i}$ gives us an estimate of $\sqrt{N}\bar{x}$.  The algorithm requires $O(1/\epsilon^2)$ measurements to achieve error tolerance of $\epsilon$.

\subsection{Computational Complexity}
\label{sec:complexity}

\emph{Computational complexity} quantifies the resources, particularly time and memory, an algorithm requires to complete a computational task.  For quantum algorithms, the most relevant dimensions of computational complexity are time, the number of qubits (qubit complexity) and the number of gates (gate complexity) required to complete a computation.  Time complexity is the most commonly cited metric.

Computational complexity is often expressed as a function of the size of the input, $N$, using \emph{Big O} notation \citep{nielsen2002quantum}. The most popular measure of computational complexity is the \emph{upper bound} $O(g(\cdot))$, which indicates that required resources are bounded from above by a function $g(\cdot)$ of the relevant parameters, such as the size $N$.  For example, naive matrix multiplication of two $N\times N$ matrices has time complexity of $O(N^3)$. Sometimes we use $\tilde{O}$, pronounced as ``soft-O'', to indicate that $\tilde{O}(g(\cdot)) = O(g(\cdot)\log^k g(\cdot))$, for some finite $k$.
Other measures of complexity that sometimes appear in the quantum computing literature are \emph{lower bound} complexity $\Omega(\cdot)$ (pronounced as ``Big Omega") and asymptotically tight complexity $\Theta(\cdot)$ (pronounced as ``Big Theta") where lower and upper bound coincide.

The set of all problems that a quantum computer can solve in polynomial time -- i.e. time polynomial in the size of the input $N$ -- with an error probability of at most 1/3 (by convention) comprises the \emph{complexity class BQP}.  The complexity class BQP contains some problems that a classical computer cannot resolve in less than exponential time (time exponential in $N$) -- NP-hard problems.  However, the class BQP does not contain NP-complete problems -- the NP-hard problems that, if solved in polynomial time, lead to a polynomial-time solution of all the other NP-hard problems.

For an informal (although mathematical) and engaging discussion of topics in computational complexity of quantum algorithms and the complexity classes relevant to quantum computation, see \cite{aaronson2013quantum}.

\subsection{A Brief Overview of Quantum Algorithms}
\label{sec:algos_overview}

Quantum computing is a rapidly evolving field, with many new algorithms emerging every year.  For this review, we have assembled a representative selection of algorithms most relevant to statisticians and data scientists.  This section provides an overview of these algorithms, with detailed explanations and references available in Sections~\ref{sec:grover_section} through \ref{sec:qsvt}.

We start with the Grover search family of algorithms (Section~\ref{sec:grover_section}).   Grover’s search algorithm finds a labeled item in an unstructured database of size $N$ using $O(\sqrt{N})$ queries, quadratically faster than the best classical approach that requires $O(N)$ queries.   Grover proved that his algorithm is optimal – no quantum algorithm can perform unstructured search faster.  

The influence of Grover's algorithm extends far beyond unstructured search.  
Grover’s algorithm gave rise to several widely-used subroutines, including Quantum Amplitude Amplification (QAA) and the closely related Quantum Amplitude Estimation (QAE).  These subroutines appear in many quantum algorithms, for example to amplify the amplitude of the desired result and suppress undesirable byproducts at the end of a quantum computation. The core ideas of Grover's search algorithm power quantum Monte Carlo integration and the efficient estimation of the statistical properties of a function on an unstructured set, such as its mean, median, minimum, and maximum.  Additionally, Grover’s search algorithm inspired a popular version of quantum walks.

Quantum walks (Section~\ref{sec:quantum_walks}) are quantum analogues of classical random walks or Markov chains, widely used in randomized classical algorithms.   Quantum walks mix quadratically faster than classical random walks.
This quadratic speedup can offer efficiencies in applications such as quantum Markov chain Monte Carlo.

The next important family of quantum algorithms, presented in Section~\ref{sec:Quantum Linear Systems}, leverages the quantum Fourier transform (QFT) – the quantum analogue of the classical discrete Fourier transform.  QFT powers many important algorithms of interest to statisticians and data scientists, such the quantum linear systems algorithm, quantum matrix inversion, and quantum PCA.  QFT exploits quantum parallelism – the ability to apply a function to multiple elements of a vector in parallel - to deliver exponential speedups to the quantum algorithms it powers.   For a vector of size $N$, QFT requires only $O(\text{poly}\log(N))$ calls to the function, provided the vector is encoded into a quantum state using amplitude encoding.   Classical discrete Fourier transform requires $O(N\log N)$ calls to transform the classically encoded vector.    

Arguably the most famous application of QFT is Shor’s factoring algorithm, one of the first quantum algorithms with direct potential real-world consequences – the breaking of the encryption system based on prime factorization.   Most encryption systems today that keep financial transactions and other sensitive information secure rely on the fact that factoring a large number $N$ is exponentially hard for classical computers.  Shor demonstrated that, using QFT, it is possible to perform the task in time polylogarithmic in $N$.   
Since Shor’s algorithm does not have a direct application in statistics and many clear reviews of the algorithm are available \citep[see, e.g.,][]{nielsen2002quantum}, we omit it in this review.  We focus instead on applications of QFT to fast linear algebra, critical to many data science applications.  

Quantum linear algebra applications leverage QFT via another quantum subroutine, quantum phase estimation (QPE) –- an algorithm to record an estimate of a phase, such as the phase $\theta$ in an eigenvalue of the form $e^{i\theta}$ of a unitary operator, in a quantum state in computational basis.   A common use of QPE is to extract the eigenvalues of a Hermitian matrix (or the singular values of a rectangular matrix embedded in a Hermitian matrix).  The quantum linear systems algorithm, first proposed by Harrow, Hassidim, and Lloyd (HHL) \citep{harrow2009quantum} uses QPE to perform fast matrix inversion, which has a wide range of statistical applications, such as regression analysis.   Other applications of QPE include Quantum Principal Component Analysis (QPCA) and fast gradient estimation.   QFT, at the heart of these algorithms, powers their exponentially faster runtime (under certain conditions detailed in Section~\ref{sec:Quantum Linear Systems}) compared with their classical counterparts.

In Section~\ref{sec:Hamiltonian Simulation}, we describe Hamiltonian simulation algorithms, which simulate the evolution of complex quantum systems, including those of direct practical importance such as biologically active molecules.  Advances in Hamiltonian simulation could change the face of agriculture, materials, and energy; they are potentially even more consequential than Shor’s factoring algorithm.  Many quantum algorithms rely on Hamiltonian simulation as a building block.  For example, the HHL algorithm for quantum linear systems uses Hamiltonian simulation together with QPE to perform matrix inversion.   The computational complexity of quantum linear systems algorithms has dramatically improved since HHL’s first proposal, and most of these advances occurred because of more efficient Hamiltonian simulation techniques.

Another family of quantum algorithms of interest to statisticians are the quantum optimization algorithms (Section~\ref{sec:Quantum Optimization}), which find a quantum state that minimizes a cost function. Many quantum optimization algorithms leverage ideas of Adiabatic Quantum Computation (AQC), a quantum computing framework based on
the \emph{Adiabatic Theorem}.   The Adiabatic Theorem states that a system in the eigenstate corresponding to the smallest eigenvalue of its Hamiltonian\footnote{Physicists refer to the state corresponding to the smallest eigenvalue of its Hamiltonian as the \emph{lowest-energy state} or \emph{ground state} of the system.} will stay in this state if the system changes sufficiently slowly -- i.e.~in such a way that a finite eigenvalue gap between the two lowest eigenvalues persists throughout the evolution.  In AQC, an easy-to-initialize quantum state slowly transforms into the desired state through gradual Hamiltonian evolution.  AQC is, in effect, analog quantum computing; nevertheless, on a noiseless quantum computer, it is equivalent to circuit-based quantum computing \citep{aharonov2008adiabatic}. 

The Quantum Approximate Optimization Algorithm (QAOA) is the most famous optimization algorithm inspired by AQC.  QAOA is a hybrid quantum-classical variational algorithm, which optimizes a quantum state using a classical outer loop.  In QAOA, a specific sequence of classically-controlled parametrized gates prepares a variational quantum state; the cost function takes the form of a quantum observable.   The classical outer loop adjusts the parameters based on the measurements of the cost function.  QAOA can be a powerful method to obtain approximate solutions to combinatorial optimization problems, such as MaxCut -- the NP-hard problem of cutting a graph in two parts, such that the number of edges between the parts is as large as possible. QAOA and other hybrid quantum-classical algorithms play to the strengths of quantum and classical algorithms to deliver efficient optimization \citep{mcclean2016theory,mcardle2019variational}.

Section~\ref{sec:qsvt} presents Quantum Singular Value Transformation (QSVT), an algorithm to perform polynomial transformations of the singular values of a matrix.    This versatile method represents the cutting edge of quantum algorithm development.  It serves as a unifying framework encompassing many existing quantum algorithms, such as Grover search, Hamiltonian simulation, and matrix inversion, provides a consistent way to develop efficient versions of these algorithms, and enables the systematic development of new algorithms. For example, the QSVT-based version of quantum matrix inversion is the most efficient known algorithm for this task. QSVT is a generalization of Quantum Signal Processing (QSP), a method inspired by signal processing used in nuclear magnetic resonance -- an important and well-developed technology used in many industries from mining to medicine.  QSP embeds a quantum system in a larger system to perform a non-linear/non-unitary transformation of the subsystem.   QSVT extends the method to general rectangular matrices.     The flexible paradigm may result in novel algorithms of interest to statisticians being developed in the near future.

\subsection{Quantum Machine Learning}\label{sec:quantumML}

Quantum machine learning is a very active area of quantum computing research.  Results in quantum learning theory point to classes of problems where quantum computing can deliver significant advantages \cite[see, e.g.][]{arunachalam2017guest}, including polynomially faster learning rates.  Additionally, evidence suggests that NISQs  (Section~\ref{sec:nisqs}) may be able to deliver quantum advantage in machine learning over classical counterparts.   We refer readers interested in a more detailed review of quantum machine learning to  \cite{ciliberto2018quantum,schuld2018supervised,adcock2015advances}; and the informal review by \cite{dunjko2020non}.

The most direct application of quantum computers to machine learning is the development of quantum neural networks -- artificial neural networks designed to benefit from quantum superposition and entanglement and encoded in quantum states that are difficult to sample from classically \citep[see, e.g.][and references therein]{low2014quantum,rebentrost2018quantum,schuld2019quantum,schuld2020circuit,abbas2020power,bausch2020recurrent,park2020geometry}.  A critical challenge in development of quantum neural networks is that quantum transformations are fundamentally linear, and non-linearity is seen as particularly important in successful classical neural networks. \cite{schuld2016prediction} demonstrated that the vast majority of early proposals for quantum neural networks did not meet the non-linearity requirements of artificial neural networks. But recent models are overcoming this disadvantage using, for example, quantum measurement \citep[see e.g.][]{romero2017quantum,wan2017quantum} or kernel functions \citep{farhi2018classification,blank2020quantum,liu2021rigorous} to include non-linearity.  \cite{huang2021power} develop a class of kernel models that can provide rigorously demonstrable speedup over classical models in the presence of noise, which means they are potentially achievable on the current generation of quantum computers.

Faster optimization is another way to leverage quantum computers in machine learning.  Many machine learning methods use optimization techniques for parameter learning. Quantum Optimization (Section~\ref{sec:algos_overview}), which includes annealing \citep{kadowaki1998quantum,kadowaki2021greedy} and adiabatic methods, approximate optimization, and hybrid quantum-classical optimization, has the potential to deliver quadratic or polynomial speedups  \citep{aaronson2009need,mcclean2021low} for a variety of machine learning tasks.  For example, \cite{miyahara2018quantum} perform mean-field VB via quantum annealing, an alternative to gradient descent optimization.  Hybrid quantum-classical variational algorithms, which combine the strengths of quantum and classical computers, can speed up variational methods \cite[see, e.g.,][]{farhi2018classification,mcclean2018barren,mitarai2018quantum}.  Hybrid approaches are particularly attractive in the near term, because they may help to harness the power of the current generation of quantum computers.

Bayesian computation has long called for scalable techniques.
Markov chain Monte Carlo (MCMC) has been the main workforce in Bayesian statistics, but it is also well-known that MCMC can be too slow in many modern applications.
Quantum Markov chains (Section~\ref{sec:algos_overview}) hold the promise to speed up MCMC \cite[see, e.g.][]{szegedy2004quantum,chowdhury2017quantum,orsucci2018faster}. 
Quantum computation has also been exploited to speed up Variational Bayes \citep{lopatnikova.tran:2021:quantum} - another popular technique for Bayesian computation.

When fault-tolerant quantum computers become available, machine learning methods may benefit from fast quantum linear algebra. For example, the algorithms to solve systems of linear equations, such as the HHL algorithm \citep{harrow2009quantum} and its updates (Section~\ref{sec:hhl}), enable fast matrix inversion used widely in machine learning models \citep{wiebe2012quantum,lloyd2013quantum,rebentrost2014quantum,lloyd2016quantum,cong2016quantum,kerenidis2016quantum,childs2017quantum,rebentrost2018svd,wang2019accelerated,kerenidis2020quantum}.  Under certain conditions, such as when quantum access to data is provided and the matrices to be inverted are sparse or low-rank, quantum computers can deliver exponential speedup relative to classical computers.  Direct applications include linear regression for data fitting \citep{wiebe2012quantum} and prediction \citep{schuld2016prediction}, ridge \citep{yu2019improved} and logistic \citep{liu2019quantum} regression. \cite{lopatnikova.tran:2021:quantum} use quantum matrix inversion to speed up the estimation of natural gradient for Variational Bayes (VB).  Related algorithms such as the quantum PCA and the QSVT algorithms (Section~\ref{sec:algos_overview}) have also been influential; e.g., quantum PCA using parameterized quantum circuits can support face recognition tasks \citep{xin2021experimental}.\footnote{\cite{gilyen2018quantum,tang2018quantum,chia2020sampling} have recently demonstrated that, if data are made available to classical computers in structures similar to those required for efficient quantum computations, then some of the quantum linear systems algorithms can be de-quantized -- i.e.~significant efficiencies can be obtained using randomized classical algorithms.  These quantum-inspired algorithms can bring intriguing efficiencies to machine learning; however, currently these algorithms suffer from disadvantaged scaling in critical parameters, such as condition numbers and sparsity of matrices, which may render them impractical in the near term.}

Quantum machine learning and classical machine learning have benefited from cross-over ideas.  Classical machine learning algorithms have incorporated quantum-insired structures \citep{gilyen2018quantum,tang2018quantum,chia2020sampling}.  Similarly, classical machine learning ideas, such as kernel methods, helped better understand the nature of quantum neural networks and improve their design \citep{schuld2019quantum}.

\section{Programming Quantum Computers}
\label{sec:programming_qcs}

Quantum algorithm designers have the opportunity to perform proof-of-concept computations on real quantum computers, available from a number of companies as a cloud-based service.   In this section we review the state of the art of quantum computing technology and outline ways to access quantum computers.

\subsection{Physical Implementation on NISQs and Beyond}
\label{sec:nisqs}

Today's quantum computers are small and noisy, similar to classical computers in the 1950s.  The largest quantum computers today comprise up to 100 qubits, with error rates at best around $1\%$ and \emph{coherence time} -- the duration of time qubits can represent quantum states with sufficient accuracy -- up to 100 microseconds.  In the 1950s, classical computers suffered a similar problem.  Made of vacuum tubes or mechanical relays, the bits in the early classical computers tended to flip at random, introducing errors.  To perform computation, error correction code, based on redundant bits, was required.   Modern classical computing technology is so advanced that bits are extremely stable without error correction.  But quantum computers still require redundant qubits to compensate for the decoherence errors.  As a result, a system of 100 qubits may have an order of magnitude fewer \emph{logical} qubits, units acting as qubits for algorithm implementation.   The connectivity between qubits -- i.e.~our ability to apply two or three-qubit gates with high accuracy -- and also how long it takes to apply a gate also play an important role.\footnote{For a popular account of the role of noise in quantum computing and the importance of quantum error correction, see e.g.~\cite{cho2020biggest}.}  

Most quantum algorithms that offer provable speedups over classical counterparts require much larger, fault tolerant quantum computers.   Two metrics can help identify the quantum resources an algorithm might require:  the size of the qubit registers and the \emph{circuit depth} of the algorithm.   \emph{Circuit depth} refers to the number of gates required to implement a quantum algorithm on a quantum computer sequentially.  Higher circuit depth algorithms require higher coherence times.  

But even though today's quantum computers are relatively small and noisy, we may be crossing over to the era when these computers surpass classical computers for some classes of problems.   In 2018, \cite{preskill2018quantum} called these noisy computers with 50+ qubits NISQs -- noisy intermediate-scale quantum systems.   Some algorithms such as the Quantum Approximate Optimization Algorithm, described in Section~\ref{sec:algos_overview}, or the Variational Quantum Eigensolver \citep{peruzzo2014variational} -- an algorithm important in simulating quantum physical systems -- can be implemented on NISQs.  For a discussion of quantum algorithms on NISQs, see e.g.~\cite{bharti2021noisy} and references therein.

The most popular quantum computing technologies today are based on superconductors and cold ions.  Superconductor-based quantum computers have been able to achieve highest qubit counts and fast gate times, but the gates and qubits on these computers are noisy.  Ion-based quantum computers offer slower gate times, but much higher qubit fidelities and greater connectivity.  Other qubit types include silicon qubits, nitrogen-vacancy qubits, and optical qubits.  

\subsection{Quantum Programming}
\label{sec:programming}

Physical quantum computers are available today through a number of companies, such as IBM, Amazon AWS, Google, Rigetti, and others, as a cloud-based service.  Some of the cloud quantum computing providers, such as IBM and Rigetti, offer access to their in-house quantum computers; other, like Amazon AWS and Google, aim to provide access and programming tools to run programs on third-party quantum computers, such as those built by IonQ, Honeywell, and Rigetti.   Additionally, the quantum computing providers also offer classical simulation of quantum algorithms, which can be used to test these algorithms.  To operate the quantum computers and simulators, each cloud service offers a software-development kit (SDK).   

At the time of writing, by far the most popular quantum SDK is IBM’s Qiskit (pronounced as ``{\bf'kiss}-kit’’).  Qiskit enables quantum software developers to run quantum code on IBM Quantum Experience, IBM’s cloud quantum computing service.  At present, IBM is the only platform that provides free quantum computing access to the general public.   It also provides premium access to its latest, higher-fidelity, higher-qubit-count quantum computers.   Qiskit, like most other quantum SDKs, is Python-based and open-source.  Like all other current quantum programming packages, Qiskit has an assembly language at its core; it manipulates individual logical qubits, gates, and quantum circuits.  The growing library of higher-level tools and application packages at this stage are also written in quantum assembly language.\footnote{When classical computing technology was less mature, assembly language coding was much more popular.  Today, the vast majority of classical software designers operate at a higher level of abstraction.}  Qiskit provides extensive documentation, tutorials, including many quantum programming examples, and an active developer community.   Statisticians and data scientists, who have read this review up to this point, should be able follow Qiskit programming tutorials at \texttt{https://qiskit.org/textbook/} with relative ease.  

Rigetti Computing also provides a cloud computing service to access quantum computers and simulators developed in-house.  The SDK called Rigetti Forest uses a custom instruction language Quil, particularly strong at facilitating hybrid quantum/classical computing.   Like Qiskit and the vast majority of other quantum programming tools, Quil is based on open-source Python packages. The Python library \texttt{pyQuil} is a library of higher-level Quil applications.

The other significant industry players, such as Amazon, Google, and Microsoft provide cloud computing access to third-party quantum computers.  Each of these players have their own SDKs and cloud quantum computing access tools.  Amazon has recently launched Amazon Braket, a quantum computing service within their broader on-demand computing service Amazon Web Services (AWS).  Braket comes with its own script, access to third-party quantum computers such as those made by IonQ, Rigetti, and Oxford Quantum Computing, as well as quantum annealers by D-Wave, and classical quantum-computing simulators.    Google Quantum Computing Service uses Cirq, its open source Python-based framework, and provides access to third-party quantum computers, such as those from IonQ and Honeywell.   Microsoft’s quantum computing service Azure Quantum is a part of its Microsoft Azure on-demand computing services.  It has its own language Q\#, and provides access to third-party quantum computers.

\subsection{Quantum Gates and Other Primitives}\label{sec:gates}

Quantum algorithms are implemented physically using \emph{quantum circuits} -- i.e.~series of basic \emph{quantum gates} -- analogous to classical circuits.  Quantum gates are unitary transformations applied to one, two, or three qubits at a time.    \footnote{A general quantum transformation requires exponentially many quantum gates \citep{knill1995approximation}.  The art of writing quantum algorithms is in finding ways to perform useful transformations efficiently with respect to all resources -- time, qubits, and gates.}  This section provides an overview of some of the most common gates.  For a more extended discussion of quantum circuits, see, e.g.~\cite{nielsen2002quantum} and \cite{kitaev2002classical}; for a general theory of quantum circuits, see, e.g.~\cite{aharonov1998quantum}.

\subsubsection*{Hadamard Gate}

The most widely used quantum gate is the Hadamard gate.  In the computational basis $\{\ket{0},\ket{1}\}$, it is represented by the matrix $H$:
\begin{align}
    H = \frac{1}{\sqrt{2}}\begin{pmatrix} 1 & 1\\ 1 & -1 \end{pmatrix}.
\end{align}
When applied to a qubit in the basis state $\ket{0}$ it creates a uniform superposition of states $\ket{0}$ and $\ket{1}$, often denoted as $\ket{+}$:
$\ket{+}=H\ket{0} = \frac{1}{\sqrt{2}}(\ket{0}+\ket{1})$. Similarly, $\ket{-}=H\ket{1} = \frac{1}{\sqrt{2}}(\ket{0}-\ket{1})$.  Application of the Hadamard gate to each qubit of an $n$-qubit register creates a uniform superposition of the $N = 2^n$ possible computational basis states
\begin{align}
    H^{\otimes n}\ket{0\ldots0} = \frac{1}{\sqrt{N}}\sum_{i=0}^{N-1}\ket{i},
\end{align}
where $\ket{i}$ is a shorthand for the computational basis state of $n$-qubits that can be interpreted as an $n$-bit integer $i$; the notation $H^{\otimes n}$ means that the gate $H$ is applied to each of $n$ qubits once (as compared with $H^n$, which means $n$ sequential applications of $H$ to the same qubit).  The operator $H^{\otimes n}$ that creates a uniform superposition on $n$ qubits is sometimes called the \emph{Walsh-Hadamard transform}.

\subsubsection*{Single-Qubit Rotations}

To understand qubit rotations, we reparametrize the qubit in (\ref{eq:q}) in terms of angular coordinates of a unit sphere
\begin{align}
    \ket{q} = \cos (\theta/2) \ket{0} + e^{{\rm i}\phi} \sin (\theta/2)\ket{1}.
\end{align}
The unit sphere representing a qubit is called the \emph{Bloch sphere}.  It is a two-dimensional object embedded in a three dimensional space.  A single-qubit unitary transformation can be decomposed into a sequence of basic rotations around $x$, $y$, and $z$ axes
\begin{align}
    R_x(\varphi) & = e^{-{\rm i}X\varphi/2} = \begin{pmatrix} \cos \varphi/2 & -{\rm i}\sin \varphi/2 \\ -{\rm i}\sin \varphi/2 & \cos \varphi/2\end{pmatrix} \label{eq:R_x}\\
    R_y(\varphi) & = e^{-{\rm i}Y\varphi/2} = \begin{pmatrix} \cos \varphi/2 & -\sin \varphi/2 \\ \sin \varphi/2 & \cos \varphi/2\end{pmatrix},\label{eq:R_y} \\
    R_z(\varphi) & = e^{-{\rm i}Z\varphi/2} = \begin{pmatrix} e^{-{\rm i}\varphi/2} & 0 \\ 0 & e^{{\rm i}\varphi/2}\end{pmatrix},\label{eq:R_z}
\end{align}
where $X =  \begin{pmatrix} 0 & 1\\ 1 & 0 \end{pmatrix}$, $Y =  \begin{pmatrix} 0 & -{\rm i}\\ {\rm i} & 0 \end{pmatrix}$, and $Z = \begin{pmatrix} 1 & 0\\ 0 & -1 \end{pmatrix}$ are \emph{Pauli matrices} (sometimes also written as $\sigma_x$, $\sigma_y$, and $\sigma_z$).

\subsubsection*{The NOT Gate}

The Pauli matrix $X$ has the effect of a NOT gate: $X\ket{0} = \ket{1}$ and $X\ket{1} = \ket{0}$.

\subsubsection*{Controlled-NOT Gate}

The Controlled-NOT gate is an example of a two-qubit gate.  It applies the NOT gate to the second qubit only if the first qubit is in the state $\ket{1}$.  As a matrix in the computational basis of two qubits, $\{\ket{00},\ket{01},\ket{10},\ket{11}\}$, Controlled-NOT ($C\text{-}X$) gate is
\begin{align*}
    C\text{-}X = \begin{pmatrix} 1 & 0 & 0 & 0 \\ 0 & 1 & 0 & 0 \\ 0 & 0 & 0 & 1 \\ 0 & 0 & 1 & 0 \end{pmatrix},
\end{align*}
so that $C\text{-}X \ket{00} = \ket{00}$, $C\text{-}X \ket{01} = \ket{01}$, $C\text{-}X \ket{10} = \ket{11}$ and $C\text{-}X \ket{11} = \ket{10}$.  An alternative way to write down the Controlled-NOT gate is using the bra-ket notation:
\begin{align}
    C\text{-}X = \ketbra{0}\otimes I + \ketbra{1}\otimes X. \label{eq:cx_tensor}
\end{align}
When used as a part of an operator, the tensor product $\otimes$ means that different parts of the operator apply to different qubits (or, more generally, qubit registers).  Consider a two-register state $\ket{\psi}\ket{\phi} \equiv \ket{\psi}\otimes \ket{\phi}$, then the operator $A\otimes B$ performs the operation $A$ on the state $\ket{\psi}$ in the first register and the operation $B$ on the state $\ket{\phi}$ in the second register:
\begin{align}
    (A\otimes B)\ket{\psi}\ket{\phi} = (A\ket{\psi})\otimes (B\ket{\phi}).
\end{align}
It is straightforward to verify that the operator $C\text{-}X$ in (\ref{eq:cx_tensor}) applied to two-qubit states performs the desired Controlled-NOT operation.

As an example, we can use the Controlled-NOT gate to create a maximally entangled two-qubit state, $\frac{1}{\sqrt{2}}(\ket{00}+\ket{11})$, starting with two qubits initialized to $\ket{00}$.  First, we apply a Hadamard gate $H$ to the first qubit: $H\ket{00} = \frac{1}{\sqrt{2}}(\ket{0}+\ket{1})\ket{0} = \frac{1}{\sqrt{2}}(\ket{00}+\ket{10})$.  Then, we apply $C\text{-}X$ to the register of two qubits: $C\text{-}X[\frac{1}{\sqrt{2}}(\ket{00}+\ket{10})] = \frac{1}{\sqrt{2}}(\ket{00}+\ket{11})$.  The $C\text{-}X$ gate flips the second qubit only when the first qubit is in the state $\ket{1}$, resulting in the desired state.

The three-qubit extension of the Controlled-NOT gate is the Toffoli gate, also known as the $CCNOT$ gate.  It applies $NOT$ to the third qubit conditionally on the state of the first two qubits being $\ket{11}$.  

Basic gates such as rotations, the NOT gate, CNOT gate, or the Hadamard gate can be implemented on existing quantum computers.

\subsubsection*{Auxiliary Qubits}

Many algorithms require supplementary qubits to support computation in addition to the qubit registers encoding data (Section~\ref{sec:encoding} discusses data encoding in detail).   These qubits are called \emph{auxiliary} or \emph{ancilla} qubits.   The addition of a single auxiliary qubit effectively doubles the Hilbert space:  An $n$-qubit register spans a $2^n$-dimensional Hilbert space; the addition of an auxiliary qubit expands this Hilbert space to $2^{n+1}$ dimensions.   Therefore, the addition of auxiliary qubits embeds the data registers in a larger space enabling, for example, non-linear transformation of the data registers (for an example of a non-linear transformation with the help of an auxiliary qubit, see \emph{Postselection} later in this Section).

\subsubsection*{Controlled Rotation}
\label{sec:ctrl-r}

Controlled-rotation is the application of a series of gates that act on an auxiliary qubit conditionally on the state of one or more other qubits.  For example, consider a state $\ket{x}$, which encodes an $n$-bit binary string $x$ on a register of $n$ qubits.  Append an auxiliary qubit to create the state $\ket{x}\ket{0}$.  A popular form of controlled rotation is
\begin{align}
    C\text{-}R_y(f(x)) =  \ketbra{x}\otimes e^{-iY f(x)}, \label{eq:c-r}
\end{align}
where $R_y$ is a single-qubit rotation operator introduced in \eqref{eq:R_y}, and $f(x)$ is a function of $x$ that is reasonably simple to compute.  In matrix notation, the operator $e^{-iY \phi}$ has the effect of $    e^{-iY \phi} = \Big[ \begin{array}{cc}
            \cos \phi & -\sin \phi \\
            \sin \phi & \cos \phi \end{array} \Big],$ 
so that the effect of the operator $ C\text{-}R_y(f(x))$ on the
state $\ket{x}$ and the auxiliary qubit is
\begin{align}
     C\text{-}R_y(f(x)) \ket{x}\ket{0} = \ket{x}\big(\cos f(x) \ket{0} + \sin f(x) \ket{1}\big).  \label{eq:ctrl-r}
\end{align}

A common example of $f(x)$ is arcsine of $x/C$, where $C$ is a constant selected so that $|x/C|\leq 0.5$.  Arcsine is efficient to compute using an expansion based on the inverse square root \citep[see, e.g.][and references therein]{haner2018optimizing}.  Quantum algorithms exist to approximate the arcsine function, inspired by classical reversible algorithms for the inverse square root -- used, e.g., in gaming \citep{lomont2003fast}.

Controlled rotation supports many important quantum algorithms such as quantum Monte Carlo (Section~\ref{sec:QMC}), the application of a Hermitian operator (Section~\ref{sec:hermitian}), or the HHL quantum linear systems algorithm (Section~\ref{sec:hhl}).

\subsubsection*{Controlled Unitary}
\label{sec:ctrl-u}

Controlled-unitary $C\text{-}U$ is an operation applied to a multi-qubit register conditional on the state of an auxiliary qubit:
\begin{align}
    C\text{-}U & = \ketbra{0} \otimes I + \ketbra{1} \otimes U.
\end{align}

The controlled-unitary operation exists for some, but not all unitaries \citep{lloyd2014quantum}.  

When the operator $U$ applies to a single-qubit register, $C\text{-}U$ is implemented using a decomposition of the operator $U$ such that $U = AXBXC$ and $ABC=I$.  Then, substituting CNOT ($C\text{-}X$) for NOT ($X$), we get the desired controlled-unitary.

\subsubsection*{Postselection}

Another important algorithmic building block is \emph{postselection}, where a quantum state is kept or discarded conditionally on the result of a quantum measurement of a part of the state (usually an auxiliary qubit).  Postselection enables nonlinear quantum transformations at the cost of having to discard quantum states where the measurement did not yield the desired result.  

For example, let $\ket{\psi}$ be a quantum state such that $\ket{\psi} = \sum_x\psi_x\ket{x}$, where the states $\{\ket{x}\}$ form an orthonormal basis.  We can apply a controlled rotation in Eq.~(\ref{eq:ctrl-r}) to the state $\ket{\psi}$ and an auxiliary qubit. The result of the controlled rotation is 
\begin{align}
    C\text{-}R_y\ket{\psi}\ket{0} = \sum_x \psi_x\ket{x}\big(\cos f(x) \ket{0} + \sin f(x) \ket{1}\big).
\end{align}
We now measure the auxiliary qubit.  If the measurement yields $1$, we keep the state; if $0$, we discard it and then repeat the preparation of the state $\ket{\psi}$, the controlled rotation, and the measurement until the measurement yields $1$.   The state we keep will be proportional to $\sum_x \psi_x \sin f(x) \ket{x}$ -- a nonlinear transformation of the original state $\ket{\psi}$.  Postselection plays an important role in algorithms such as the application of a Hermitian operator (Section~\ref{sec:hermitian}), the HHL quantum linear systems algorithm (Section~\ref{sec:hhl}), or the quantum singular value transformation (Section~\ref{sec:qsvt}).

\subsubsection*{Oracles}

\emph{Oracles} are not gates, but they too are important algorithmic building blocks -- in both classical and quantum computation.  Oracles are ``black box'' parts of algorithms that solve certain problems in a single operation.   They are calls of a function that do not take into account the structure of the function.   

Quantum algorithms employ two types of oracles -- classical oracles, which provide a classical solution to a given problem, and quantum oracles, which make the solution available to the quantum computer as a quantum state.

\section{Grover's Search and Descriptive Statistics on a Quantum Computer}
\label{sec:grover_section}

One of the most influential quantum algorithms is the search algorithm by Lov Grover \citep{grover1996fast}. Grover considers the problem of searching for a binary string $x_0$ among $N$ strings, provided that there is an oracle binary function such that $f(x) = 0$ if $x \neq x_0$ and $f(x_0) = 1$.  The classical search algorithm over an unstructured space requires $O(N)$ oracle calls; Grover's quantum algorithm requires $O(\sqrt{N})$ calls, offering a quadratic improvement.  

Grover's algorithm has been influential because it led to the development of a class of practically important applications of quantum computers, including those for efficient estimation of statistical quantities such as the sample mean, median, or minimum (maximum) of a function over a discrete domain.   Quantum Amplitude Amplification, based on Grover's algorithm, is a widely used subroutine to many other quantum algorithms.  It works by amplifying the amplitude of the correct result in a quantum state holding the result in a superposition with byproducts of computation.

\subsection{Grover's Search Algorithm}
\label{sec:grover_algo}

Assume, without loss of generality, that $N = 2^n$, where $n$ is an integer.  The starting state is a uniform superposition of all possible $n$-bit strings $x$ created on $n$ qubits:
\begin{align}\label{eq: uniform super state}
    \ket{s} = \frac{1}{\sqrt{N}}\sum_x \ket{x}.
\end{align} 
One of these bit strings is $x_0$ -- the target.  Grover's algorithm proceeds in a series of iterative steps, each amplifying the amplitude of the state holding $x_0$ by $2/\sqrt{N}$.  After $O(\sqrt{N})$ iterations this amplitude -- and the corresponding measurement probability -- becomes of order unity.   

To understand Grover's algorithm, it is helpful to think of the state $\ket{s}$ as a superposition of the target state $\ket{x_0}$ and its orthogonal complement $\ket{s'}$:
\begin{align}
     \ket{s'}=\frac{1}{\sqrt{N-1}}\sum_{x\neq x_0}\ket{x},\;\;\;\;\;\; \braket{s'}{x_0}=0,
\end{align}
so that 
\begin{align}
    \ket{s} & = \frac{\sqrt{N-1}}{\sqrt{N}}\ket{s'}+\frac{1}{\sqrt{N}}\ket{x_0}. \label{eq:s_sp_x0}
\end{align} 
When viewed as vectors in a Hilbert space, the starting state $\ket{s}$ is very close to $\ket{s'}$.  Let $\theta/2$ be the angle between $\ket{s}$ and $\ket{s'}$.  Then, by Eq.~(\ref{eq:s_sp_x0}), $\sin (\theta/2) = 1/\sqrt{N}$.  

At each iteration of Grover's algorithm, the quantum state of the system is pushed slightly toward $\ket{x_0}$ and away from $\ket{s'}$ by an angle $\theta$. The state stays in the two-dimensional plane spanned by $\ket{x_0}$ and $\ket{s'}$, and the result of each iteration is a superposition of $\ket{x_0}$ and $\ket{s'}$. After $t$ iterations such that $\sin^2 ((t+\frac{1}{2})\theta) \approx 1$, the algorithm results in a quantum state which, when measured in the computational basis, yields $x_0$ with a probability close to 1.   The number of iterations is $t \approx \frac{\pi}{4}\sqrt{N}$ or $O(\sqrt{N})$.

Let $U_G$ represent one Grover iteration.  Each iteration consists of two unitary operators $U_s$ and $U_f$: $U_G = U_sU_f$.  The first operator is $U_f = I - 2\ketbra{x_0}$, a reflection with respect to $\ket{s'}$ in the plane spanned by $\ket{x_0}$ and $\ket{s'}$ as it maps $\ket{x_0}$ into $-\ket{x_0}$.  Application to the initial state $\ket{s}$ yields
\begin{align}
    U_f \ket{s} = (I - 2\ketbra{x_0})\ket{s} = \ket{s} - \frac{2}{\sqrt{N}}\ket{x_0} = \ket{s'} - \frac{1}{\sqrt{N}}\ket{x_0}.
\end{align}
The second operator is $U_s = 2\ketbra{s}-I$, a reflection around the state $\ket{s}$ which, applied after the operator $U_f$, results in an increased amplitude of $\ket{x_0}$:
\begin{align}
    U_G\ket{s} = U_s U_f \ket{s} = (2\ketbra{s}-I)(\ket{s} - \frac{2}{\sqrt{N}}\ket{x_0}) = (1-\frac{4}{N})\ket{s} + \frac{2}{\sqrt{N}}\ket{x_0}.
\end{align}
So, the application of $U_G$ to the state $\ket{s}$ has amplified the amplitude of the target state $\ket{x_0}$. Both operators $U_f$ and $U_s$ can be implemented on a quantum computer.  

The implementation of operator $U_f$ requires an auxiliary qubit and quantum oracle access $O_f$ to the function $f(x)$.  Given a value $X$ encoded in a quantum state, the oracle $O_f$ records the value $f(x)$ in a quantum state in the following way.   Let $\ket{x}\ket{y}$ be a quantum system that represents the value $x$ and, in the auxiliary qubit, some value $y$.  The quantum oracle $O_f$ applied to $\ket{x}\ket{y}$ mod-adds the value $f(x)$ to the value of the auxiliary qubit: $O_f\ket{x}\ket{y} = \ket{x}\ket{y \oplus f(x)}$, where $\oplus$ indicates mod-addition.  It is easy to show that, if we start with the auxiliary qubit in the Hadamard $\ket{-}$ state, we have $O_f\ket{x}\ket{-} = (U_f\ket{x})\ket{-}$.

The operator $U_s$ can be implemented using a decomposition into three parts:  an uncomputation of $\ket{s}$ to recover the ground state $\ket{0}$, a sign-flip on the ground state, and a re-computation of $\ket{s}$.  Let $\mathcal{A}$ be the operator that creates the state $\ket{s}$ when applied to the ground state: $\ket{s} = \mathcal{A}\ket{0}$.  In the case when $\ket{s}$ is a uniform superposition of all $n$-bit strings as in \eqref{eq: uniform super state}, we know that $\mathcal{A} = H^{\otimes n}$, but we will use the more general form here because it will help us to derive the Quantum Amplitude Amplification algorithm in Section~\ref{sec:qaa}.   Let $U_0 = I - 2\ketbra{0}{0}$, an operator that changes the sign of the $\ket{0}$ state.  Then,
\begin{align}
    -\mathcal{A}U_0\mathcal{A}^{\dagger} = - \mathcal{A}( I - 2\ketbra{0})\mathcal{A}^{\dagger}= 2\mathcal{A}\ketbra{0}\mathcal{A}^{\dagger}-\mathcal{A}I\mathcal{A}^{\dagger} = 2\ketbra{s} - I = U_s, \label{eq:u_s_three_parts}
\end{align}
where we used the fact that the operator $\mathcal{A}$ is unitary, and therefore, $\mathcal{A}^{\dagger} = \mathcal{A}^{-1}$. Equation (\ref{eq:u_s_three_parts}) demonstrates the implementation of $U_s$ using the the three parts mentioned above. 

With minor modifications, it is possible to apply Grover's algorithm to the situation with more than one desired state \citep{boyer1998tight}. 

\cite{brassard2002quantum} provide an intuitive explanation for how Grover's algorithm delivers the quadratic improvement in computational efficiency.   Consider a classical randomized algorithm that succeeds with probability $p$, where $p \ll 1$.  After $j$ repetitions of the algorithm, if $j$ is small enough, the cumulative probability of success is approximately $jp$.  In this classical case, the probability of success increases by a constant increment $p$ at each iteration.   Grover's algorithm increases the \emph{amplitude} of the desired state $\ket{x_0}$ by an approximately constant increment at each iteration.  The probability that a quantum measurement of the resulting quantum state yields the outcome $x_0$ is proportional to the squared amplitude of $\ket{x_0}$, resulting in a quadratically faster increase in probability at every iteration of Grover's algorithm.

\subsection{Quantum Amplitude Amplification (QAA)}
\label{sec:qaa}

Many quantum algorithms deliver the result of computation in a quantum superposition state $\ket{\psi}$ of ``good'' and ``bad'' results.  The extraction of the ``good" result often requires repeated measurements of auxiliary qubits.  For example, the quantum algorithm in Section~\ref{sec:hermitian} (application of a Hermitian operator to a quantum state) produces the desired result only when the measurement of the auxiliary qubit yields 1.  If the probability of measuring $1$ is $p$, then it takes roughly $1/p$ measurements on average to extract the desired result.  In some cases the probability $p$ may be quite small, making the measurement step a substantial computational overhead.

\cite{brassard2002quantum} generalize Grover's algorithm to mitigate this problem and reduce the computational burden of repeated measurements from $O(1/p)$ to $O(1/\sqrt{p})$.  Let $\mathcal{A}$ be the algorithm that creates the state $\ket{\psi}$: $\ket{\psi}=\mathcal{A}\ket{0} = \sum_{x}a_x\ket{x}$, a superposition of ``good'' and ``bad'' results. Consider a validation function $\chi$ that returns $1$ if $x$ is a ``good'' result, and $0$ if a ``bad'' result.  The aim of the QAA algorithm is to amplify the amplitudes of the subspace of ``good'' results in order to increase the probability that a measurement yields those results.  

The algorithm leverages the fact that the quantum state $\ket{\psi}$ can be decomposed as
\begin{align}
    \ket{\psi} = a_1\ket{\psi_1}+\sqrt{1-a_1^2}\ket{\psi_2},\label{eq:psi_psi1_psi2} 
\end{align}
where $\ket{\psi_1}$ is a projection of $ \ket{\psi}$ onto the ``good'' subspace spanned by the states representing the ``good'' results and $\ket{\psi_2}$ is the projection onto its complement -- the ``bad'' subspace.  The probability $p$ of a measurement of $\ket{\psi}$ yielding a ``good'' state is $p = |a_1|^2$.

An iteration of QAA, represented as an operator $Q$, increases the amplitude of the ``good'' state $\ket{\psi_1}$.   By analogy with Grover's search algorithm, the operator $Q$ is
\begin{align}
    Q = -\mathcal{A}U_0\mathcal{A}^\dagger U_\chi,
\end{align}
where
\begin{align}
    U_0 = I - 2\ketbra{0},\;\;\;\;\;U_\chi  = I - 2\ketbra{\psi_1},
\end{align}
and the algorithm $\mathcal{A}$ is assumed to be reversible, i.e.~containing no measurements.

Repeated applications of $Q$ gradually increase the amplitude of the ``good'' $\ket{\psi_1}$ component.  After $t$ measurements, where $t \approx \frac{\pi}{4|a_1|} = \frac{\pi}{4\sqrt{p}}$, the probability that a measurement of $Q^t\ket{\psi}$ yields a ``good'' component is of order unity.  

Both QAA and Grover's algorithms are periodic.  Once the minimum error is reached at $t \approx \frac{\pi}{4|a_1|}$, repeated applications of the search operator $Q$ start pushing the state of the system away from the target state, and the error starts to increase until $t \approx \frac{3\pi}{4|a_1|}$.  After maximum error is reached, repeated applications of $Q$ start pushing the system closer to the target state again. In the absence of prior knowledge of $|a_1|$, the periodicity makes it difficult to determine the stopping point optimally. The literature refers to this problem as the ``souffl\'{e} problem,'' referring to the way the dessert rises during baking, but starts to deflate if baked too long (the analogy would have been more apt if the souffl\'{e} inflated and deflated periodically with baking time).  Solutions to the problem include fixed-point quantum search \citep{yoder2014fixed} and variable time amplitude amplification \citep{ambainis2012variable}.

\subsection{Quantum Amplitude Estimation (QAE)}
\label{sec:qae}

As discussed in the previous section, the QAA algorithm is periodic with respect to repeated application of the amplification step $Q$.  When the amplification step $Q$ is applied $t$ times, $Q^t$, the amplification error cycles between its minimum and maximum with a period of $t = \frac{\pi}{|a_1|}$, where $a_1$ is the amplitude of the ``good'' subspace in (\ref{eq:psi_psi1_psi2}).  \cite{brassard2002quantum} leverage the fact that the period of $Q^t$ is a function of the absolute value of the amplitude $|a_1|$ to estimate this amplitude using Quantum Phase Estimation, an influential algorithm we describe in Section~\ref{sec:qpe}.

The QAE algorithm starts with a register of $n$ qubits containing $\ket{\psi} = \mathcal{A}\ket{0}$ and an auxiliary register of $n$ qubits, initialized to a uniform superposition $\frac{1}{\sqrt{N}}\sum_{j=0}^{N-1}\ket{j}$, where $N=2^n$.  Let $\Lambda_N(U)$ be a controlled unitary operator that applies multiple copies of a unitary $U$ conditional on the state of the auxiliary qubit register: 
\begin{align}
    \Lambda_N(U) \ket{j}\ket{\psi} = \ket{j}(U^j\ket{\psi}),
\end{align}
where the state $\ket{j}$ acts as a reference.  The operator $\Lambda_N(Q)$ applied to the state $\frac{1}{\sqrt{N}}\sum_{j=0}^{N-1}\ket{j}\ket{\psi}$ creates a superposition of states with a range of repeated quantum amplitude amplification steps $Q^j$.   Quantum Phase Estimation enables the extraction of the phase of $Q^j$.  Quantum Phase Estimation records $\hat{y}$, an $n$ bit approximation of $y$ such that $|a_1|^2 = \sin^2 (\pi \frac{y}{N})$, in the computational basis in the auxiliary register.  Measurement of the auxiliary register yields the outcome $\ket{\hat{y}}$ with probability 
of at least $8/\pi^2$.   With $\hat{a}_1 = \sin^2 (\pi \frac{\hat{y}}{N})$, the estimation error bound after $t$ iterations of the algorithm is:
\begin{align}
    |\hat{a}_1 - a_1| & \leq 2\pi\frac{a_1(1-a_1)}{t}+\frac{\pi^2}{t^2}.  \label{eq:qae_error_est}
\end{align}

\subsection{Estimating the Mean of a Bounded Function}
\label{sec:mean}

The QAE algorithm provides an efficient way to estimate the mean of a bounded function (see Section \ref{sec:readout} for a method to estimate the mean amplitude of a state that leverages the Swap Test).

Let $F: \{0,\ldots,N-1\} \to X$, where $X \subset [0,1]$, be a black-box function.   \cite{brassard2011optimal} propose a method to approximate the mean value of $F$, $\mu = \frac{1}{N}\sum_x F(x)$; see also \cite{heinrich2002quantum}.  The idea is to create a state of the form:
\begin{align}
    \ket{\psi} = \alpha\ket{\psi_0} + \beta\ket{\psi_1},\label{eq:psi_for_mean}
\end{align}
such that $\ket{\psi_0}$ and $\ket{\psi_1}$ are orthogonal and $|\beta|^2 = \frac{1}{N}\sum_{i=0}^{N-1}F(i) = \mu$.  If the creation of state $\ket{\psi}$ requires $O(1)$ oracle calls to $F$, then QAA can yield an estimate of $|\beta|^2$ with precision $\epsilon$ in $O(1/\epsilon)$ oracle calls regardless of the size $N$.  

Creation of state $\ket{\psi}$ in (\ref{eq:psi_for_mean}) proceeds in a few steps and requires three registers.  The first, $n$-qubit register $\ket{ }_n$ encodes index values $i$ in the computational basis; the second, $m$-qubit register $\ket{ }_m$ is an auxiliary register that temporarily holds an $m$-bit approximation of $F(i)$ -- i.e.~an approximation of $F(i)$ written down using $m$ bits; the third, single-qubit register $\ket{ }_1$ is an auxiliary register that helps create orthogonal states $\ket{\psi_0}$ and $\ket{\psi_1}$.  

Let $\mathcal{A}$ be an algorithm that encodes an $m$-bit approximation of $F(i)$ in an $m$-qubit auxiliary register initialized to $\ket{0}_m$:
\begin{align}
    \mathcal{A}\ket{i}_n\ket{0}_m = \ket{i}_n\ket{F(i)}_m.
\end{align}
Using a control rotation operator $C\text{-}R$ (see Section~\ref{sec:ctrl-r}), we transfer the values $F(i)$ into the amplitudes of the quantum state
\begin{align}
    C\text{-}R\ket{F(i)}_m\ket{0}_1 = \ket{F(i)}_m\big(\sqrt{1-F(i)}\ket{0}_1 + \sqrt{F(i)}\ket{1}_1 \big).
\end{align}
The operator $A = (\mathcal{A}^{-1} \otimes I_1)(I_m\otimes I_n \otimes C\text{-}R)(\mathcal{A}\otimes I_1)$, where $I_m$, $I_n$, and $I_1$ are identity operators acting on the $m$- and $n$-qubit registers and the auxiliary qubit, respectively, applies an extension of $\mathcal{A}$ to all the three registers and then uncomputes the stored values of $F(i)$.   When applied to a state where the $n$-qubit register holds a uniform superposition of $n$-bit strings $\ket{i}$, the operator $A$ produces the desired state $\ket{\psi}$:
\begin{align}
    \ket{\psi} = A\big(\frac{1}{\sqrt{N}}\sum_{i=0}^{N-1}\ket{i}_n\big)\ket{0}_m\ket{0}_1 = \frac{1}{\sqrt{N}}\sum_{i=0}^{N-1}\ket{i}_n\ket{0}_m\big(\sqrt{1-F(i)}\ket{0}_1 + \sqrt{F(i)}\ket{1}_1 \big).
\end{align}
Discarding the $m$-qubit auxiliary register and rearranging, we have:
\begin{align}
    \ket{\psi} = & \frac{1}{\sqrt{N}}\sum_{i=0}^{N-1}\ket{i}_n\big(\sqrt{1-F(i)}\ket{0}_1 + \sqrt{F(i)}\ket{1}_1 \big) \label{eq:mean_f} \\ \nonumber
    = & \frac{\sqrt{N-\sum_{j}F(j)}}{\sqrt{N}} \Big[\frac{1}{\sqrt{N-\sum_{j}F(j)}}\sum_{i=0}^{N-1}\sqrt{1-F(i)}\ket{i}_n\ket{0}_1\Big] \\ \nonumber
    & + \frac{\sqrt{\sum_{j}F(j)}}{\sqrt{N}} \Big[\frac{1}{\sqrt{\sum_{j}F(j)}} \sum_{i=0}^{N-1}\sqrt{F(i)}\ket{i}_n\ket{1}_1\Big] \\ \nonumber
    = & \alpha\ket{\psi_0} + \beta\ket{\psi_1},
\end{align}
where the expressions in the square brackets are the properly normalized and orthogonal states $\ket{\psi_0}$ and $\ket{\psi_1} $. The coefficients in front of the brackets are $\alpha$ and $\beta$ respectively, so that $|\beta|^2 = \frac{1}{N}\sum_{i=0}^{N-1}F(i) = \mu$ can be efficiently approximated by QAE, described in Section~\ref{sec:qae}.  The creation of the quantum state $\ket{\psi}$ requires only two calls to the quantum oracle $\mathcal{A}$ (one to initialize the $m$-qubit auxiliary register containing the values $\ket{F(i)}$ and the other to uncompute it).

\cite{montanaro2015quantum} extends the \citeauthor{brassard2011optimal} method to functions $F$ with non-negative output, $X\subset \mathbb{R}_{\geq 0}$.   The method decomposes the function $F$ into $k$ components $F_{(x_{i-1},x_{i}]}$, for $i = 1,..,k$, whose output falls within disjoint intervals $(x_{i-1},x_{i}]\subset X$ of the codomain of $F$.  The mean is then estimated for each function $F_{(x_{i-1},x_{i}]}$, and the overall mean of the function of $F$ is constructed from the means of $F_{(x_{i-1},x_{i}]}$.   \cite{montanaro2015quantum} extends the method further to functions with output that does not have to be non-negative but has a bounded variance.

\subsection{Minimum (or Maximum) of a Function over a Discrete Domain}

The fastest classical algorithm for finding the minimum (or maximum) value of in an unsorted table of $N$ items requires $O(N)$ oracle calls;  with the help of a quantum computer this requirement decreases to $O(\sqrt{N})$ through repeated application of the QAA \citep{durr1996quantum}.

Let $F: \{0,\ldots,N-1\} \to X$, $X \subset \mathbb{R}$ be a black-box function.  The algorithm outputs $y^*$, the index corresponding to the minimum value of $F$.   
The algorithm starts by selecting uniformly at random an index value $y$ such that $0 \leq y \leq N-1$.  Assuming without loss of generality that $N = 2^n$ for an integer $n$, two registers of $n$ qubits each can support a quantum state of the form $\frac{1}{\sqrt{N}}\sum_{j=0}^{N-1}\ket{j}\ket{y}$.  The first register encodes a uniform superposition of all indices $j$, the second register encodes the value $y$, both in the computational basis (i.e.~in the form of binary strings).  

Mark every item with $F(j) < F(y)$ using an efficient oracle -- i.e.~an oracle able to perform the operation in $O(\log N)$ calls to the function $F$ (see subsection \emph{Oracles} in Section~\ref{sec:gates}).  The marked items represent the ``good'' state in QAA (Section~\ref{sec:qaa}).   Run the algorithm to amplify this ``good'' state.  Next, uniformly at random select $y'$ until $F(y') < F(y)$ and re-run the algorithm with $y'$ instead of $y$.  \cite{durr1996quantum} show that, after $22.5\sqrt{N}+1.4\log_2^2 N$ calls to $F$, the desired state is reached with probability at least $1/2$.

\subsection{Median and $k$th Smallest Value}

\cite{nayak1999quantum} generalized the algorithm of \cite{durr1996quantum} to find not just the smallest value, but the $k$th smallest value of a discrete function $F$, including the $N/2$th smallest value which is the median. %
Denote by $\text{rank}(F(i))$ the rank of $F(i)$ in $\{F(0),...,F(N-1)\}$ ordered in non-decreasing order.
Let $\Delta \geq 1/2$ be the accuracy parameter.  Given $k$, $1\leq k\leq N$, the problem is to find the value $F(i)$ such that $\text{rank}(F(i))$ is the smallest value in $(k-\Delta,k+\Delta)$. This is referred to as the $\Delta$-approximation of the $k$th smallest element of $F$.

The algorithm relies on two subroutines.
The first subroutine implements a function $K(l)$ that returns `yes' if $F(l)$ is in the $\Delta$-approximation of the $k$th smallest element; `$<$' if the rank of $F(l)$ is at most $k - \Delta$; and `$>$' if the rank of $F(l)$ is at least $k+\Delta$. 
This subroutine can be implemented based on the counting algorithm of \cite{brassard1998quantum} (see Section~\ref{sec:counting}).
The second subroutine implements the sampler $S(i,j)$ that chooses an index $l$ uniformly at random such that $F(i) < F(l) < F(j)$.  The sampler can be implemented based on the generalized search algorithm of \cite{boyer1998tight}.

The $k$th smallest value algorithm works as follows:  For convenience, define $F(-1) = -\infty$ and $F(N) = \infty$, 
\begin{enumerate}
    \item $i \leftarrow -1$; $j\leftarrow N$
    \item $l \leftarrow S(i,j)$
    \item If $K(l)$ returns `yes', output $F(l)$ (and/or the index $l$) and stop.  Else, if $K(l)$ returns `$<$', $i \leftarrow l$, go to step 2.  Else, if K(l) returns `$>$', $j \leftarrow l$, go to step 2.
\end{enumerate}
\cite{nayak1999quantum} prove that the expected number of iterations before termination is $O(\log N)$.\footnote{More precisely, let $n = \sqrt{N/\Delta}+\sqrt{k(N-k)}/\Delta$.  The expected number of iterations is $O(\log n)$.}

\subsection{Counting}
\label{sec:counting}
In some applications, the problem of interest is not to \emph{find} the solution but to \emph{count} how many solutions exist. 
Consider the function $F : \{0,\ldots,N-1\} \to X=\{0,1\}$. We are interested in counting the number of indices $x$ such that $F(x)=1$.
Then, the algorithm for estimating the mean of $F$ is effectively an algorithm for counting how many indices $x$ such that $F(x) = 1$ there are.

\subsection{Quantum Monte Carlo}\label{sec:QMC}

The techniques used in the previous sections provide a way to estimate the mean of a function with respect to a probability distribution \citep{low2017optimal,rebentrost2018finance}.   Classically, the most popular method for estimating an expectation is Monte Carlo simulation.   In this method, samples are drawn from the probability distribution and the function is evaluated for each sample;  the average of these outputs, $\tilde{\mu}$, is the Monte Carlo estimate of the true mean $\mu$.    Chebyshev's inequality guarantees that, for $k$ independent draws from the probability distribution, the probability that the estimate is far from the real mean $\mu$ is bounded
\begin{align}
    \text{Pr}[|\hat{\mu}-\mu|\geq \epsilon] \leq \frac{\sigma^2}{k\epsilon^2},
\end{align}
where $\sigma^2$ is the variance of the function with respect to the probability distribution. In other words, in order to estimate $\mu$ up to an additive error $\epsilon$, $k  = O(\sigma^2/\epsilon^2)$ samples -- and function evaluations -- are required.

Quantum Amplitude Estimation holds the promise of reducing the required number of function evaluations to $O(\sigma/\epsilon)$ to achieve the same error bound. That is, Quantum Monte Carlo provides a quadratic speedup over classical Monte Carlo. 
The quantum method requires two efficient quantum oracles.  The first oracle $\mathcal{P}$ helps to prepare a state that encodes the probability distribution in a quantum state -- called a \emph{quantum sample state}.  The second oracle $\mathcal{F}$ applies the function whose mean is to be estimated with respect to the probability distribution. 

To define the quantum sample state, let $X$ be a finite $N$-dimensional set, and $p(x)$ a probability distribution over $X$, such that $\sum_{x\in X} p(x) = 1$.  Let the basis set of an $N$-dimensional Hilbert space, $\{\ket{x}\}$, represent the elements $x$ of $X$.  The quantum state $\ket{p}$ of the form
\begin{align}
    \ket{p} = \sum_{x \in X} \sqrt{p(x)} \ket{x}
\end{align}
is the quantum sample with respect to the distribution $p$.   The state $\ket{p}$ has the property that, by the Born rule, the probability that a  measurement of all qubits supporting the state yields $x$ is $p(x)$.

Using these two oracles $\mathcal{P}$ and $\mathcal{F}$ and an auxiliary qubit we have:
\begin{align}
    \mathcal{P}:& \ket{0} \mapsto \ket{p} = \sum_{x \in X} \sqrt{p(x)} \ket{x}, \\
    \mathcal{F}:& \ket{p}\ket{0} = \sum_{x \in X} \sqrt{p(x)} \ket{x} (\sqrt{1-f(x)}\ket{0} + \sqrt{f(x)}\ket{1}). \label{eq:mean_pi_f}
\end{align}
The quantum state in (\ref{eq:mean_pi_f}) has a structure parallel to that of the quantum state in (\ref{eq:mean_f}), and we can apply to it a similar technique to isolate the expectation value of $f$ with respect to the distribution $p$, $\mathbb{E}_p[f] = \sum_{x\in X} p(x) f(x)$. If $\mathcal{F}$ is \emph{efficiently computable} (i.e., with a sub-polynomial number of operations), then the algorithm yields $\hat{f}$, an estimate of $\mathbb{E}_p[f]$ within the error bound $\epsilon$ with $O(\sigma/\epsilon)$ function evaluations.

\subsubsection*{Further Discussion on Quantum Sample Preparation}

The quantum Monte Carlo method described above delivers a quadratic speedup over classical methods provided the quantum oracles $\mathcal{P}$ and $\mathcal{F}$ can be implemented efficiently.   If the function $f$ is easily computable, then $\mathcal{F}$ has an efficient implementation \citep{low2017optimal,rebentrost2018finance}.  But implementing the oracle $\mathcal{P}$ to create a quantum sample can be challenging.  For an arbitrary probability distribution, preparing a quantum sample state is computationally equivalent to solving the graph isomorphism problem and is exponentially hard \citep{plesch2011quantum,chakrabarti2019quantum}.  For efficiently integrable distributions, such as the normal distribution, the method proposed by \cite{grover2002creating} has been popular.   However, \cite{herbert2021problem} recently demonstrated that the \cite{grover2002creating} method is limited to situations simple enough to be solved without the use of the Monte Carlo method -- classical or quantum.  Because quantum Monte Carlo delivers quadratic rather than exponential speedup over classical methods, the modest computation overhead of the \cite{grover2002creating} method negates the quantum gains.

Efficient preparation of quantum samples for distributions of practical interest is, at the time of writing this review, an open problem in quantum algorithm design.   A machine-learning approach based on empirical data has been proposed by \cite{zoufal2019quantum}.    \cite{vazquez2021efficient} propose to view the probability distribution function $p$ as a function, implemented similarly to function $f$, using a controlled rotation of an auxiliary qubit. This method works for simpler distributions, with an efficiently-computable $p(x)$.     For more complex, high-dimensional distributions \cite{kaneko2021quantum} propose to create quantum samples using pseudorandom numbers.  \cite{an2021quantum} quantize the classical method of multilevel Monte Carlo to find approximate solutions to stochastic differential equations (SDEs), particularly for applications in finance.

It may also be possible to achieve relatively efficient quantum sampling using a sequence of slowly varying quantum walks (see Section \ref{sec:quantum_walks}) -- the quantum equivalents of Markov chains \citep{wocjan2008speedup,wocjan2009quantum}.  The method, called Quantum Markov chain Monte Carlo and discussed in more detail in Section~\ref{sec:qmcmc}, is analogous to classical Markov chain Monte Carlo (MCMC), and can be used to generate a Markov chain with a given equilibrium distribution $\pi$. 

\section{Quantum Markov Chains}
\label{sec:quantum_walks}

This section introduces Quantum Markov chains, often called \emph{quantum walks} in the quantum computing literature, the quantum equivalents of classical Markov chains which are widely used in probability and statistics. 
Quantum walks can provide polynomial speed-ups for a wide variety of problems from estimating the volume of convex bodies \citep{chakrabarti2019quantum} to option pricing \citep{an2021quantum}, search for marked items \citep{magniez2011search}  and active learning in artificial intelligence \citep{paparo2014quantum}.\footnote{For special classes of problems, such as a subclass of black-box graph traversal problems, quantum walks deliver exponential speeds up over any classical algorithm \citep{childs2003exponential}.  Generally, quantum walks are universal for quantum computation \citep{childs2009universal} (i.e. any sequence of gates can be expressed as a quantum walk).} 
However, because of quantum interference,
quantum Markov chains behave substantially differently from their classical counterparts.
For example, quantum Markov chains do not admit any equilibrium distribution, but the average of the states of the chain does (see below for a formal definition). This makes applications of quantum Markov chains in statistics different from those of classical chains, and can lead to new interesting and important applications.

Questions of interest to statisticians are: how to quantize a Markov chain, i.e. how to implement a Markov chain in a quantum computer? What are the properties of the resulting quantum Markov chain? How can this quantum Markov chain be used in statistical applications such as MCMC sampling?
In this section, we review the two most popular approaches for quantizing  a Markov chain: \emph{coin walks} and \emph{Szegedy walks} \citep{szegedy2004quantum,watrous2001quantum}.  Coin walks quantize Markov chains on unweighted graphs.  Szegedy walks work on weighted directional graphs.  We focus on Markov chains with a discrete state space, but it is also theoretically possible to quantize a continuous-space Markov chain.\footnote{In practice, quantum computers have finite precision and can only encode finite (if high-dimensional) sets, making discrete-space quantum samples most relevant for quantum algorithm applications \citep{chakrabarti2019quantum}. }   For a detailed and thorough review of quantum walks, see \cite{venegas2012quantum}.

\subsection{Coin Walks}
Consider a Markov chain with a discrete state space. It can be represented on a graph $G=(V,E)$: the vertices $V$ represent the states; after each time step, the chain stays at the current vertex or jumps to one of its adjacent vertices according to a transition probability.
This creates a random walk on the graph.

Let $H_V$ and $H_E$ be Hilbert spaces whose basis states encode the vertices in $V$ and the edges in $E$, respectively.
Define a shift operator $S$ on $H_V\otimes H_E$ that determines the next vertex $u$ given the current vertex $v$ and the edge $e$, i.e., $S\ket{e}\ket{v}=\ket{e}\ket{u}$.
Define a \emph{coin operator} $C$ to be a unitary transformation on $H_E$. Then, $U=S(C\otimes I)$ implements one step of the random walk on graph $G$. If the initial state is $\ket{\psi_0}$, the state after $t$ steps is
\[\ket{\psi_t}=U^t\ket{\psi_0}.\]
The dynamic of this quantum random walk is governed by the coin operator $C$. 
Because of the quantum interference and superposition effect, the distribution of $\ket{\psi_t}$ behaves very differently from the classical Markov chain.  
Denote by $P_t(v|\psi_0)$ the probability of finding $\ket{\psi_t}$ at a node $v\in V$. The probability distribution $P_t(\cdot|\psi_0)$ does not converge \citep{venegas2012quantum}, but its average does. More precisely, let 
\[\bar P_t(v|\psi_0)=\frac{1}{t}\sum_{s=1}^tP_s(v|\psi_0), \]
then $\bar P_t(\cdot|\psi_0)$ converges and this stationary distribution can be determined.
With a suitable definition of mixing time, it is shown that the mixing time of a quantum walk is quadratically faster than a classical random walk - 
a property that attracted attention of researchers seeking to speed up algorithms based on Markov chains.

\subsection{Szegedy Walks}
\label{sec:szegedy}

\cite{szegedy2004quantum}, based on the earlier work of \cite{watrous2001quantum}, proposed another approach to quantize Markov chains. 
Consider a Markov chain operating  on a bipartite graph.
Let $X$ and $Y$ be two finite sets, and matrices $P$ and $Q$ describe the probabilities of jumps from elements of $X$ to elements of $Y$ and $Y$ to $X$, respectively.   The  elements of $P$ and $Q$, $p_{x,y}$ and $q_{y,x}$, are transition probabilities and, as such, are non-negative and normalized so that $\sum_{y\in Y}p_{x,y}=1$ and $\sum_{x\in X}q_{y,x}=1$.  A Markov chain that maps $X$ to $X$ with a transition matrix $P$ is equivalent to a bipartite walk where $q_{y,x} = p_{y,x}$ for every $x,y \in X$ or, equivalently, $Q=P$.

To quantize the bipartite random walk, 
define a two-register quantum system spanned by $\ket{x}\ket{y}$ with $x\in X$, $y\in Y$.
We start with two unitary operators,
\begin{align}
     U_P: \ket{x}\ket{0} \mapsto \sum_{y\in Y}\sqrt{p_{x,y}}\ket{x}\ket{y},\;\;\;\;\;
     V_Q: \ket{0}\ket{y} \mapsto \sum_{x\in X}\sqrt{q_{y,x}}\ket{x}\ket{y},
\end{align}
which are quantum equivalents of the transition matrices $P$ and $Q$.  The quantization is based on the observation that the Grover ``diffusion'' operator $U_f$ (from Section~\ref{sec:grover_algo}) is similar to a step of a random walk over a graph -- a transition from each state to all the other $N$ states.  In matrix form, the operator $U_f$ (up to an overall negative sign) is such that its off-diagonal elements equal $\frac{2}{N}$ and the diagonal elements are $-1 + \frac{2}{N}$.  This unitary operator effectively distributes quantum probability mass from each node to all the other nodes -- a property that led to naming this operator a ``diffusion'' operator.

Using the operators $U_P$ and $V_Q$ we define operators similar to Grover's diffusion operators:
\begin{align}
    \mathcal{R}_1  = 2 U_PU_P^\dagger - I,\;\;\;\;\;
    \mathcal{R}_2  = 2 V_QV_Q^\dagger - I,   
\end{align}
where the identity operator $I$ acts on both registers.
The quantum walk operator $W$ is defined as the product of the two diffusion operators
\begin{align}
    W = \mathcal{R}_2\mathcal{R}_1.
\end{align}

For a Markov chain from $X$ to $X$, the expression can be simplified by replacing $R_2$ with $SR_1S$, where $S$ is the swap operator, which swaps the two registers:
\begin{align}
    S & = \sum_{x,y} \ketbra{y,x}{x,y}.
\end{align}
This operator is self-inverse, so that $S^2 = I$ and $SR_1S^{-1} = SR_1S$.  For a Markov chain we can write
\begin{align}
    W = S(2 U_PU_P^\dagger - I) S(2 U_PU_P^\dagger - I),
\end{align}
so that some researchers \cite[see e.g.][]{chakrabarti2019quantum} define a step of a quantized Markov chain as
\begin{align}
    W_{MC} = S(2 U_PU_P^\dagger - I).\label{eq:qw_unitary}
\end{align}

A Szegedy quantum walk, similar to a coin walk, is a unitary process and, as such, does not converge to a stationary  distribution. Instead the quantum walk ``cycles through'' the stationary distribution $\pi$ of $P$, similarly to the way Grover's search (Section~\ref{sec:grover_algo}) or QAA (Section~\ref{sec:qaa}) pass through the desired state with a certain period (QAE, described in Section~\ref{sec:qae}, exploits this periodicity).  The quantum state analogous to the stationary distribution $\pi$, $\ket{\pi} = \sum_x \sqrt{\pi(x)}\ket{x}$, is the highest-eigenvalue eigenstate of the Szegedy quantum walk operator $W$ \citep{orsucci2018faster}; the eigenvalue of eigenstate $\ket{\pi}$ equals $1$, i.e. $W\ket{\pi} = \ket{\pi}$. This important property of 
the Szegedy quantum walk can be exploited to develop quantum Markov chain Monte Carlo to sample from $\pi$; see Section \ref{sec:qmcmc}.

\subsection{Quantum Markov Chain Monte Carlo}
\label{sec:qmcmc}
Can a quantum walk be used to derive a quantum Markov chain Monte Carlo algorithm for sampling from a target probability distribution? The answer is yes. Consider a distribution $\pi$ over a discrete space $X$; the problem is to prepare a quantum sample $\ket{\pi} = \sum_{x\in X} \sqrt{\pi(x)}\ket{x}$.

Let $P$ be the transaction matrix of a classical ergodic Markov chain with the stationary distribution $\pi$; $P$ can be derived based on, e.g., the Metropolis algorithm.
Let $W(P)$ be the Szegedy quantum walk with respect to $P$. Then $\ket{\pi}$ is the unique eigenstate of $W(P)$ with the eigenvalue 1. All other eigenstates have an eigenphase which is at least quadratically larger than the spectral gap $\delta$ -- the difference between the top and the second highest eigenvalues.  This property allows one to use phase estimation (or phase detection) to distinguish $\ket{\pi}$ from the other eigenstates of $W(P)$. 

However, the mixing time of the Szegedy quantum walk
is $O(1/\sqrt{\delta\pi_{min}})$ steps in general with $\pi_{min} = \min_{x\in X}\pi(x)$  \citep{aharonov2007adiabatic,montanaro2015quantum}. This can be problematic when the size $N$ of $X$ is large.  
One way to reduce the dependence of mixing time on $\pi_{min}$ is to employ a slowly varying series of quantum walks to reach the desired quantum sample state \citep{wocjan2008speedup,wocjan2009quantum}. This is similar to the idea of annealed sampling.   
Let $P_0,\ldots,P_r$ be classical reversible Markov chains with stationary distributions $\pi_0,\ldots,\pi_r$, where $\pi_r=\pi$, such that each chain has a relaxation time at most $\tau$ \citep{montanaro2015quantum}.  Then given an easy-to-prepare state $\ket{\pi_0}$, e.g.~the uniform state $\frac{1}{\sqrt{N}}\sum_{x\in X}\ket{x}$, and the condition that $\braket{\pi_i}{\pi_{i+1}}\geq p$ for some $p > 0$ and $\forall i=1,..,r-1$, for any $\epsilon > 0$, there is a quantum algorithm which results in a quantum sample $\ket{\tilde{\pi}_r}$ such that $\|\ket{\tilde{\pi}_r}-\ket{{\pi}_r}\|\leq \epsilon$. The algorithm uses $O(r\sqrt{\tau}\log^2(r/\epsilon)(1/p)\log(1/p))$ quantum walk steps.   \cite{chakrabarti2019quantum} generalized this approach and used it to design a quantum MCMC algorithm to speed up evaluation of volume of convex bodies.  
Further speedup has been proposed by \cite{magniez2011search} and \cite{orsucci2018faster}.  Additionally, even more efficient methods exist to reflect about the states $\ket{\pi_i}$, with a runtime that does not depend on $r$ \citep{yoder2014fixed}.

Applications of quantum Monte Carlo and quantum Markov chain Monte Carlo are rich and varied.  They include, for example, speeding up classical annealing approaches to combinatorial optimization problems \citep{somma2008quantum}, search  \citep{magniez2011search}, speeding up learning agents \citep{paparo2014quantum}, derivative pricing \citep{rebentrost2018finance}, and risk analysis \citep{woerner2019quantum}.

\section{Quantum Linear Systems, Matrix Inversion, and PCA}\label{sec:Quantum Linear Systems}
Consider a system of linear equations $Ax = b$, where $A$ is an $M \times N$ matrix, $b$ is an $M \times 1$ input vector, and $x$ is an $N \times 1$ solution vector, provided it exists. For a square $N \times N$ well-conditioned matrix $A$, the solution to the classical system of linear equations is $x = A^{-1}b$. The system of linear equations powers many applications in statistics and machine learning, especially in high and ultra-high dimensional settings such as deep learning.
The quantum analog of the system of linear equations takes the form  $A\ket{x} = \ket{b}$, where the quantum states $\ket{x}$ and $\ket{b}$ represent vectors $x$ and $b$ in amplitude encoding (Section \ref{sec:encoding}).  The solution is the quantum state $\ket{x} = \frac{1}{C}A^{-1}\ket{b}$, where $C$ is a constant to ensure normalization of the state $\ket{x}$.   \cite{harrow2009quantum} discovered a quantum algorithm, known as the HHL algorithm, to solve the quantum system of linear equations in time that scales with $\log N$, provided that the matrix $A$ is sparse. This offers an exponential speedup over the best classical algorithms, which require a runtime of at least $O(N)$.  The exponential advantage of the HHL algorithm stems from its use of the Quantum Fourier Transform (QFT) -- a foundational quantum routine and the engine at the core of many quantum algorithms including Shor's famous factoring algorithm \citep{shor1994algorithms}.  

The QFT directly exploits quantum parallelism -- the ability to apply a function to all elements of a vector simultaneously if this vector is encoded in a quantum state.  The QFT powers the HHL algorithm through another influential quantum subroutine -- Quantum Phase Estimation (QPE), an algorithm that enables recording of a quantum phase $\theta$, for example in an eigenvalue $e^{i\theta}$ of a unitary matrix, into a computational basis state within an error $\epsilon$ with a high probability (Section~\ref{sec:qpe}). 

This section presents the QFT (Section~\ref{sec:qft}) and its many uses in other quantum algorithms, such as the application of a Hermitian (rather than an unitary) operator to a quantum state (Section~\ref{sec:hermitian}), finding the solution of a system of linear equations (Section~\ref{sec:hhl}), fast gradient computation (Section~\ref{sec:gradient}), and quantum principal component analysis (Section~\ref{sec:qpca}). Even though more efficient ways to perform some of these computations have been discovered recently (see, e.g.~Section~\ref{sec:qsvt}), the QFT remains an influential and pedagogical quantum subroutine.

\subsection{Quantum Fourier Transform (QFT)}
\label{sec:qft}

The QFT \citep{coppersmith1994approximate} transforms a state $\ket{x} = \sum_{m=0}^{N-1} x_m\ket{m}$, where $\ket{m}$ are binary-encoded basis vectors in the computational basis, so that
\begin{align}
    \text{QFT: } \ket{x} & \mapsto \ket{y} = \sum_{k=0}^{N-1} y_k \ket{k}, \\
    y_k & = \frac{1}{\sqrt{N}}\sum_{m=0}^{N-1} x_m e^{i2\pi k \frac{m}{N}}.
\end{align}
QFT is the quantum equivalent of the classical discrete Fourier transform where a vector $x = (x_0,..,x_{N-1})$ is transformed into a vector $y = (y_0,..,y_{N-1})$.  That is, QFT transforms a superposition state $\ket{x}$ into a new superposition state $\ket{y}$ whose amplitudes $y_k$ are the classical discrete Fourier transforms of the amplitudes $x_m$.

QFT exploits quantum computers' ability to encode $N$-dimensional states using $d = \lceil \log N\rceil$ qubits.  The structure of the Fourier transform allows the operation to be performed as a series of $\mathcal{O}(\log^2 N)$ Hadamard gates, controlled rotations, and swap gates -- an exponential improvement in efficiency compared with $\mathcal{O}(N\log N)$ operations required by classical fast Fourier transform.  Because of state preparation and readoff, it is difficult to benefit from the quantum speedup of QFT for estimating the Fourier coefficients.  However, QFT serves as a powerful module in other algorithms, such as Shor's factoring algorithm, the HHL linear systems algorithm, and many others.

Consider $\ket{m}$, a basis state of the Hilbert space containing $\ket{x}$.  Assume without loss of generality that the dimension of the Hilbert space $N = 2^d$.  As shown below, it turns out that the QFT of state $\ket{m}$ is a tensor product of single-qubit states, possible to create in a quantum computer using a series of one- and two-qubit gates.

The QFT of the state $\ket{m}$ is
\begin{align}
    QFT\ket{m} = \frac{1}{\sqrt{N}}\sum_{k=0}^{N-1}e^{i2\pi m \frac{k}{2^d}}\ket{k},
\end{align}
where $\ket{k}$, $k=0,..,N-1$ represent basis states of an $N=2^d$-dimensional Hilbert space.  These states can be chosen to be binary numbers from $0$ to $N-1$ expressed in the computational basis such that, if $k$ is expressed as a binary string $k_1k_2...k_d$, where $k_j = \{0,1\}$, each state $\ket{k}$ is a tensor product state of $d$ qubits in state $\ket{0}$ or $\ket{1}$:
\begin{align}
    \ket{k} = \ket{k_1}\otimes\ket{k_2}\otimes\ldots\otimes\ket{k_d},
\end{align}
and $k = \sum_{j=1}^{d}k_j2^{(d-j)}$.
Using this notation, QFT$\ket{m}$ becomes
\begin{align}
    QFT\ket{m} & = \frac{1}{2^{d/2}}\sum_{k_1,k_2,..,k_d}e^{i2\pi m \frac{\sum_{j=1}^dk_j2^{(d-j)}}{2^d}}\ket{k_1}\ket{k_2}...\ket{k_d} = \frac{1}{2^{d/2}}\sum_{k_1,k_2,..,k_d}\bigotimes_{j=1}^de^{i2\pi m\frac{k_j}{2^j}}\ket{k_j} \nonumber \\
    & = \frac{1}{\sqrt{2}^{d}}(\ket{0}+e^{i2\pi \frac{m}{2}}\ket{1})\otimes (\ket{0}+e^{i2\pi \frac{m}{2^2}}\ket{1})\otimes\ldots\otimes(\ket{0}+e^{i2\pi \frac{m}{2^d}}\ket{1}),\label{eq:qft_expand}
\end{align}
a separable state of $d$ qubits.  The exponent $e^{i2\pi \frac{m}{2^j}}$ effectively extracts the binary ``decimal'' of $m$ represented as a binary string $m= m_1m_2..m_d$ so that $e^{i2\pi \frac{m}{2^j}} = e^{i2\pi m_1..m_{d-j}.m_{d-j+1}..m_{d}} = e^{i2\pi 0.m_{j}..m_{d}}$, since $e^{i2\pi k}=1$ for any integer $k$.  QFT of $\ket{m}$ then simplifies to
\begin{align}\label{eq:QFT output}
    QFT\ket{m} = &\frac{1}{\sqrt{2}^{d}}(\ket{0}+e^{i2\pi 0.m_1m_2..m_{d-1}m_d}\ket{1})\otimes(\ket{0}+e^{i2\pi 0.m_2..m_{d-1}m_d}\ket{1})\otimes\ldots \nonumber\\ 
    &\ldots\otimes(\ket{0}+e^{i2\pi 0.m_d}\ket{1}),
\end{align}
a state that can be created via a series of relatively simple single-qubit and two-qubit gates.   It is easy to see that the QFT operator is linear and applies similarly to a linear superposition of states $\ket{m}$, i.e. any state $\ket{x}$.

\subsection{Quantum Phase Estimation (QPE)}
\label{sec:qpe}

Let $U$ be a unitary operator with eigenstates $\ket{u}$.  Because the operator $U$ is unitary, its eigenvalues take the form $e^{i2\pi\theta}$, where $\theta \in [0,1)$, and $U\ket{u} = e^{i2\pi\theta}\ket{u}$, as we discuss in Section~\ref{sec:q-statesandops}.

Quantum Phase Estimation (QPE) \citep{kitaev1995quantum} is an algorithm to estimate, within a finite precision, the phase $\theta$ of the operator $U$ and record its binary approximation in a quantum state in the computational basis.   QPE is a building block in many algorithms, particularly those requiring the application of a Hermitian (rather than unitary) operator to a quantum state (Section~\ref{sec:hermitian}).

The linchpin of QPE is the control-unitary gate $C\text{-}U$, which applies the unitary $U$ conditional on the state of an auxiliary qubit  (Section~\ref{sec:q-statesandops}).  Consider the state $\ket{0}\otimes\ket{u}$, where $\ket{0}$ represents the auxiliary qubit.  Applying a Hadamard gate to the auxiliary qubit yields the state $\frac{1}{\sqrt{2}}(\ket{0}+\ket{1})\otimes \ket{u}$.   The controlled-unitary operator $C\text{-}U$ acting on the state applies the operator $U$ to state $\ket{u}$ if the auxiliary qubit is in state $\ket{1}$ and does nothing if the auxiliary qubit is in the state $\ket{0}$:
\begin{align}
    C\text{-}U\big[\frac{1}{\sqrt{2}}(\ket{0}+\ket{1})\otimes \ket{u}\big] = \frac{1}{\sqrt{2}}(\ket{0}+e^{i2\pi \theta} \ket{1})\otimes \ket{u}.
\end{align}
Even though the operator $U$ acts on the state $\ket{u}$, it is the auxiliary qubit state that ends up being modified because the operator is controlled on the state of this qubit.   This effect is called \emph{phase kickback}.

To capture the $n$-bit approximation of $\theta$, $\tilde{\theta} = 0.\theta_1\theta_2...\theta_n$ with $\theta_j \in \{0,1\}$, QPE requires $n$ auxiliary qubits.  Each auxiliary qubit is initialized to $\ket{0}$ and then a Hadamard gate is applied to each qubit to yield the state
\begin{align}
    \frac{1}{2^{n/2}}(\ket{0}+\ket{1})\otimes(\ket{0}+\ket{1})\otimes \ldots \otimes (\ket{0}+\ket{1})\otimes \ket{u}.
\end{align}

Next, we apply a series of controlled-unitary gates, $C_j\text{-}U^{2^{j-1}}$, which apply the unitary operator $U^{2^{j-1}}$ to state $\ket{u}$ conditional to the state of the auxiliary qubit $j$.  This results in the state
\begin{align}
    & \frac{1}{2^{n/2}}(\ket{0}+e^{i2\pi \theta 2^0}\ket{1})\otimes(\ket{0}+e^{i2\pi \theta 2^1}\ket{1})\otimes \ldots \otimes (\ket{0}+e^{i2\pi \theta 2^{n-1}}\ket{1})\otimes \ket{u} \nonumber \\
    & = \frac{1}{2^{n/2}}(\ket{0}+e^{i2\pi \frac{2^n\theta}{2^n} }\ket{1})\otimes(\ket{0}+e^{i2\pi \frac{2^n\theta}{2^{n-1}}}\ket{1})\otimes \ldots \otimes (\ket{0}+e^{i2\pi \frac{2^n\theta}{2}}\ket{1})\otimes \ket{u}  \nonumber \\
    & = \Big( \frac{1}{2^{n/2}}\sum_{k=0}^{2^n-1}e^{i2\pi (2^n\theta) \frac{k}{2^n}}\ket{k} \Big)\otimes \ket{u}, \label{eq:qpe_intermediate}
\end{align}
where $k$s are integers represented as $n$-bit strings $k_1...k_n$ by the qubits in the auxiliary register as $\ket{k} = \ket{k_1}\otimes..\otimes \ket{k_n}$.  

If $\theta$ is an $n$-bit number, so that $\theta = \tilde{\theta}$, then $\theta 2^n$ is an integer.  In this case, the state in the auxiliary register of (\ref{eq:qpe_intermediate}) is the QFT of the $n$-qubit state $\ket*{{\theta}2^n}$ (c.f.~Eq.~\ref{eq:qft_expand} with $m=\theta 2^n$).  This state represents ${\theta}2^n$ as a binary integer encoded in the computational basis.  However, in general, $\theta$ is a number with greater than $n$ bits, so that $\theta \neq \tilde{\theta}$.  In this case, $\theta 2^n$ is not an integer.  Splitting $\theta$ into its $n$-bit approximation $\tilde{\theta}$ and a residual $\delta$, $\delta = \theta - \tilde{\theta}$, we can write $\theta 2^n = \tilde{\theta}2^n + \delta 2^n$, where $\tilde{\theta}2^n$ is the integer part of $\theta 2^n$. 

In the last step of QPE, we recover $\tilde{\theta}2^n$ using the inverse QFT applied to the auxiliary register:
\begin{align}
    QFT^\dagger \Big( \frac{1}{2^{n/2}}\sum_{k=0}^{2^n-1}e^{i2\pi (2^n\theta) \frac{k}{2^n}}\ket{k} \Big) & = \frac{1}{2^{n}}\sum_{y=0}^{2^n-1}\sum_{k=0}^{2^n-1}e^{-i2\pi k \frac{y}{2^n}}e^{i2\pi (2^n\theta) \frac{k}{2^n}}\ket{y} \nonumber \\
    & = \frac{1}{2^{n}}\sum_{y=0}^{2^n-1}\sum_{k=0}^{2^n-1}e^{i2\pi (2^n\tilde{\theta}-y) \frac{k}{2^n}}e^{i2\pi\delta k}\ket{y}\label{eq:qpe_superposition}
\end{align}
Probability in the superposition state in (\ref{eq:qpe_superposition}) is peaked around the state $\ket*{\tilde{\theta}2^n}$, which encodes $\tilde{\theta}2^n$ in the computational basis.

The last step is to measure the auxiliary register in the computational basis.   If $\delta = 0$, i.e. if $\theta$ is an $n$-bit number, then the measurement yields $\theta2^n$ with probability $1$.  If $0 < |\delta| \leq \frac{|\theta|}{2^{n}}$, then the measurement yields $\tilde{\theta}2^n$ with probability $4/\pi^2$ or greater \citep{cleve1998quantum}.\footnote{By using $O(\log (1/\epsilon))$ qubits and discarding the qubits above $n$, it is possible to increase the probability of measuring $\tilde{\theta}2^n$ to $1-\epsilon$.}

\subsection{Applying a Hermitian Operator}
\label{sec:hermitian}

Quantum gates are unitary, but it is often useful to apply a Hermitian operator, rather than a unitary operator, to a quantum state.  This can be done in two steps:  QPE (Section~\ref{sec:qpe}) and a controlled rotation (Section~\ref{sec:ctrl-r}).

Let $\mathcal{H}$ be a Hermitian operator on an $N$-dimensional Hilbert space, such that $N=2^n$ for an integer $n$. The goal is to apply the operator $\mathcal{H}$ to an $N$-dimensional state $\ket{\psi}$.
Because the operator $\mathcal{H}$ is Hermitian, there exists an orthonormal basis $\{\ket{u_i} \}_{i=1}^N$ such that $\mathcal{H}\ket{u_i} = \lambda_i\ket{u_i}$.  The scalars $\lambda_i$ are (real) eigenvalues of $\mathcal{H}$.  In quantum notation, the expression $\mathcal{H} = \sum_{i=1}^N \lambda_i \ketbra{u_i}$ reflects the fact that $\mathcal{H}$ is diagonal in the basis of its eigenvectors.   The target state $\mathcal{H}\ket{\psi}$ then takes the form
\begin{align}
    \mathcal{H} \ket{\psi} = \sum_{i=1}^N \lambda_i \ket{u_i}\braket{u_i}{\psi} = \sum_{i=1}^N \lambda_i\beta_i \ket{u_i},
\end{align}
where the scalar coefficients $\beta_i = \braket{u_i}{\psi}$ equal the inner products of the state $\ket{\psi}$ with the eigenstates $\ket{u_i}$.

For any Hermitian operator $\mathcal{H}$, there exists a unitary operator $U = e^{-i\mathcal{H}t}$, where $t$ is a scalar constant.  Because the operator $U$ is unitary, it can be constructed as a series of quantum gates and applied to a quantum state prepared on $n$ qubits. Alternatively, it is often more efficient to interpret the operator $U$ as an evolution operator (Section~\ref{sec:q-statesandops}) by Hamiltonian $\mathcal{H}$ over time $t$ and to apply it using a suitable Hamiltonian simulation algorithm (Section~\ref{sec:Hamiltonian Simulation}).  We will use the operator $U$ to create the target state $\mathcal{H}\ket{\psi}$ on a quantum computer.    

The first step is to express the state $\ket{\psi}$ as a linear combination of eigenstates of $\mathcal{H}$ using the identity operator $I = \sum_{i=1}^N \ketbra{u_i}$ 
\begin{align}
    \ket{\psi} = \sum_{i=1}^N \ket{u_i}\braket{u_i}{\psi} = \sum_{i=1}^N \beta_i \ket{u_i}.\label{eq:psi_exp}
\end{align}
Note that the state $\ket{\psi}$ in (\ref{eq:psi_exp}) does not undergo a transformation; instead, it is simply re-written in the eigenbasis $\{\ket{u_i} \}_{i=1}^N$.  The eigenvectors $\{\ket{u_i} \}_{i=1}^N$ are also eigenvectors of $U$, with corresponding eigenvalues $e^{i\lambda_it}$.   

The algorithm exploits this property to extract the $m$-bit approximations of eigenvalues $\lambda_i$ using QPE.  QPE requires a $m$-qubit auxiliary register, where $m$ is the desired binary precision of $\lambda_i$. QPE takes the two-register state $\ket{\psi}\ket*{0}$ as an input; here, we denote the initialized auxiliary register as $\ket{0}$, a shorthand for $\ket{0\ldots 0}$ -- an $m$-qubit register with each qubit initialized to $\ket{0}$.  For each eigenstate $\ket{u_i}$, QPE records $\tilde{\lambda}_i $, the $m$-bit approximation of $\lambda_i$, in the auxiliary register:  
\begin{align}
    QPE\ket{\psi}\ket*{0} = \sum_{i=1}^N \beta_i QPE \ket{u_i}\ket*{0} = \sum_{i=1}^N \beta_i \ket{u_i}\ket*{\tilde{\lambda}_i}, \label{eq:qpe1}
\end{align}
exploiting the linearity of QPE.

After applying QPE, the next step is to perform a controlled rotation, as described in Section~\ref{sec:ctrl-r}.  The controlled rotation requires an additional auxiliary qubit and uses the $m$-qubit register holding $\tilde{\lambda}_i$ as the reference register.  The $\arcsin(\frac{\tilde{\lambda}_i}{C})$ function acts as $f(x)$ in the definition of controlled rotation(Equation \ref{eq:c-r} in Section~\ref{sec:ctrl-r}), where $C$ is chosen so that $\frac{|\tilde{\lambda}_i|}{C}\leq 1$ for all $\tilde{\lambda}_i$ (\cite{haner2018optimizing} demonstrated that arcsine is efficiently computable).  We obtain:
\begin{align}
    C\text{-}R_y\sum_{i=1}^N \beta_i \ket{u_i}\ket*{\tilde{\lambda}_i}\ket{0} = \sum_{i=1}^N \beta_i \ket{u_i}\ket*{\tilde{\lambda}_i}\big(\sqrt{1-\Big(\frac{\tilde{\lambda}_i}{C}\Big)^2} \ket{0} + \frac{\tilde{\lambda}_i}{C} \ket{1} \big). \label{eq:lambda-rotation}
\end{align}

The next step is to measure the auxiliary qubit in the computational basis.  If the measurement yields $0$, the quantum state on all registers is discarded and the computation is performed again.  If the measurement yields $1$ then the resulting quantum state is:
\begin{align}
    \frac{1}{C_1}\sum_{i=1}^N \tilde{\lambda}_i\beta_i \ket{u_i}\ket*{\tilde{\lambda}_i}\ket{1},  \label{eq:herm_intermediate}
\end{align}
where $C_1 = C\sqrt{\sum_{i=1}^N |\tilde{\lambda}_i \beta_i|^2}$ is the normalization constant.  The number of measurements (and recomputations) required to achieve the desired state can be reduced using amplitude amplification \citep{brassard2002quantum}, described in Section~\ref{sec:qaa}.

The last step is to uncompute the register $\ket*{\tilde{\lambda}_i}$ in order to return it to the ground state $\ket*{0}$.  Discarding this register and the auxiliary qubit yields
\begin{align}
    \frac{1}{C_1}\sum_{i=1}^N \tilde{\lambda}_i\beta_i \ket{u_i} = \frac{1}{C_1} \mathcal{H} \ket{\psi},
\end{align}
the desired result up to a normalization constant.

Similar techniques can be used to implement smooth functions of sparse Hermitian operators \citep{subramanian2019implementing}.
\cite{rebentrost2019quantum} used the application of a Hermitian operator to a quantum state in order to perform gradient descent on a homogeneous polynomial. Homogeneous polynomials have the property that the application of a gradient operator is equivalent to the application of a linear operator.   Let $f(\mathbf{x})$ be a homogeneous polynomial of $ \mathbf{x} = (x_1,\ldots,x_N)^T$.  Then there exists an operator $D(\mathbf{x})$ such that  $\nabla f(\mathbf{x}) = D(\mathbf{x})\mathbf{x}$. Because of this property, it is possible to estimate the gradient of $f(\mathbf{x})$ using the techniques described in this section.    

The gradient descent algorithm starts with an initial guess vector $\mathbf{x}^{(0)}$.  \cite{rebentrost2019quantum} encode this vector in a quantum state $\ket{\mathbf{x}^{(0)}}$ and then use the method described in this section to apply the operator $D(\mathbf{x^{(0)}})$ to the state $\ket{\mathbf{x}^{(0)}}$ in order to evaluate the gradient  $\nabla f(\mathbf{x})$.   For homogeneous polynomials, the operator $D(\mathbf{x^{(0)}})$ has a relatively simple structure, which makes it possible to simulate $e^{-iDt}$ efficiently on a quantum computer using simulation techniques from the quantum principal component analysis method \citep{lloyd2014quantum} described in Section~\ref{sec:qpca}. The efficient computation of  $e^{-iDt}$ makes it possible to use QPE as in (\ref{eq:qpe1}) and, therefore, use the Hermitian operator method described in this section to evaluate gradients for homogeneous polynomials.

\subsection{The HHL Linear Systems Algorithm}
\label{sec:hhl}

The linear systems algorithm by \cite{harrow2009quantum} exploits the ability to apply a Hermitian operator to a quantum state in order to solve linear systems of the form $A x = b$, where $x$ and $b$ are $N$-dimensional vectors and $A$ is an $N \times N$ matrix.  Finding the solution $x$ requires the inversion (or pseudoiversion) of the matrix $A$, which is computationally expensive for a high-dimensional matrix.\footnote{Inversion of an $N\times N$ matrix on a classical computer requires $O(N^d)$ operations, where $2< d\leq 3$.}

To introduce the core of the algorithm, we assume $A$ is a Hermitian matrix and generalize it at the end.  We also assume $N = 2^n$, where $n$ is an integer.   We initialize an $n$ qubit register and encode the state $b$ in amplitude encoding:
\begin{align}
    \ket{b} = \frac{1}{\|b\|}\sum_{i=1}^Nb_i\ket{i},
\end{align}
where states $\ket{i}$ are $n$-qubit states in computational basis; the state of each qubit in the register can (but does not have to) correspond to the binary encoding of integers $i$; the amplitudes $b_i$ are the elements of vector $b$; $\|b\|$ is the normalization constant, $\|b\| = \sqrt{\sum_{i=1}^N|b_i|^2}$.

If $A$ is a Hermitian matrix and is invertible, the solution to the quantum linear system $A \ket{x} = \ket{b}$ is a quantum state $\ket{x}$ such that
\begin{align}
    \ket{x} = \frac{1}{C_1}A^{-1}\ket{b},
\end{align}
where the constant $C_1$ ensures normalization of $\ket{x}$.  To streamline notation, without loss of generality, we assume in this section that $C_1 = 1$.

Let $\{\alpha_i\}_{i=1}^N$ be the set of eigenvalues of matrix $A$ and $\{\ket{a_i}\}_{i=1}^N$ be the set of corresponding eigenstates.  The eigenstates of the matrix $A^{-1}$, which is Hermitian since $A$ is Hermitian, are also $\{\ket{a_i}\}_{i=1}^N$, with eigenvalues $\{\frac{1}{\alpha_i}\}_{i=1}^N$.  Using the identity matrix expressed in terms of the eigenvectors of $A$, $I = \sum_{j=1}^N \ketbra{a_j}$, we transform the expression for $\ket{x}$ into
\begin{align}
    \ket{x} = A^{-1}\frac{1}{\|b\|}\sum_{i=1}^Nb_i\ket{i} = A^{-1}\frac{1}{\|b\|}\sum_{i=1}^N\sum_{j=1}^N b_i\ket{a_j}\braket{a_j}{i} =\frac{1}{\|b\|} \sum_{i=1}^N\sum_{j=1}^N b_i\frac{1}{\alpha_j}\ket{a_j}\braket{a_j}{i}.\label{eq:x_sol}
\end{align}
The result in (\ref{eq:x_sol}) resembles the result of the Hermitian operator routine from Section~\ref{sec:hermitian} with one difference:  in the controlled-rotation of the auxiliary qubit in (\ref{eq:lambda-rotation}), $\arcsin \frac{C}{\tilde{\alpha}_i}$ replaces $\arcsin \frac{\tilde{\lambda}_i}{C}$.   Here, the value $\tilde{\alpha}_i$ is the approximation of $\alpha_i$ obtained by QPE, and the constant $C$ is such that $\frac{C}{|\tilde{\alpha}_i|} \leq 1$.  If some values $\tilde{\alpha}_i$ are very small or 0, regularization techniques, similar to those in classical matrix inversion, can provide stability (for example, \cite{tikhonov1963solution} regularization).  The small values of $\tilde{\alpha}_i$ can be discarded or collected in a separate state for further analysis. We refer the reader to \cite{harrow2009quantum} and \cite{dervovic2018quantum} for details.

The routine generalizes to the case where $A$ is non-Hermitian.   In this case, in place of $A$, we use the Hermitian matrix $\mathbf{I}A$, where $\mathbf{I}$ is the \emph{isometry} superoperator such that:
\begin{align}
    \mathbf{I}A = \begin{bmatrix}
    0 & A \\
    A^{\dagger} & 0
    \end{bmatrix}.
\end{align}
The eigenvectors and eigenvalues of the matrix $\mathbf{I}A$ are closely related to right and left eigenvectors and singular values of matrix $A$, $u_k$, $v_k$, and $\alpha_k$, respectively.  Following \cite{harrow2009quantum}, we append an auxiliary qubit and define $\ket*{a^\pm_k} = \frac{1}{\sqrt{2}}(\ket{0}\ket{u_k}\pm\ket{1}\ket{v_k})$, where $\ket{u_k} = \sum_{i=1}^N u_{ik}\ket{i}$ and $\ket{v_k} = \sum_{j=1}^M v_{jk}\ket{j}$, $k = 1,..,K$, and $K$ is the rank of matrix $A$.  The operator $\mathbf{I}A$ takes the from $\mathbf{I}A = \ketbra{0}{1}\sum_{s=1}^S\alpha_s\ketbra{u_s}{v_s}+\ketbra{1}{0}\sum_{s=1}^S\alpha_s\ketbra{v_s}{u_s}$.\footnote{If $K < (N+M)/2$, $\mathbf{I}A$ has  $N+M-2K$ zero eigenvalues, corresponding to the basis states of the orthonormal complement to the $2K$-dimensional Hilbert space spanned by $\ket*{a^\pm_k}$.}  

The runtime of the algorithm is $O(d^4\kappa^2\log N/\epsilon)$, where $d$ is the sparsity of the matrix $A$ (the number of non-zero elements in each row or column of $A$), $\kappa = \alpha_{max}/\alpha_{min}$ is its condition number (the ratio of the largest to the smallest singular value), and $\epsilon$ is the admissible error.  If the matrix $A$ is sparse and well-conditioned, the algorithm runs exponentially faster than the best classical linear systems algorithm, with runtime scaling as $O(Nd\kappa \log(1/\epsilon))$ \citep{saad2003iterative}.  

The result of the quantum linear systems algorithm is a quantum state, which can be passed onto another quantum subroutine.  For example, \cite{schuld2016prediction} use the building blocks for the quantum linear systems algorithm embedded in another quantum algorithm to perform prediction by linear regression.   Alternatively, the quantum state can be read out for use by a classical computer, although the information in the quantum state may need to be compressed in order to preserve the computational efficiency of the algorithm.    Another approach, discussed in Section \ref{sec:qpca} is to characterize the state using quantum PCA.

Since HHL formulated the quantum linear systems problem, $\ket{x} = A^{-1}\ket{b}$, many efficient algorithms have been found, with computational complexity gradually improving to reduce the dependence on the error $\epsilon$, sparsity $d$, and condition number $\kappa$ of the matrix $A$.  \cite{ambainis2012variable} replaced QAA with variable-time amplitude amplification to reduce the dependence on the condition number.  \cite{clader2013preconditioned} proposed to precondition the matrix $A$ using sparse approximate inverse preconditioning and reduced the dependence on both condition number and error.  \cite{kerenidis2016quantum} and \cite{wossnig2018quantum} replaced Hamiltonian simulation using optimized Product Formula (Section~\ref{sec:productformula}) as in \cite{berry2007efficient} with a Hamiltonian simulation based on a quantum walk (Section~\ref{sec:hs_qw}), making it possible to deliver exponential speedup to low-rank (rather than sparse) problems. \cite{childs2017quantum} decomposed $A^{-1}$ as a linear combination of unitaries (Section~\ref{sec:lcu}) and eschewed QPE, achieving a logarithmic dependence on the error.  \cite{subacsi2019quantum} and \cite{an2019quantum} proposed an adiabatic-inspired method.  \cite{gilyen2019optimizing} used their QSVT method (Section~\ref{sec:qsvt}) lowering the computational complexity to the multiplicative lower bounds in all variables and eliminating the dependence on size entirely, so that the overall complexity of solving the linear systems problem is $O(\kappa \log (\kappa/\epsilon))$.

\subsection{Fast Gradient Computation}
\label{sec:gradient}

The QFT at the core of most algorithms in this section also enables efficient evaluation of gradients.  The quantum gradient algorithm proposed by \cite{jordan2005fast} (and refined by \cite{gilyen2019optimizing}) calculates an approximate $N$-dimensional gradient $\nabla f(\mathbf{x})$ of a function $f: \mathbb{R}^N \to \mathbb{R}$ at point $\mathbf{x}\in\mathbb{R}^N$ using a single evaluation of $f$.  For comparison, standard classical techniques require $N+1$ function evaluations. The algorithm is suitable for problems where the function evaluations are computationally taxing, but the number of dimensions is only moderately high, because it requires $O(N)$ qubits (and $O(N)$ measurements if the result is to be used in a classical computation).   

The algorithm uses a \emph{phase oracle} (see a brief discussion of oracles in Section \ref{sec:gates}).   An oracle is an algorithmic ``black box'' assumed to perform a computational task efficiently.  A quantum phase oracle $O_g$ adds a phase to a quantum state such that, given a function $g(y)$, we have
\begin{align}
    O_g \ket{y} = e^{ig(y)}\ket{y}. 
\end{align}

The algorithm stems from two observations.  The first observation is that, if $f$ is twice-differentiable then, for a vector $\delta$ with a sufficiently small norm, the expansion of $f(\mathbf{x}+\delta)$ in the vicinity of $\mathbf{x}$ takes the form $f(\mathbf{x}+\mathbf{\delta}) = f(\mathbf{x})+\nabla f \cdot \mathbf{\delta} +\mathcal{O}(||\mathbf{\delta}||^2)$.  The second observation is that the phase oracle for $f(\mathbf{x}+\mathbf{\delta})$ takes a convenient form.   To define the phase oracle, we take $g = 2\pi D f$, where where $D>0$ is a scaling factor necessary for all values of $2\pi D f$ on the relevant domain to be less than $2\pi$.   When acting on a quantum state $\ket{\delta}$ initialized to hold the value of $\delta$, the phase oracle adds a phase that depends on the value of $f$: $O_{2\pi D f}: \ket*{\mathbf{\delta}} \mapsto e^{2\pi i D f(\mathbf{x}+\mathbf{\delta})}\ket*{\mathbf{\delta}}$.   Using the expansion of  $f(\mathbf{x}+\mathbf{\delta})$ we can write approximately $O_{2\pi D f}: \ket*{\mathbf{\delta}} \mapsto e^{2\pi i D f(\mathbf{x})}e^{2\pi i D \nabla f\cdot\mathbf{\delta}}\ket*{\mathbf{\delta}} $.  The phase contains the dot product of the gradient $\nabla f$ and the differential $\delta$, enabling the extraction of the gradient using inverse QFT (Section~\ref{sec:qft}).  The algorithm works under the assumption that the third and fourth derivatives of $f$ around $\mathbf{x}$ are negligible.  

The algorithm starts with a uniform superposition $\ket{\psi}=\frac{1}{\sqrt{|G_x^d|}}\sum_{\mathbf{\delta}\in G^d_x}\ket{\mathbf{\delta}}$ over the points of a sufficiently small discretized $N$-dimensional grid $G_x^N$ around $\mathbf{x}$.  Each state $\ket{\delta}$ reflects coordinates recorded in the computational basis (in an $N\times m$-qubit register, where $m$ is the binary precision).  
A single call to the phase oracle $O_{2\pi D f}$ creates the state
\begin{align}
    O_{2\pi D f}\ket{\psi} & = \frac{e^{2\pi i D f(\mathbf{x})}}{\sqrt{|G_x^d|}}\sum_{\mathbf{\delta}\in G^d_x}e^{2\pi i D \nabla f\cdot\mathbf{\delta}}\ket*{\mathbf{\delta}}, 
\end{align}
ready for the application of inverse QFT, which extracts $\tilde{\nabla f}$, an $m$-bit approximation of the gradient $\nabla f$.  The output of the algorithm is an $N\times m$-qubit state that records the coordinates of the gradient $\tilde{\nabla f}$ in the computational basis.  

For a step-by-step description of Jordan's algorithm, we refer the reader to the paper by \cite{gilyen2019optimizing} who review the algorithm and modify it to take advantage of central-difference formulas.  

\subsection{Quantum Principal Component Analysis (QPCA)}
\label{sec:qpca}

Algorithms such as HHL (Section~\ref{sec:hhl}) yield a result in the form of an unknown quantum state that has to be characterized.  The most straightforward way to characterize the state is to create multiple copies and take measurements to enable statistical analysis of the state.  However, this approach can be computationally taxing and inefficient, particularly for non-sparse but low-rank quantum states.  \cite{lloyd2014quantum} propose an alternative method that uses multiple copies of a quantum system to perform Principal Component Analysis (PCA) of the system, i.e.~to extract its principal components -- the eigenstates corresponding to largest eigenvalues.  Quantum PCA (QPCA) performs the task for any unknown low-rank $N$-dimensional quantum state in $O(\log N)$ runtime, exponentially faster than any existing classical algorithm.  The algorithm leverages the \emph{density matrix} formalism of quantum theory, an alternative way to describe quantum states, introduced in Section~\ref{sec:q-statesandops}.  

Let $\rho$ be a density matrix describing a quantum state.  The density matrix is Hermitian and has real eigenvalues $r_j$ corresponding to eigenstates $\ket{\chi_j}$, with $j = 1,..,N$.  The goal of the quantum PCA (QPCA) algorithm is to extract the eigenstates and eigenvalues of $\rho$.  Given $\ket{\psi}$, an $N$-dimensional quantum state, and an $m$-qubit register of auxiliary qubits, QPCA performs the transformation
\begin{align}
    QPCA \ket{\psi}\ket{0} \mapsto \sum_j \psi_j\ket{\chi_j}\ket*{\tilde{r}_j}, \label{eq:qpca}
\end{align}
where $\tilde{r}_i$ are $m$-qubit approximations of the eigenvalues $r_i$ and $\psi_i = \braket{\chi_i}{\psi}$.  The state in (\ref{eq:qpca}) has a structure similar to that of the intermediate result (\ref{eq:herm_intermediate}) of the algorithm to apply a Hermitian operator to quantum state (Section~\ref{sec:hermitian}).  This is because the QPCA algorithm treats the density matrix $\rho$ as a Hermitian operator that can be applied to an arbitrary quantum state.  The trouble is, for the algorithm in Section~\ref{sec:hermitian} to be efficient, the Hermitian operator $\mathcal{H}$ has to be sufficiently structured for the controlled-unitary with the unitary operator $e^{-i\mathcal{H}t}$ to be realized; the density matrix $\rho$ may lack such structure.   

The critical insight at the core of the QPCA is that it is possible to construct the controlled-unitary $C\text{-}U$ with $U=e^{-i\rho\Delta t}$ for any density matrix $\rho$ using a series of swap operations, provided the increment $\Delta t$ is sufficiently small.  It is then possible to use Product Formula (Section~\ref{sec:productformula}) to develop the controlled unitary with $U=e^{-i\rho t}$ based on the controlled unitary with $U=e^{-i\rho\Delta t}$, where $t = n\Delta t$.

Consider the application of $e^{-i\rho\Delta t}$ to an arbitrary $N$-dimensional state described by a density matrix $\sigma$: 
\begin{align}
    e^{-i\rho \Delta t}\sigma e^{i\rho \Delta t} & = \sigma -i\Delta t [\rho,\sigma] + O(\Delta t^2).
\end{align}
\cite{lloyd2014quantum} demonstrate that this operation is equivalent to applying the swap operator to the tensor product state $\rho\otimes \sigma$ and a subsequent partial trace $\tr_P$ of the first variable:
\begin{align}
    \tr_P e^{-iS\Delta t}\rho\otimes \sigma e^{iS\Delta t} & = (\cos^2 \Delta t)\sigma +(\sin^2 \Delta t)\rho - i \sin\Delta t \cos \Delta t [\rho,\sigma] \nonumber \\
    & = \sigma - i\Delta t [\rho,\sigma]s+O(\Delta t^2) \label{eq:trace_rho_sig}.
\end{align}
The swap operator $S$ is represented by a sparse matrix and $e^{-iS\Delta t}$ is computable efficiently.

The derivation of (\ref{eq:trace_rho_sig}) uses the identity
\begin{align}
    e^{-iS\Delta t} = \cos (\Delta t) I -i \sin (\Delta t) S,
\end{align}
so that the expression $ e^{-iS\Delta t}\rho\otimes \sigma e^{iS\Delta t}$ can be rewritten as:
\begin{align}
    [\cos (\Delta t) I -i \sin (\Delta t) S] \rho\otimes \sigma [\cos (\Delta t) I + i \sin (\Delta t) S]. 
\end{align}
Partial trace over the $(\cos^2 \Delta t)$ term yields $\tr_P I \rho\otimes \sigma I = \sigma$.   Partial trace over the $(\sin^2 \Delta t)$ term yields $\tr_P S \rho\otimes \sigma S = \rho$. Partial trace over the $\sin\Delta t \cos \Delta t$ term results in:
\begin{align}
    \tr_P S \rho\otimes \sigma I & = \sum_i \bra{i}_P S \rho\otimes \sigma\ket{i}_P = \sum_i \sum_{j,k} \bra{i}_P S \rho\otimes \sigma\ket{j,k}\bra{j,k}\ket{i}_P  \nonumber \\
    & = \sum_i \sum_{j,k}\sum_{l,m} \bra{i}_P \ket{l,m}\bra{l,m}S \rho\otimes \sigma\ket{j,k}\bra{j,k}\ket{i}_P \nonumber \\
    & = \sum_i \sum_{j,k}\sum_{l,m} \delta_{i,l}\ket{m}\bra{l,m}S \rho\otimes \sigma\ket{j,k}\bra{k}\delta_{j,i}\nonumber  \\
    & = \sum_i \sum_{m,k}\ket{m}\bra{i,m}S \rho\otimes \sigma\ket{i,k}\bra{k} \nonumber \\
    & = \sum_i \sum_{m,k}\ket{m}\bra{m,i}\rho\otimes \sigma\ket{i,k}\bra{k} \nonumber \\
    & = \sum_i \sum_{m,k}\ket{m}\bra{m}\rho\ket{i}\bra{i} \sigma\ket{k}\bra{k} \nonumber \\
    & = \sum_{m,k}\ket{m}\bra{m}\rho\sigma\ket{k}\bra{k} = \rho\sigma.
\end{align}

The transformation in (\ref{eq:trace_rho_sig}) therefore results in $e^{-i\rho \Delta t}\sigma e^{i\rho \Delta t}$.   Given multiple copies of $\rho$, the transformation can be applied repeatedly to provide an $\epsilon$-approximation to $e^{-i\rho t}\sigma e^{i\rho t}$, which requires $n=O(t^2\epsilon^{-1}|\rho-\sigma|^2)\leq O(t^2\epsilon^{-1})$ steps (a consequence of Suzuki-Trotter theory, see Section~\ref{sec:productformula}).   The procedure is quite flexible.  For example, using matrix inversion in \cite{harrow2009quantum}, it is possible to implement $e^{-ig(\rho)}$ for any ``simply computable'' function $g(\rho)$.   

Armed with $e^{-i\rho t}$, the next step is to apply QPE in order to record $m$-bit approximations of the eigenvalues $r_i$ in auxiliary register (Section~\ref{sec:qpe}) transforming any initial state $\ket{\psi}\ket{0}$ into the desired state $\sum_i\psi_i\ket{\chi_i}\ket*{\tilde{r}_i}$.

\section{Hamiltonian Simulation}\label{sec:Hamiltonian Simulation}

\subsection{Overview and Preliminaries}

Introduced briefly in Section \ref{sec:q-statesandops}, the Hamiltonian of an $N$-dimensional quantum system is a Hermitian operator that guides the evolution of the system over time.  Let $\mathcal{H}$ be a Hamiltonian. A system in the initial state $\ket{\psi}$ evolves over time according to the unitary operator $ e^{-i\mathcal{H}t}$
\begin{align}
    \ket{\psi(t)} = e^{-i\mathcal{H}t}\ket{\psi},
\end{align}
provided the Hamiltonian $\mathcal{H}$ stays unchanged during the period of evolution.  

The time evolution of the state $\ket{\psi}$ can be expressed in terms of eigenvalues and eigenstates of the Hamiltonian $\mathcal{H}$.  Let $\ket{\lambda}$ be the eigenstates of the Hamiltonian $\mathcal{H}$ with eigenvalues $\lambda$: $\mathcal{H}\ket{\lambda}=\lambda\ket{\lambda}$.   Then we can express the initial state $\ket{\psi}$ in terms of the eigenstates $\ket{\lambda}$:
\begin{align}
    \ket{\psi} & = \sum_\lambda \braket{\lambda}{\psi}\ket{\lambda} \equiv \sum_\lambda \psi_\lambda \ket{\lambda},\label{eq:initial_psi}
\end{align}
where $\psi_\lambda \equiv \braket{\lambda}{\psi}$.  The state $\ket{\psi(t)}$ evolved under $\mathcal{H}$ over time $t$ takes the form
\begin{align}
    \ket{\psi(t)} & = \sum_\lambda\psi_\lambda e^{-i\lambda t}\ket{\lambda}.
\end{align}

Hamiltonian simulation is an approximation of the evolution of a system using the evolution of a different system -- usually one that is simpler or easier to control using a simple set of controllable operations.   In his classic work on universal quantum simulation, \cite{lloyd1996universal} draws a parallel between Hamiltonian simulation and parallel parking a car, which is possible even though a car is only able to move backward and forward.  Using a simpler, controllable quantum system, Hamiltonian simulation approximates the result of the evolution over time $t$ of a more complex quantum system.   

The goal of Hamiltonian simulation is to evolve the initial state $\ket{\psi}$ in (\ref{eq:initial_psi}) in a way that creates pre-factors $e^{-i\lambda t}$ in front of each $\ket{\lambda}$, within $\epsilon$-error.  The creation of these pre-factors is equivalent to the evolution of $\ket{\psi}$ under Hamiltonian $\mathcal{H}$.

The ability to simulate Hamiltonian evolution efficiently on a quantum computer will not only revolutionize molecular engineering but also allow us to tackle computationally hard problems such as combinatorial optimization and high-dimensional linear systems of equations.  Because of its central role in quantum computing applications, Hamiltonian simulation is an active and rapidly evolving field.   Additionally, Hamiltonian simulation is BQP-complete \citep{low2019hamiltonian,berry2015hamiltonian} (see Section~\ref{sec:complexity} for a brief discussion of computational complexity).  In the words of \cite{low2019hamiltonian}, Hamiltonian simulation is a ``universal problem that encapsulates all the power of quantum computation.''

\subsubsection*{Simulatable Hamiltonians and Simulation Limits}

A \emph{simulatable} Hamiltonian is one that makes it possible to approximate the unitary evolution operator $e^{-i\mathcal{H}t}$ by a quantum circuit efficiently, i.e.~to an accuracy at most polynomial in the precision of the circuit and in time at most polynomial in evolution time $t$ \citep{aharonov2003adiabatic}. There is no Hamiltonian simulation algorithm able to simulate the evolution under a general Hamiltonian in $poly(\|\mathcal{H}\|t,\log N)$, where $\|\mathcal{H}\|$ is the spectral norm (defined in Eq.~\ref{eq:spectralnorm}) \citep{childs2009limitations}.  Additionally, there is no general algorithm to simulate a general sparse Hamiltonian in time less than linear in $\|\mathcal{H}\|t$ -- a statement known as the \emph{No Fast-Forwarding Theorem} \citep{berry2007efficient}.

\subsubsection*{Hamiltonian Input Models}

Efficient Hamiltonian simulation algorithms exploit the structure of the Hamiltonian.\footnote{Just like it is not possible to implement an arbitrary unitary efficiently, it is not possible to efficiently to simulate an arbitrary Hamiltonian \citep{childs2009limitations}. }  Efficient simulation algorithms have been proposed for time-independent Hamiltonians such as Hamiltonians that are linear combinations of \emph{local} terms -- terms acting on a small number of qubits \citep{lloyd1996universal,berry2007efficient}, \emph{sparse} Hamiltonians that have at most  $poly log(N)$ entries in each row \citep{aharonov2003adiabatic,berry2007efficient,berry2012black,berry2014exponential}, Hamiltonians that comprise a \emph{linear combination of unitaries (LCU)} \citep{childs2012hamiltonian}, and \emph{low-rank} Hamiltonians \citep{berry2012black,wang2018quantum}.

Hamiltonian simulation algorithms specify input models that make the terms of the Hamiltonian accessible to the quantum computer.   The most popular input models include \emph{black-box}, \emph{sparse access}, \emph{QROM}, and \emph{LCU} models.

The \emph{black-box} input model, proposed by \cite{grover2000synthesis} and refined by \cite{sanders2019black}, uses a unitary oracle $O_H$ that returns matrix terms $\mathcal{H}_{jk}$ and their indices $j,k$ in a binary format:
\begin{align}
    O_H\ket{j,k}\ket{z} = \ket{j,k}\ket{z\oplus \mathcal{H}_{jk}},
\end{align}
where $\oplus$ represents the bit-wise XOR.  The oracle $O_H$ represents a quantum algorithm that performs the encoding of Hamiltonian terms into a quantum state.  For example, \cite{sanders2019black} propose an algorithm that performs $O_H$ using quantum computing primitives, such as Toffoli gates.

If the Hamiltonian is sparse and has at most $d$ nonzero entries in any row, then the \emph{sparse access} input model helps drive further efficiency \citep{aharonov2003adiabatic}.   This model uses two unitary oracles, $O_H$ -- the black-box Hamiltonian oracle -- and $O_F$ the address oracle.  
Oracles
\begin{align}
    O_H\ket{j,k}\ket{z} & = \ket{j,k}\ket{z\oplus \mathcal{H}_{jk}}\\
    O_F\ket{j,l} & = \ket{j}\ket{f(j,l)}
\end{align}
where $f(j,l)$ is a function that gives the column index of the $l$th non-zero element in row $j$.  The sparse access model has been the most popular input model for sparse Hamiltonian simulation.   \cite{aharonov2003adiabatic} defined a Hamiltonian as \emph{row-sparse} if the number of non-zero entries in each row is  $O(\text{poly}\log N)$ or lower.   A Hamiltonian is called \emph{row-computable} if there exists an efficient (i.e. requiring $O(\text{poly}\log N)$ or fewer operations) -- quantum or classical -- algorithm that, given a row index $i$, outputs a list $(j,H_{ij})$ of all non-zero entries in that row.   Row-sparse and row-computable Hamiltonians are simulatable, provided they have a bounded spectral norm $\|H\| \leq O(\text{poly}\log N)$.\footnote{This statement is called \emph{the sparse Hamiltonian lemma} due to \cite{aharonov2003adiabatic}.}

For Hamiltonians decomposed into a linear combination of unitaries, the input model provides the constituent unitaries and their weights.  

An increasingly popular input model is the QROM input model, based on the classical QROM structure \citep{kerenidis2016quantum,chakraborty2018power}, outlined in Section \ref{sec:encoding}, that provides efficient quantum access to Hamiltonian terms.  This input structure is used in Hamiltonian simulation based on quantum singular value transformation \citep{gilyen2019quantum}, found to be a unifying framework for the top quantum algorithms of the last two decades \citep{martyn2021grand}.

\subsubsection*{Matrix Norms}

For reference, we include the most common matrix norms used in Hamiltonian simulation literature.   The matrix norms arise in normalizations of quantum states encoding Hamiltonian terms and in estimations of computational complexity.   For rigorous definitions of matrix norms and the relationships between them see, e.g.,  \cite{childs2009limitations}.

The \emph{spectral norm} $\|\mathcal{H}\|$ of Hamiltonian $\mathcal{H}$ is defined as
\begin{align}
    \|\mathcal{H}\| = \max_{\|v\|=1} \|\mathcal{H}v\| = \sigma_{max}(\mathcal{H}), \label{eq:spectralnorm}
\end{align}
where, for a vector $v$, $ \|v\|$ represents the Euclidean $2$-norm; $\sigma_{max}(\mathcal{H})$ denotes the largest singular value.  Spectral norm is the matrix $2$-norm, induced by the vector $2$-norm; it sometimes denoted as $\|\mathcal{H}\|_2$.  Similarly, the matrix \emph{1-norm} induced by the vector $1$-norm is $ \|\mathcal{H}\|_1 = \max_{j}\sum_i |\mathcal{H}_{ij}|$.

The \emph{max norm} is 
\begin{align}
    \|\mathcal{H}\|_{max} = \max_{i,j}|\mathcal{H}_{ij}|.
\end{align}

The \emph{Frobenius norm}, also called the \emph{Hilbert-Schmidt norm}, is 
\begin{align}
        \|\mathcal{H}\|_F = \sqrt{\sum_{i,j}|\mathcal{H}_{ij}|^2} = \sqrt{\sum_k\sigma_k^2(\mathcal{H})},
\end{align}
where $\sigma_k(\mathcal{H})$ are the singular values of $\mathcal{H}$.

\subsubsection*{Overview of Hamiltonian Simulation Algorithms}

As a critical potential application of quantum computers and a linchpin of other algorithms, such as quantum linear systems algorithms, Hamiltonian simulation is an active area of algorithm design.  In this section, we review a few influential methods of Hamiltonian simulation.  We start with a foundational method \emph{Product Formula} that splits the Hamiltonian into easy-to-simulate parts and then uses sequences of small subsystem simulations to approximate whole-system Hamiltonian evolution (Section~\ref{sec:productformula}).  We then overview Hamiltonian simulation by quantum walk, an influential method that works for general Hamiltonians and is highly efficient for sparse Hamiltonians (Section~\ref{sec:hs_qw}).    We also summarize the method of linear combination of unitaries for Hamiltonians that can be expressed as linear combinations of unitary operators (Section~\ref{sec:lcu}).  And last, we demonstrate how to use quantum signal processing in combination with the quantum walk method to perform Hamiltonian simulation with computational efficiency that corresponds multiplicatively to all known lower bounds (Section~\ref{sec:hs_qsp}).

\begin{table} 
\small 
\caption{Hamiltonian simulation algorithms. The table demonstrates the gradual improvement of query complexity of Hamiltonian situation with respect to the essential parameters of the simulation:  the dimension of the Hilbert space $N$, admissible error $\epsilon$, and the sparsity parameter $d$, equal to the number of non-zero elements in each row (column) of the Hamiltonian.  } 
\begin{center}
\footnotesize
\begin{tabular}{>{\arraybackslash}p{3cm}>{\arraybackslash}p{2.7cm}>{\arraybackslash}p{4.2cm}>{\arraybackslash}p{4.6cm}}
\hline
Algorithm & Citation & Method & Query Complexity \phantom{abcdef} \\[0.2cm]
\hline
Lie-Suzuki-Trotter Product Formula & \cite{lloyd1996universal} & Finite sum of $l$ local Hamiltonians (split into $r$ parts) &  $O(lr)$ \\[0.2cm]
\hline
Adiabatic Hamiltonian evolution & \cite{aharonov2003adiabatic} &  Adiabatic evolution for row-sparse, row-computable Hamiltonians & $O(\text{poly}(\log(N),d)(\|\mathcal{H}\|t)^2/\epsilon$ \\[0.2cm]
\hline
Optimized Product Formula & \cite{berry2007efficient} &  Efficiently decompose Hamiltonian into a sum of local Hamiltonians & $O(d^4(\log^*N\|\mathcal{H}t\|)^{1+o(1)})$  \\[0.2cm]
\hline
Quantum walk-based Hamiltonian simulation & \cite{berry2012black}  &  Combine quantum walk with quantum phase estimation for general Hamiltonians with black-box access &  $ O(d^{2/3}[(\log\log d)\|\mathcal{H}\|t]^{4/3}/\epsilon^{1/3})$ \\[0.2cm]
\hline
Linear combination of unitaries (LCU) & \cite{childs2012hamiltonian} &  Decompose a Hamiltonian into a finite linear combination of unitaries &   $ O(d\|\mathcal{H}\|_{max}t/\sqrt{\epsilon})$ \\[0.2cm]
\hline
Taylor-series based & \cite{berry2015simulating} & Taylor series expansion of $e^{-i\mathcal{H}t}$ & $O(\tau\log (\tau/\epsilon)/\log\log(\tau/\epsilon))$, where $\tau = d^2\|\mathcal{H}\|_{max}t$  \\[0.2cm]
\hline
BCK & \citet*{berry2015hamiltonian} & Combine quantum walk with linear combination of unitaries & $O(\tau\log (\tau/\epsilon)/\log\log(\tau/\epsilon))$, where $\tau = d\|\mathcal{H}\|_{max}t$  \\[0.2cm]
\hline
Quantum signal processing & \cite{low2017optimal} &  Combine quantum walk with quantum signal processing & $O(\tau+\frac{\log(1/\epsilon)}{\log\log(1/\epsilon)})$ matches theoretical additive lower bound \cite{berry2015hamiltonian}   \\[0.2cm]
\hline
\end{tabular}
\end{center}
\end{table}

\subsection{Product Formula}
\label{sec:productformula}

One of the foundational methods of Hamiltonian simulation is the Lie-Trotter-Suzuki Product Formula \citep{suzuki1990fractal,suzuki1991general}.  This approach works for time-independent local or, more generally, separable Hamiltonians.   Many Hamiltonians\footnote{Particularly local Hamiltonians that usually apply to real physical systems.} $\mathcal{H}$ can be expressed as a sum of $l$ Hamiltonians $\mathcal{H}_j$: $\mathcal{H} = \sum_{j=1}^{l} \mathcal{H}_j$.  If the time evolution operation $e^{-i\mathcal{H}_jt}$ is relatively easy to apply to a quantum state, for example because it represents a simple sequence of rotation gates, then it would be beneficial to express the time evolution $e^{-i\mathcal{H}t}$ by the Hamiltonian $\mathcal{H}$ as a function of time evolution operators $e^{-i\mathcal{H}_jt}$ of the constituent Hamiltonians $\mathcal{H}_j$.  

In general $e^{-i\mathcal{H}t} \neq \prod_{j=1}^l e^{-i\mathcal{H}_jt}$, but we can follow \cite{suzuki1990fractal,suzuki1991general} and adopt the classical Lie-Trotter formula for exponentiated matrices:
\begin{align}
    e^{-i\mathcal{H}t} = e^{-i\sum_{j=1}^l \mathcal{H}_jt} = \lim_{r\to\infty}\big(\prod_{j=1}^le^{-i\mathcal{H}_jt/r}\big)^r.
\end{align}
For a finite $r$, the product method approximates the time evolution by $\mathcal{H}$ using a sequence of time evolutions over short segments of time $t/r$  by constituent Hamiltonians $\mathcal{H}_j$.  

This method is flexible and robust, but it requires $O(lr)$ queries to the Hamiltonians $\mathcal{H}_j$.  For an $N$-dimensional systems, the number $l$ of simple-to-apply constituent Hamiltonian evolution operators $e^{-i\mathcal{H}_jt}$ can be of $O(N)$. This can be efficient in many situations, but would not yield exponential speedup relative to classical methods.  Additionally, the algorithm scales superlinearly in simulation time $t$, which is above the optimal linear dependence on $t$.  The approach also suffers from poor scaling in the sparseness $O(d^4)$.

\subsection{Hamiltonian Simulation by Quantum Walk}
\label{sec:hs_qw}

\cite{childs2010relationship} proposed a way to simulate Hamiltonian evolution using a quantum walk (Section~\ref{sec:szegedy}).  This method became a part of many efficient and influential algorithms, such as the \cite{berry2015hamiltonian} and \cite{low2019hamiltonian} algorithms, considered the state of the art in Hamiltonian simulation at the time of writing this review.

The method uses two properties of quantum walks.   First, it is possible to construct a quantum walk operator from a Hamiltonian, provided the elements of the Hamiltonian can be encoded in a quantum state.   Second, each eigenvector of the quantum walk unitary corresponds to an eigenvector of the simulated Hamiltonian $\ket{\lambda}$, with eigenvalues $\pm e^{\pm i \arcsin \lambda}$. \cite{childs2010relationship} demonstrates that it is possible to combine these two properties of quantum walks with QPE (Section~\ref{sec:qpe}) to perform Hamiltonian simulation.

Hamiltonian simulation via a quantum walk proceeds in a duplicated Hilbert space.  For a Hamiltonian acting on a space spanned by $N = 2^n$ qubits, first an auxiliary qubit is appended doubling the dimension of the Hilbert space.  Then another register of $n+1$ qubits is appended, expanding the Hilbert space for the quantum walk.   To implement the walk, an operator $T$ is defined that acts on the doubled Hilbert space:
\begin{align}
    T = \sum_{j=0}^{N-1}\sum_{b\in \{0,1\}}(\ketbra{j}\otimes\ketbra{b})\otimes \ket{\varphi_{jb}},
\end{align}
where the first register is the register that spans the Hilbert space of $\mathcal{H}$, the second register is the auxiliary qubit, and the state $\ket{\varphi_{jb}}$ is across the third and fourth registers that duplicate the Hilbert space of $\mathcal{H}$ and the auxiliary qubit.  The state $\ket{\varphi_{jb}}$ encodes the absolute values of the non-zero elements of $\mathcal{H}$ as follows:
\begin{align}
    \ket{\varphi_{j1}} & = \ket{0}\ket{1} \\
    \ket{\varphi_{j0}} & = \frac{1}{\sqrt{d}}\sum_{l\in F_j}\ket{l}\Big(\sqrt{\frac{\mathcal{H}_{jl}^*}{X}}\ket{0} + \sqrt{1-\frac{|\mathcal{H}_{jl}^*|}{X}}\ket{1}\Big), 
\end{align}
where $X \geq \|\mathcal{H}\|_{max}$ and $F_j$ is the set of nonzero elements of $\mathcal{H}$ in column $j$.   

The unitary $U$ corresponding to a Szegedy quantum walk (Section \ref{sec:szegedy}) takes the form
\begin{align}
    U & = iS(2TT^\dagger-\mathbb{I}),
\end{align}
where the swap operator $S$ swaps the two registers $(S\ket{a_1}\ket{a_2}\ket{b_1}\ket{b_2} = \ket{b_1}\ket{b_2}\ket{a_1}\ket{a_2})$ and $\mathbb{I}$ is identity that acts on both registers.  

The random walk unitary $U$ has properties that make it an effective building block of Hamiltonian simulation:  The eigenstates of $U$ are related to eigenstates of $\mathcal{H}$ and the eigenvalues of $U$ are close to the desired prefactor of Hamiltonian simulation $e^{-it\lambda }$:
\begin{align}
    U \ket{\mu_\pm} & = \mu_\pm \ket{\mu_\pm} \\
    \ket{\mu_\pm} & = (T + i\mu_\pm ST)\ket{\lambda}\ket{0} \\
    \mu_\pm & = \pm e^{\pm i\arcsin \frac{\lambda}{\|\mathcal{H}\|_1}}.
\end{align}

\cite{childs2010relationship} was first to propose the use of the quantum walk phase
$e^{\pm it\arcsin \lambda}$ to construct the exponent $e^{-it\lambda}$ that simulates Hamiltonian evolution.  His Hamiltonian simulation algorithms uses QPE (Section~\ref{sec:qpe}) and a special transformation $F_t = e^{-i t \sin \phi} \ket{\theta,\phi}$ to induce with high fidelity the phase $e^{-i\tilde{\lambda}t}$, where $\tilde{\lambda}$ is an approximation of $\lambda$ from QPE.

\cite{berry2015hamiltonian} further optimized the method by using a linear combination of unitaries (LCU) to construct the prefactor $e^{-it\lambda}$ from $e^{\pm it\arcsin \lambda}$.  Instead of relying on QPE and a functional transformation, \cite{berry2015hamiltonian} decompose $e^{-it\lambda}$ into a series of exponents of $e^{\pm it\arcsin \lambda}$:
\begin{align}
    \sum_{m=-\infty}^\infty J_m(z)\mu_\pm^m & = e^{iz\frac{\lambda}{\|\mathcal{H}\|_1}},
\end{align}
where $J_m(z)$ are Bessel functions of the first kind.   As a result, effectively we have:
\begin{align}
    \sum_{m=-\infty}^\infty J_m(-t)U^m & = e^{-it\frac{\mathcal{H}}{\|\mathcal{H}\|_1}}. \label{eq:hamsim_qw_expansion}
\end{align}
The sum in (\ref{eq:hamsim_qw_expansion}) truncated to $|m| \leq k$ is an efficient approximation of the desired phase $e^{-it\frac{\mathcal{H}}{\|\mathcal{H}\|_1}}$.  Success probability for $U^m$ decays with $m$, limiting the efficiency of this algorithm.  As will be discussed in the next section, \cite{low2017optimal} circumvent this problem by proposing an alternative route from the prefactors $e^{\pm it\arcsin \lambda}$ to $e^{-it\lambda}$.

The query complexity of Hamiltonian simulation based on a quantum walk is linear in $t$ and in $d$, the sparseness parameter.  

Hamiltonian simulation is also a way to apply a unitary, since for any unitary there is a corresponding Hamiltonian.  The quantum-walk approach enables the implementation of an $N\times N$ unitary in $\tilde{O}(N^{2/3})$ queries, with typical unitaries requiring $\tilde{O}(\sqrt{N})$ queries \citep{berry2012black}.

\subsection{Linear Combination of Unitaries}
\label{sec:lcu}

The linear combination of unitaries (LCU) method uses a series of controlled unitaries combined with multi-qubit rotations to apply a complex unitary $U$ such that
\begin{align}
    U = \sum_j \beta_j V_j, 
\end{align}
where the weights $\beta_j > 0$.

Let $B$ be a unitary operator such that
\begin{align}
    B\ket{0} = \frac{1}{\sqrt{s}}\sum_j \sqrt{\beta_j}\ket{j},
\end{align}
where $s = \sum_j\beta_j$.

Using the sequence of gates below:

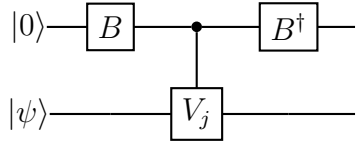
\begin{center}
    \begin{tikzpicture}
        \node[scale=1.0] {
            \begin{quantikz}
                \ket{0} & \gate{B}     & \ctrl{1} & \gate{B^\dagger} & \qw \\
                \ket{\psi} & \qw  & \gate{V_j} & \qw & \qw
            \end{quantikz}
        };
    \end{tikzpicture}
\end{center}

obtain
\begin{align}
    \frac{1}{s}\ket{0}U\ket{\psi}+\sqrt{1-\frac{1}{s^2}}\ket{\Psi^\perp},
\end{align}
where the state $\ket{\Psi^\perp}$ spans the subspace defined by the auxiliary in state $\ket{1}$, i.e. it is orthogonal to any state of the form $\ket{0}\ket{\bullet}$, including $\ket{0}U\ket{\psi}$.

Use oblivious amplitude amplification to amplify the state $\ket{0}U\ket{\psi}$.  Unlike original QAA (Section~\ref{sec:qaa}), oblivious amplitude amplification (developed by \cite{berry2014exponential} based on work by \cite{marriott2005quantum}) provides a way to amplify an a priori unknown state.  

Linear combination of unitaries enables Hamiltonian simulation through expansion of $e^{-i\mathcal{H}t}$, for example as a Taylor series \citep{berry2015simulating} or a series of quantum walk steps \citep{berry2015hamiltonian}.  For the Taylor series expansion, the Hamiltonian is expressed as a linear combination of unitaries $\mathcal{H} = \sum_l \alpha_l\mathcal{H}_l$, where each constituent Hamiltonian $H_l$ is unitary.  Then the Taylor expansion of $e^{-i\mathcal{H}t}$ is a linear combination of unitaries
\begin{align}
    e^{-i\mathcal{H}t} \approx \sum_{k=0}^{K}\frac{(-it)^k}{k!}\alpha_{l_1}\ldots\alpha_{l_k}\mathcal{H}_{l_1}\ldots \mathcal{H}_{l_k}.
\end{align}

Alternatively, \cite{berry2015hamiltonian} approximate the evolution operator $e^{-i\mathcal{H}t}$ as a series of quantum walk steps as shown in (\ref{eq:hamsim_qw_expansion}), where powers of the quantum walk operator $U$ are unitary.

The expansion of the quantum evolution operator $e^{-i\mathcal{H}t}$ as a series of quantum walk steps is one of the most efficient Hamiltonian simulation methods to date, requiring a number of queries proportional to $O(\tau \log (\tau/\epsilon)/\log\log(\tau/\epsilon))$, where $\tau = d\|\mathcal{H}\|_{max}t$, to the input oracles $O_H$ and $O_F$.  The dependence on time is nearly linear, close to the lower bound set by the No Fast-Forwarding theorem \citep{berry2007efficient}.  The dependence of sparsity $d$ is also nearly linear, while the dependence on the simulation error $\epsilon$ is sublinear.  The computational complexity only depends on the dimension of the system indirectly, through the matrix norm of the Hamiltonian $\|\mathcal{H}\|_{max} $.   Additionally, gate complexity of this method -- the number of gates required to implement it -- is only slightly larger than query complexity.  A small disadvantage of the method is that, because it is based on quantum walks, it requires duplicating the register of qubits simulating the quantum systems. It is possible to extend the method to time-dependent Hamiltonians \citep{berry2020time}.

\subsection{Hamiltonian Simulation by Quantum Signal Processing (QSP)}
\label{sec:hs_qsp}

\cite{low2017optimal} proposed an alternative way to perform Hamiltonian simulation using a method they called ``quantum signal processing'' (QSP), because some of their methodologies are analogous to filter design in classical signal processing. (We discuss QSP in greater detail in Section~\ref{sec:qsp}.) The time complexity of the method matches the proved lower bounds for Hamiltonian simulation of sparse Hamiltonians.  The method uses a single auxiliary qubit to encode the eigenvalues of the simulated Hamiltonian, and then transforms these eigenvalues using a sequence of rotation gates applied to the auxiliary qubits.  Using specialized sequences of rotation gates, the method makes it possible to perform polynomial functions of degree $d$ on the input using $O(d)$ elementary unitary operations.  

QSP borrows ideas from quantum control -- the area of quantum computing that deals with extending the life of qubits and improving the fidelity of quantum computation.  \cite{low2017optimal} point out that Hamiltonian simulation is a mapping of a physical system onto a different physical system possible to control more precisely.    Because of this connection between Hamiltonian simulation and quantum control, it is possible to create robust quantum simulations using the QSP framework.

Let $W$ be a quantum walk unitary that encodes the Hamiltonian $\mathcal{H}$ as in Section \ref{sec:hs_qw}.  Consider a function $e^{ih(\theta)} = e^{-it\sin \theta}$, where $\theta = \arcsin \frac{\lambda}{\|\mathcal{H}\|_1}$, so that $e^{ih(\theta)} = e^{it\frac{\lambda}{\|\mathcal{H}\|_1}}$.   QSP provides an efficient way to calculate an finite approximation to the Fourier transform of $e^{ih(\theta)}$.  

\cite{low2017optimal} observe that $e^{ih(\theta)}$ splits into real and imaginary parts, $A(\theta)$ and $C(\theta)$, respectively
\begin{align}
    A(\theta) +i C(\theta) = e^{ih(\theta)} = e^{-it \sin \theta},
\end{align}
and find their Fourier transforms using the Jacobi-Anger expansion
\begin{align}
    \cos(\tau \sin \theta) & = J_0(\tau) + 2 \sum_{k\,\, \text{even} >0}^\infty J_k(\tau) \cos (k\theta) \\
    \sin (\tau \sin \theta) & = 2 \sum_{k\,\,  \text{odd} >0}^\infty J_k(\tau) \sin (k\theta), 
\end{align}
where $J_k(z)$ are Bessel functions of the first kind.  

The method turns out to have query complexity that corresponds multiplicatively to all known lower bounds.

\section{Quantum Optimization}\label{sec:Quantum Optimization}

Optimization is the process of finding $x$ such that $f(x)$, where $f: \mathbb{R}^ \mapsto \mathbb{R}$, is minimized.  Optimization plays an important role in many statistical methods, including most machine learning methods.   A subset of optimization problems -- combinatorial optimization problems of finding an $n$-bit string that satisfies a number of Boolean conditions -- represent some of the most computationally difficult tasks for classical computers to solve.  These problems, which belong to the computational complexity class NP, include practical tasks such as manufacturing scheduling or finding the shortest route for delivery to multiple locations.  To solve NP-hard problems, classical computers need time or memory that scale, in the worst case, exponentially with $n$.  Further, some of the problems belong to the \emph{NP-complete} subset of NP-hard problems.  These problems can be converted into one other in time polynomial in $n$.  An efficient solution to a single NP-complete problem would solve all problems in the computational class NP.

While it is not likely that quantum computers can solve NP-complete problems, they can solve some NP-hard problems (such as prime factorization) and efficiently deliver controlled approximations to some NP-complete problems.  Quantum Approximate Optimization Algorithm (QAOA) is the most famous quantum algorithm for the approximate solution of combinatorial optimization problems.  The relative efficiency of QAOA for a general combinatorial problem compared with the best classical algorithm to solve the problem is not known \citep[see, e.g.][]{zhou2020quantum}; however, we do know that it is not possible to simulate QAOA on a classical computer \citep{farhi2016quantum}. Additionally, QAOA, as well as other hybrid quantum-classical variational algorithms (Section~\ref{sec:variational}) may be robust enough to harness the power of NISQs, the noisy quantum computers available today, and to deliver practical quantum computing advantage in the near term \citep{mcclean2016theory}.

\subsection{Adiabatic Quantum Computing (AQC)}
\label{sec:quant_adiabatic}

As discussed in Section \ref{sec:q-statesandops}, a closed quantum system that evolves according to a Hamiltonian $\mathcal{H}$ with eigenstates $\ket{\lambda}$ and eigenvalues $\lambda$ will evolve from a starting state $\ket{\psi(0)}$ into the state $\ket{\psi(t)} = \sum_\lambda \psi_0e^{-i\lambda t}\ket{\lambda}$, where $\psi_0 = \braket{\lambda}{\psi(0)}$, the inner product of $\ket{\psi(0)}$ and $\ket{\lambda}$.  This property implies that, if the starting state $\ket{\psi(0)}$ is one of the eigenstates of $\mathcal{H}$, $\ket{\lambda}$, $\ket{\psi(0)} = \ket{\lambda}$, then the evolution under the Hamiltonian $\mathcal{H}$ will leave the state unchanged, $\ket{\psi(t)} = e^{-i\lambda}\ket{\lambda} \propto \ket{\psi(0)}$, since the overall phase has no impact on quantum states and is ignored.   This is the essence of the \emph{Adiabatic Theorem} of \cite{born1928beweis}: if the Hamiltonian of the system is changed slowly enough, the system will stay in its \emph{ground state} 
(the eigenstate corresponding to the lowest eigenvalue of the Hamiltonian), provided the \emph{spectral gap} (the gap between the lowest and the second lowest eigenvalues) is maintained throughout the evolution. 

Adiabatic quantum computing (AQC) algorithms \citep{farhi2000quantum} leverage the Adiabatic Theorem.  An AQC algorithm starts in an easy-to-prepare starting state $\ket{S}$, usually the ground state of the Hamiltonian $\mathcal{H}_S$.  The goal of the algorithm is to arrive at the ground state of the ending Hamiltonian $\mathcal{H}_E$.  To transform the initial state $\ket{S}$ into the desired state $\ket{E}$, the system is evolved slowly by applying the unitary evolution operator corresponding to a Hamiltonian $\mathcal{H}(t)$ that slowly transitions from $\mathcal{H}_S$ to $\mathcal{H}_E$ by increasing the weight $\beta(t)$:
\begin{align}
    \mathcal{H}(t) = \mathcal{H}_S (1-\beta(t)) + \mathcal{H}_E \beta(t).
\end{align}
If the spectral gap is preserved throughout the transformation from $\mathcal{H}_S$ to $\mathcal{H}_E$, the starting state evolves into the ground state of $\mathcal{H}_E$.   The squared inverse of the spectral gap bounds the rate at which the Hamiltonian can evolve from $\mathcal{H}_S$ to $\mathcal{H}_E$ which, in turn, bounds the runtime efficiency of the AQC algorithm \citep{reichardt2004quantum}.  \cite{aharonov2008adiabatic} proved that, in the absence of noise and decoherence, adiabatic quantum computation is theoretically equivalent to circuit based computation.

In practice, AQC has both advantages and disadvantages compared with circuit-based computation.  An attractive property of adiabatic quantum computation is that, if the spectral gap is large enough (compared with the inverse of the evolution time $1/t$), the computation is robust even in the presence of noise.  A disadvantage is that, for many systems, there are stringent physical limits on how fast the effective Hamiltonian can change; because of this, the required time for full transition from $\mathcal{H}_S$ to $\mathcal{H}_E$ may be slower than available system coherence time.    

A recent review article by \cite{albash2018adiabatic} provides more detail about AQC.\footnote{A related way of performing optimization is quantum annealing \citep{apolloni1989quantum}.  Here the optimization starts in an arbitrary initial state and then explores the cost function landscape until it finds the ground state of the system.  Quantum annealing is the method deployed by the company D-Wave. }  AQC may contribute most powerfully in chemistry \citep[see, e.g.][]{babbush2014adiabatic,veis2014adiabatic} and may unlock applications in other fields, such as machine learning \citep{neven2008training,denchev2012robust,seddiqi2014adiabatic,potok2021adiabatic} and combinatorial optimization \citep{farhi2000quantum,farhi2001quantum,choi2011different,choi2020effects}.

\subsection{Quantum Approximate Optimization Algorithm (QAOA)}
\label{sec:qaoa}

The Quantum Approximate Optimization Algorithm (QAOA) proposed by \cite{farhi2014quantum} finds computationally-efficient approximations for a class of NP-hard combinatorial optimization problems.  Combinatorial optimization is equivalent to searching for string $z$ of length $n$ that satisfies $m$ clauses.   The objective function is the number of clauses string $z$ satisfies:
\begin{align}
    C(z) = \sum_{k=1}^{m}C_k(z),
\end{align}
where $C_k(z) = 1$ if $z$ satisfies clause $k$ and $0$ if it does not.  The algorithm finds a string $z$ -- among the $2^n$ possible $n$-bit strings -- such that $C(z)$ is close to $\text{max}_z C(z)$.

The core idea of the algorithm is that it is possible to create a parameterized quantum superposition of $2^n$ states that represent all possible binary strings $z$ such that the expectation of the objective function $C$ for this state is maximized for a set of parameters.   The quantum state corresponding to the maximum expectation of $C$ contains the approximate solution to the combinatorial optimization with a high probability.

The algorithm starts with a uniform superposition of all possible $n$-bit binary strings $z$ in the computational basis
\begin{align}
    \ket{s} = \frac{1}{\sqrt{2^n}}\sum_z \ket{z}.
\end{align}

Two types of parameterized unitary operators are applied to this state.  The first unitary operator has the form
\begin{align}
    U(C,\gamma) = e^{-i\gamma C} = \prod_{k=1}^m e^{-i\gamma C_k},
\end{align}
where $\gamma$ is a scalar parameter, $\gamma \in [0,2\pi)$.
The second unitary operator is based on an operator of the form
\begin{align}
    B = \sum_{j=1}^n \sigma_j^x,
\end{align}
where $\sigma_j^x$ is a Pauli operator (Section~\ref{sec:q-statesandops}) applied to qubit $j$, $j = 1,..,n$.  The unitary operator $U(B,\beta)$ equals
\begin{align}
    U(B,\beta) = e^{-i\beta B} = \prod_{j=1}^{n}e^{-i\beta \sigma^x_j},
\end{align}
where $\beta \in [0,\pi)$.\footnote{The uniform state $\ket{s}$ is the highest-eigenvalue eigenstate of $B$.} 

Alternating application of the parameterized unitary operators $U(C,\gamma)$ and $U(B,\beta)$ to the initial state $\ket{s}$ creates a parameterized quantum state
\begin{align}
    \ket{\bm{\gamma},\bm{\beta}} = U(B,\beta_p)U(C,\gamma_p) \ldots U(B,\beta_1)U(C,\gamma_1) \ket{s},
\end{align}
where $\bm{\gamma} = (\gamma_1,...\gamma_p)$ and $\bm{\beta} = (\gamma_1,...\gamma_p)$ are parameter vectors.  The parameter $p$ reflects the resulting quantum circuit depth, which, as \cite{farhi2014quantum} show, controls the quality of the approximation, converging to the exact result in the limit $p \to \infty$.  In effect, QAOA applies Product Formula (Section \ref{sec:productformula}) to the AQC algorithm (Section \ref{sec:quant_adiabatic}).

The optimal approximation is the quantum state $ \ket{\bm{\gamma}^*,\bm{\beta}^*}$ maximizing the expectation value of the objective function $C$ 
\begin{align}
    \argmax_{\bm{\gamma},\bm{\beta}} \bra{\bm{\gamma},\bm{\beta}} C \ket{\bm{\gamma},\bm{\beta}}. 
\end{align}
The maximization proceeds in a classical outer loop using gradient descent, Nelder-Mead, or other methods.   At step $t$, the classical computer controls parameterized unitary gates to create the state $\ket{\bm{\gamma}^{(t)},\bm{\beta}^{(t)}}$ and collects results of measuring $C$ in this state.  The results of the measurements of $C$ feed into the parameter update $(\bm{\gamma}^{(t)},\bm{\beta}^{(t)}) \mapsto (\bm{\gamma}^{(t+1)},\bm{\beta}^{(t+1)})$.  The unitary gates, updated with the new set of parameters and applied to the reset initial state $\ket{s}$, create the updated quantum state $\ket{\bm{\gamma}^{(t+1)},\bm{\beta}^{(t+1)}}$.  The algorithm stops when a stopping criterion, such as an increase in the expectation value of $C$ below a given threshold, is reached.\footnote{\cite{marsh2020combinatorial} observe that QAOA is a form of a quantum walk with phase shifts and use this observation to generalize the algorithm.}

Because the objective function $C$ is defined in general terms, the algorithm lends itself to a range of combinatorial approximation problems.   \cite{farhi2014quantum} use QAOA to find an approximate solution to MaxCut, the problem of cutting a graph into two parts in a way that maximizes the reduction in the cost function.   Proposed practical industrial applications include finance \citep{fernandez2020hybrid} and wireless scheduling \citep{choi2020quantum}.

\subsection{Hybrid Quantum-Classical Variational Algorithms}
\label{sec:variational}

Hybrid quantum-classical variational algorithms introduced in Sections~\ref{sec:q-statesandops} (subsection \emph{Observables}) and \ref{sec:algos_overview} play to the strengths of both quantum and classical computers.   The structure of hybrid quantum-classical variational algorithms is similar to that of QAOA: A quantum state encodes the variational (i.e.~trial) solution; a quantum observable represents the cost function \citep{rebentrost2018quantum,mitarai2018quantum,zoufal2019quantum,schuld2020circuit,zoufal2020variational}.  A classical computer stores and updates variational parameters, which it uses to classically control the quantum gates used create the variational quantum state.  The expectation value of measurements of the quantum observable in the variational quantum state is the cost associated with the set of variational parameters that define the state.   The classical computer collects the results of these measurements and uses them to update variational parameters.   The classical computer then uses the updated variational parameters to reset the quantum gates used to prepare the next iteration of the variational quantum state.  In some cases, direct measurements of the gradient of the cost function can improve convergence of hybrid variational methods \citep{harrow2019low}.  

\subsection{Quantum Gradient Descent}

An important method in classical machine learning is gradient descent (and stochastic gradient descent) -- a popular way for training statistical models by optimizing a loss function.  When gradient descent is used for training quantum neural networks, it is often performed on a classical computer.  
The reason is that it is difficult to iterate on a quantum computer.  The vast majority of quantum computing algorithms rely on postselection (see Section~\ref{sec:gates}) to produce the desired result.  Postselection yields the desired quantum state with a fractional probability (often around 1/2); the rest of the time it yields an incorrect state that has to be discarded.  Therefore, in order to have a sufficient number of desired quantum states available at the end of the iterative process, multiple copies of the initial state have to be created, with the number of copies increasing exponentially with the number of expected iterative steps.

For optimization where the number of expected iterations is small, quantum algorithms with provable speedups have been proposed.  These algorithms work for specific simple forms of the loss function.  How to extend these algorithms to alternative classes of loss functions remains an open area for future research. The most general quantum gradient descent work is by \cite{rebentrost2019quantum} who propose a quantum algorithm to find the minimum of a homogeneous polynomial using gradient descent and Newton's methods.  The proposed quantum algorithm leverages the matrix exponentiation method used in Quantum PCA (Section~\ref{sec:qpca}), followed by Quantum Phase Estimation (Section~\ref{sec:qpe}) and a controlled rotation (Sections~\ref{sec:gates} and \ref{sec:hermitian}) of an auxiliary qubit to achieve each parameter update step.    The shortcoming of this method is that it requires, on average, the destruction of approximately three sets of quantum states for each updating steps (this number can be reduced to around two with an optimized sequence of steps), and is therefore only appropriate for approximate minimization with a small number of iterations.

\cite{sweke2020stochastic} consider stochastic gradient descent for hybrid quantum-classical optimization (Section~\ref{sec:variational}).  They argue that to obtain an unbiased estimator of an integral over a probability distribution, it is sufficient to prepare a corresponding quantum state and take a measurement, repeating the process $k$ times, where $k$, in some cases, can be as low as 1.  Where the gradient can be expressed as a linear combination of expectation values, a ``doubly-stochastic'' gradient descent is possible.  The paper considers cases where the cost function can be expressed as an observable that can be readily measured, such as energy (i.e.~the expectation value of the Hamiltonian of the system) in Variational Quantum Eigensolver \citep{peruzzo2014variational}.  
\cite{stokes2020quantum} propose a quantum algorithm to estimate the natural gradient for cost functions expressed by a block-diagonal Hamiltonian in a space spanned by parameterized unitary gates.  

\section{Quantum Eigenvalue and Singular Value Transformations}
\label{sec:qsvt}

Quantum Singular Value Transformation (QSVT) by \cite{gilyen2019quantum}, a generalization of Quantum Signal Processing (QSP) by \cite{low2017optimal}, has recently emerged as a unifying framework, encompassing all the major families of quantum algorithms as specific instances \citep{martyn2021grand}.   The algorithms leverage the fact that transformations of quantum subsystems can be non-linear even though the transformations of closed quantum systems have to be linear -- more specifically, unitary -- and reversible.

Let $\mathcal{H}$ be a Hamiltonian acting on a $2^n$-dimensional space spanned by a register of $n$ qubits.   Let $\ket{\lambda}$ be the $n$-qubit eigenstates of the Hamiltonian $\mathcal{H}$ corresponding to eigenvalues $\lambda$: $\mathcal{H}\ket{\lambda} = \lambda\ket{\lambda}$.  QSVT enables us to perform a polynomial transformation of the eigenvalues of $\mathcal{H}$:
\begin{align}
    Poly (\mathcal{H}) = \sum_\lambda Poly(\lambda)\ketbra{\lambda},
\end{align}
assuming the operator norm of the Hamiltonian obeys $\|H\| \leq 1$ to enable block-encoding of the system in a larger system.  More generally, QSVT enables polynomial transformations of singular values of a general (non-Hermitian) matrix $A$.

In this section, we introduce QSP (Section~\ref{sec:qsp}), review QSVT for Hermitian matrices (Quantum Eigenvalue Transformation in Section~\ref{sec:qet}), and describe the general form of QSVT (Section~\ref{sec:qsvt_sec}) and its universality (Section~\ref{sec:universality}) as a framework for other quantum algorithms.

\subsection{Quantum Signal Processing (QSP)}
\label{sec:qsp}

Quantum Signal Processing (QSP) is a method to enact polynomial transformations of a ``signal'' $x$ (assuming $x \in [-1,1]$) embedded in a unitary $W$ acting on a single qubit.  QSP transformations proceed as sequences of simple rotations of the qubit followed by post-selection (Section~\ref{sec:gates}) to perform polynomial transformations of $x$. To develop QSP, \cite{low2016methodology} drew inspiration from the signal processing methods of nuclear magnetic resonance (NMR) -- a powerful and important technology widely used in medicine, chemistry, petroleum industry, materials science, and physics.   

A QSP algorithm has four components.  The first component is the \emph{signal unitary} $W(x)$ -- the unitary that encodes the signal $x$ to be transformed.  The second component is the \emph{signal processing unitary} $S(\phi_j)$, usually a simple rotation by an angle $\phi_j$, an element of a tuple $\vec{\phi} = (\phi_0,...,\phi_d)$.  The algorithm proceeds as a sequence of alternating unitaries $S(\phi_j)$ and $W(x)$.  The third component is the sequence of rotational angles in the tuple $\vec{\phi}$, which determines what polynomial transformation the ``signal'' $x$ undergoes.  The forth component
is the \emph{signal basis}, e.g.~$\{\ket{+},\ket{-}\}$, used to perform a measurement at the end of the algorithm.

Let $W$ be the \emph{signal unitary} acting on a qubit or a more general two-state quantum system (we will use the generalization in the next section):
\begin{align}
    W(x) & = \begin{bmatrix} x & i\sqrt{1-x^2} \\ i\sqrt{1-x^2} & x \end{bmatrix} = e^{iX\arccos{(x)}}, \label{eq:w}
\end{align}
where $X$ is a Pauli matrix and $e^{iX\arccos{(x)}}$ represents rotation about the $x$ axis by angle $2\arccos{(x)}$.  Note that encoding of $x$ in a unitary can take multiple forms, and the form in (\ref{eq:w}) represents a specific choice, convenient for discussion of Quantum Eigenvalue Transformation and QSVT and their applications in the following sections \citep{martyn2021grand}.

Let $S(\phi)$ be a \emph{signal processing} operator.   A convenient choice is 
\begin{align}
    S(\phi) = e^{i\phi Z},
\end{align}
where $Z$ is a Pauli matrix and $e^{i\phi Z} $ represents rotation about the $z$ axis by angle $2\phi$.  In principle, $S(\phi)$ can take other forms, as long as $ S(\phi)$ does not commute with $W(x)$.  

For a specific choice of a $d$-dimensional parameter vector $\vec{\phi} = (\phi_0,\phi_1,..\phi_d)$, a series of alternating applications of $S(\phi_j)$ and $W(x)$ result in a polynomial transformation of the signal $x$:
\begin{align}
    U_{\vec{\phi}}(x) = S(\phi_0)\prod_{j=1}^dW(x)S(\phi_j) = \begin{bmatrix} P(x) & i Q(x)\sqrt{1-x^2} \\ i Q^*(x)\sqrt{1-x^2} & P^*(x) \end{bmatrix},\label{eq:qsp}
\end{align}
where $P(x)$ and $Q(x)$ are complex polynomials of degree less than or equal to $d$ and $d-1$, and with parity of $(d\mod 2)$ and $(d-1 \mod 2)$, respectively.  The polynomials satisfy $|P(x)|^2+(1-x^2)|Q(x)|^2=1$ for all $x\in [-1,1]$.  Importantly, given polynomials $P(x)$ and $Q(x)$ that satisfy the aforementioned conditions, it is possible to find $\vec{\phi}$ that satisfies (\ref{eq:qsp}).

The polynomial $P(x)$ determines the probability $p$ that the state $\ket{0}$ stays unchanged under the operation $U_{\vec{\phi}}(x)$, 
$P(x) = \bra{0}U_{\vec{\phi}}(x)\ket{0}$ and $p = |P(x)|^2$.   The choice of $\vec{\phi}$ determines the polynomial.  For example, the choice $\vec{\phi} = (0,0)$ results in $P(a) = a$; $\vec{\phi} = (0,0,0)$ results in $P(a) = 2a^2-1$, and so on, with $\vec{\phi} = (0,0,\ldots,0)$ in $d$ dimensions yielding Chebyshev's polynomials of the first kind, $T_d(a)$.

For greater expressiveness of $U_{\vec{\phi}}(x) $, consider the matrix element $\bra{+}U_{\vec{\phi}}(x)\ket{+}$, where $\ket{+} = \frac{1}{\sqrt{2}}(\ket{0}+\ket{1})$ is a state in the Hadamard basis, chosen in this case as the \emph{signal basis}.  The state $\ket{+}$ plays the role of the \emph{reference state}.  Post-selection on the reference state yields a polynomial transformation of $x$ 
\begin{align}
    \bra{+}U_{\vec{\phi}}(x)\ket{+} = \Re (P(x)) + i \Re(Q(x))\sqrt{1-x^2}.  
\end{align}

Post-selection relative to the signal basis is one of the critical steps of QSP.  After all, we are able to perform a non-linear quantum transformation \emph{because} it applies to a subsystem of a larger system.  Post-selection is a way to extract the smaller system from the larger system at the end of the computation.

\subsection{Quantum Eigenvalue Transformation}
\label{sec:qet}

\cite{gilyen2018quantum} generalized QSP (Section~\ref{sec:qsp}) to perform polynomial transformations of matrices rather than scalar signals.   They combined QSP with \emph{block encoding} and \emph{qubitization}.

\emph{Block encoding} is a way to embed a linear non-unitary operator acting on a quantum system inside a unitary operator acting on a larger system that contains the smaller quantum system.   A simple way to embed a quantum system inside a larger system is to append an auxiliary qubit.   Consider a quantum system on $n$ qubits.  Its Hilbert space is $\mathbb{C}^{2^n}$.  Appending the auxiliary qubit doubles the Hilbert space to $\mathbb{C}^{2^{n+1}}$.  If the operator to be embedded $E$ is unitary operator, then a simple block-encoding of $E$ is the control-$E$ operator, where $E$ applies to the register of $n$ qubits conditional on the state of the auxiliary qubit.

Consider a Hermitian operator $\mathcal{H}$ acting on the quantum system of $n$ qubits.  It is possible to embed this Hamiltonian in a unitary $U$ acting on the expanded system of the $n$ qubits plus the auxiliary.   For example, let $U$ take the form
\begin{align}
    U = I \otimes \mathcal{H} + iX \otimes \sqrt{1-\mathcal{H}^2}, \label{eq:embed_u}
\end{align}
where the Pauli operators $Z$ and $X$ act on the auxiliary qubit and the Hermitian operators $\mathcal{H}$ and $\sqrt{1-\mathcal{H}^2}$ act on the $n$-qubit register.  The eigenstates of  $\mathcal{H}$ are $n$-qubit states $\ket{\lambda}$ with eigenvalues $\lambda$:  $\mathcal{H}\ket{\lambda}=\lambda\ket{\lambda}$.   The operator $\sqrt{1-\mathcal{H}^2}$ is Hermitian and shares its eigenstates with the Hamiltonian $\mathcal{H}$; it is easy to demonstrate using Taylor expansion that $ \sqrt{1-\mathcal{H}^2} \ket{\lambda} = \sqrt{1-\lambda^2} \ket{\lambda}$.  As a consequence, the operator $U$ is unitary, $U^\dagger U = UU^\dagger = I$.\footnote{Note that $U$ is not a unique way to embed the Hermitian operator $\mathcal{H}$ in a unitary operator acting on a larger system.}

\emph{Qubitization} is the reduction of a multi-qubit state to a two-state qubit-like system.  For example, qubitization is at the core of Grover's search algorithm (Section~\ref{sec:grover_section}), where the ``good'' state and its complement effectively act as a two-state system.  They span a (two-dimensional) plane, invariant under Grover iterations.  

Consider the application of the unitary $U$ from (\ref{eq:embed_u}) to states of the extended system $\ket{0}\ket{\lambda}$ and $\ket{1}\ket{\lambda}$, where $ \ket{0}$ and $\ket{1} $ represent the states of the auxiliary qubit:
\begin{align}
    U\ket{0}\ket{\lambda} & = \lambda \ket{0}\ket{\lambda} + i\sqrt{1-\lambda^2}\ket{1}\ket{\lambda} \\
    U\ket{1}\ket{\lambda} & = \lambda \ket{1}\ket{\lambda} + i\sqrt{1-\lambda^2}\ket{0}\ket{\lambda}.    
\end{align}

The space spanned by $\ket{0}\ket{\lambda}$ and $\ket{1}\ket{\lambda}$ is closed under $U$ and acts as a two-state system.   The unitary $U$ has the effect of a rotation on each subspace that corresponds to the eigenstate of $\mathcal{H}$ $\ket{\lambda}$:
\begin{align}
    U & = \bigoplus_\lambda \begin{bmatrix} \lambda & i\sqrt{1-\lambda^2} \\i\sqrt{1-\lambda^2} & \lambda \end{bmatrix} \otimes \ketbra{\lambda} \\
    & = \bigoplus_\lambda e^{iX\arccos{(\lambda)}} \otimes \ketbra{\lambda}, \label{eq:embed_rot_u}
\end{align}
where $e^{iX\arccos{(\lambda)}} $ is effectively a rotation by $ 2\arccos{(\lambda)}$ about the $x$-axis of the Bloch sphere defined by $\ket{0}\ket{\lambda}$ and $\ket{1}\ket{\lambda}$ for each $\lambda$.  In each subspace spanned by $\ket{0}\ket{\lambda}$ and $\ket{1}\ket{\lambda}$, the unitary operator $U$ acts as the \emph{signal unitary} in QSP, encoding the eigenvalue $\lambda$ as the signal.

By analogy with the unitary $U$ that extends the idea of a signal unitary to a multi-qubit state, we extend the signal processing operator.  The signal processing operator is independent of the signals $\lambda$ and acts on the auxiliary qubit identically for each subspace spanned by $\ket{0}\ket{\lambda}$ and $\ket{1}\ket{\lambda}$.   Using the identity operator $I = \sum_\lambda \ketbra{\lambda}$, we can express the extended signal processing operator as
\begin{align}
    \Pi_\phi = \bigoplus_\lambda e^{i\phi Z}\otimes \ketbra{\lambda},
\end{align}
where $Z$ is a Pauli operator and the rotation $e^{i\phi Z}$ acts on the auxiliary qubit.

\cite{gilyen2018quantum} demonstrate that the sequence of alternating extended signal unitaries and signal processing operators results in an embedded polynomial transformation of the Hermitian operator $\mathcal{H}$.  For an even $d$:
\begin{equation}
    U_{\vec{\phi}}  = \Big[\prod_{k=1}^{d/2}\Pi_{\phi_{2k-1}}U^\dagger \Pi_{\phi_{2k}}U\Big] = \begin{bmatrix} Poly(\mathcal{H}) & \cdot \\ \cdot & \cdot \end{bmatrix},
\end{equation}
and similarly for an odd $d$:
\begin{equation}
    U_{\vec{\phi}} = \Pi_{\phi_1}U\Big[\prod_{k=1}^{(d-1)/2}\Pi_{\phi_{2k}}U^\dagger \Pi_{\phi_{2k+1}}U\Big] = \begin{bmatrix} Poly(\mathcal{H}) & \cdot \\ \cdot & \cdot \end{bmatrix},
\end{equation}
where, postselecting on the auxiliary qubit ending in state $\ket{0}$ (if it started in state $\ket{0}$), we obtain the desired polynomial transformation $ Poly(\mathcal{H})$
\begin{align}
    Poly (\mathcal{H}) = \sum_\lambda Poly(\lambda)\ketbra{\lambda}.\label{eq:qsvt_transform}
\end{align}
The resulting polynomial of $ \mathcal{H}$ is of order less than or equal to $d$.  Its exact form depends on the sequence of signal processing rotation angles $\vec{\phi}$, which can be computed efficiently using a version of the classical Remez exchange algorithm \citep{low2016methodology,martyn2021grand}.

\subsection{Quantum Singular Value Transformation (QSVT)}
\label{sec:qsvt_sec}

In the previous section we considered the polynomial transformation of a Hermitian operator.  For non-Hermitian linear operators, it is possible to perform an analogous polynomial singular value transformation.  

Consider a non-Hermitian operator $A$, represented by a rectangular matrix.  Any general rectangular matrix can be decomposed into a diagonal matrix of singular values $\Sigma$ and unitary matrices $W_\Sigma$ and $V_\Sigma$:
\begin{align}
    A & = W_\Sigma \Sigma V_\Sigma^\dagger.
\end{align}
The matrix $\Sigma$ contains $r$ non-negative real singular values of matrix $A$ along the diagonal.   The columns of matrices $W_\Sigma$ and $V_\Sigma $ form orthonormal bases composed of left and right singular vectors respectively.  In quantum notation we denote the left singular vectors $\{\ket{w_k}\}$ and right singular vectors $\{\ket{v_k}\}$ and express the matrix $A$ as
\begin{align}
    A & = \sum_{k=1}^r \sigma_k\ketbra{w_k}{v_k}.
\end{align}
Using a quantum computer, we can efficiently perform polynomial singular value transformation of $A$:
\begin{align}
    Poly^{(SV)}(A) = \sum_k Poly(\sigma_k)\ketbra{w_k}{v_k}.
\end{align}

The matrix $A$ can be embedded in a unitary matrix that applies to a larger quantum system similarly to the way we embedded the Hermitian matrix $\mathcal{H}$ in Section~\ref{sec:qet}.   Even though the Hilbert spaces spanned by $\{\ket{w_k}\}$ and $\{\ket{v_k}\}$ in general have different dimensions, we can create an extended quantum state by appending a single auxiliary qubit.  In this case, the Hilbert spaces spanned by $\{\ket{w_k}\}$ and $\{\ket{v_k}\}$ would be encoded on the same register of $n$ qubits, where $n$ is large enough to hold the larger of the spaces.  

The extended signal unitary $U$ then takes the form
\begin{align}
    U = \bigoplus_k \begin{bmatrix} \sigma_k & i\sqrt{1-\sigma_k^2} \\ i\sqrt{1-\sigma_k^2} & \sigma_k \end{bmatrix} \otimes \ketbra{w_k}{v_k}.
\end{align}

When the input and output spaces of $A$ are different, we have two signal processing operators
\begin{align}
    \Pi_\phi & = \bigoplus_k e^{i\phi Z} \otimes \ketbra{v_k} \\
    \tilde{\Pi}_\phi & = \bigoplus_k e^{i\phi Z} \otimes\ketbra{w_k}.
\end{align}

With these definitions we have an expression analogous to the quantum eigenvalue transformation (Section~\ref{sec:qet}), e.g.~for an even $d$:
\begin{equation}
    U_{\vec{\phi}}  = \Big[\prod_{k=1}^{d/2}\Pi_{\phi_{2k-1}}U^\dagger \tilde{\Pi}_{\phi_{2k}}U\Big] = \begin{bmatrix} Poly^{(SV)}(A) & \cdot \\ \cdot & \cdot \end{bmatrix},
\end{equation}
where the polynomial transformation $Poly^{(SV)}(A)$ is postselected on the auxiliary qubit in the state $\ket{0}$ after the unitary $U_{\vec{\phi}}$ is applied.

\subsection{QSVT and the ``Grand Unification'' of Quantum Algorithms}
\label{sec:universality}

Quantum Eigenvalue Transformation and its generalization, the Quantum Singular Value Transformation, are powerful and expressive ways to perform a wide range of operations on a quantum computer.  For some quantum problems, such as Hamiltonian simulation (Section~\ref{sec:Hamiltonian Simulation}), QSVT provides the most efficient known algorithm to date, nearly optimal relative to known lower bounds.

\cite{martyn2021grand} point out that the QSVT framework unifies all the existing quantum algorithms.  After all, a quantum algorithm is a transformation of inputs -- linear or non-linear, dimension-preserving or not.   The QSVT framework provides a unified way to encode any matrix transformation that has a polynomial expansion, and the algorithm itself is encoded in a sequence of real numbers -- the phases in the tuple $\vec{\phi}$.

Consider, for example, the search problem described in Section~\ref{sec:grover_section}.  The problem has two natural singular vectors:  the starting state $\ket{\psi_0}$ and the target state $\ket{x_0}$.   The objective is to transform the small inner product between the starting and ending state $c = \braket{x_0}{\psi_0}$ into a scalar of order unity $\bra{x_0}U\ket{\psi_0} = O(1)$.  \cite{yoder2014fixed} demonstrate that a sequence of pulse sequences can be tuned to achieve this transformation efficiently, while avoiding the Grover algorithm's ``souffl\'{e} problem'' -- the fact that the approximation error in the Grover algorithm is periodic and, if the minimum error is missed, it starts rising with every iteration until it peaks at $O(1)$ and starts decreasing again. The method proposed by \cite{yoder2014fixed}, called fixed-point quantum search, generalizes reflections on a plane of Grover's algorithm to rotations on a three-dimensional Block sphere using the framework of QSP and QSVT.  The algorithm converges in $O(\sqrt{N})$ steps and is optimal.  

The Hamiltonian simulation problem is equivalent to matrix exponentiation.  Given a Hamiltonian $\mathcal{H}$ we seek to apply $e^{-i\mathcal{H}t}$ which is equivalent to applying the sum $e^{-i\mathcal{H}t} = \cos{(\mathcal{H}t)} - i\sin{(\mathcal{H}t)}$.  The trigonometric functions $ \cos{(\mathcal{H}t)}$ and $\sin{(\mathcal{H}t)}$ have polynomial expansions, called Jacobi-Anger expansions, and therefore it is possible to cast the Hamiltonian simulation problem as a sum of two Quantum Eigenvalue Transformations.  

QSVT provides an efficient way to solve quantum linear systems problem (Section~\ref{sec:hhl}) $A\ket{x} = \ket{b}$.   In the original QLSA algorithm, \cite{harrow2009quantum} used Quantum Phase Estimation to extract singular values of $A$ (or its eigenvalues, if $A$ is Hermitian) and a controlled rotation to invert these values.  The QSVT framework does not require the explicit extraction of the singular values of $A$.   The singular value inversion is performed in a single step using a polynomial approximation to $A^{-1}$.   Assuming the singular values of $A$ are bounded, e.g.~by $1/\kappa$ from below, it is possible to construct a polynomial expansion of $A^{-1}$, as shown in \cite{martyn2021grand}, Appendix C. The query complexity of the resulting algorithm is $O(\kappa \log (\kappa/\epsilon))$; remarkably, it has no dependence on $N$, the dimension of $A$, and has a logarithmic dependence on error.  (It is important to note, however, that the block encoding of $A$ may require a number of operations that scales as a function of $N$.)

\cite{martyn2021grand} demonstrate how to apply the QSVT framework to prime factorization, phase estimation, and eigenvalue thresholding;  \cite{gilyen2018quantum} apply QSVT to Gibbs sampling, quantum walks, and the computation of machine learning primitives.  A wide variety of quantum matrix functions and quantum channel discrimination methods are being developed using QSVT.  Additionally, because QSVT is based on quantum control techniques which are also used in error correction, it is possible to create naturally robust algorithms using the QSVT frameworks.   For example, embedding a matrix in a large unitary may enable algorithm designers to ``push'' errors out into the outer blocks of the matrix -- similarly to the way deep networks distribute errors through unimportant dimensions of an overparameterized model \citep{bartlett2020benign}.  

The QSVT framework distills the great variety of quantum algorithms to a (finite) string or real numbers -- the auxiliary rotation phases $\vec{\phi}$.  All the algorithms follow the same circuit -- composed of alternating block-encoded unitaries and auxiliary rotations which, depending on the sequence of auxiliary rotations, transform the singular values by a nearly arbitrary polynomial.  The specific sequence of auxiliary rotations delivers the multitude of useful outcomes.  

The QSVT framework is powerful, but, like other quantum algorithmic frameworks, it is not without challenges.   It is not trivial to design efficient block encodings -- even for the simplest systems like harmonic oscillators. Computationally efficient methods to determine the sequence of phases $\vec{\phi}$ corresponding to a given polynomial are also an open challenge.  Nevertheless, QSVT is a significant advance towards practical, actionable quantum algorithms.

\section{Conclusion}\label{sec:conclusion}

The intersection of quantum computing and statistics has delivered -- and is likely to continue to deliver -- powerful breakthroughs that unlock interesting and practical applications.  Consider, for example, advances in quantum MCMC, in particular quantum Metropolis-Hastings.  Speeding up the Metropolis-Hastings algorithm is critical for many important practical applications but, because of its mathematical structure, the method does not generally apply to high-dimensional statistical models. For such cases, Sequential Monte Carlo (SMC) can provide an attractive alternative to MCMC.  Quantum SMC is a challenging, but potentially highly impactful research question,  which remains an open question at the time of writing.

In the classical world, Variational Bayes stands out as a computationally attractive alternative to MCMC for Bayesian computation in big model and big data settings. 
\cite{lopatnikova.tran:2021:quantum} propose a Variational Bayes method based on quantum natural gradient, which can be implemented on a quantum-classical device. How to implement a quantum Variational Bayes approach entirely on a quantum computer is an interesting research question.

Potential advances stem not just from applying quantum algorithms to machine learning, but also from borrowing insights the other way around -- from statistics and machine learning to quantum computing.  For example, as discussed in Section \ref{sec:readout}, reading out a quantum state that encodes the result of a quantum computation might require too many measurements offseting quantum efficiency.  When the result is a quantum sample state, we can interpret it as a probability distribution.   We can then adopt the idea of normalizing flow in machine learning \citep{Papamakarios:2021} and transform the quantum sample state into a new manageable quantum sample state, e.g., the uniform state, using parameterized quantum circuits with the parameters trained in classical outer-loop. The method would allow us to create as many (approximate) copies of the original state as needed, via the inverse transformation, for use in quantum tomography. 
One important area that was not reviewed in this article is quantum-inspired computation, such as quantum-inspired linear algebra \citep{gilyen2018quantum,chia2020sampling}. Quantum-inspired algorithms work on classical computers, but are designed based on quantum-inspired ideas and
can still offer significant speed-ups. We leave this topic for a future work.

\bibliographystyle{apalike}
\bibliography{QuantumAndMLReferences.bib}


\end{document}